\documentclass[twocolumn, floatfix, tighten, fleqn, useAMS, usenatbib]{aastex63}
\acceptjournal{ApJ}

\RequirePackage{silence}
\usepackage{lineno}
\usepackage{float}
\usepackage{graphicx}	
\usepackage{amsmath}	
\usepackage{amssymb}	
\usepackage{amsfonts}
\usepackage{bm}		
\usepackage{mathtools}
\usepackage{booktabs}
\usepackage{listings}
\usepackage{color}
\usepackage{threeparttable}
\usepackage{multirow}
\usepackage[perpage,symbol*,flushmargin]{footmisc}
\usepackage{ae,aecompl}
\usepackage{array}
\usepackage{soul}

\usepackage[T1]{fontenc}
\usepackage[perpage,symbol*]{footmisc}
\usepackage{ae,aecompl}

\let\oldtextbf=\textbf
\renewcommand\textbf[1]{{\boldmath\oldtextbf{#1}}}

\begin{document}

\title{The High Fraction of Thin Disk Galaxies Continues to Challenge $\Lambda$CDM Cosmology}
\shorttitle{Galaxy shape distribution in $\Lambda$CDM and reality} 

\shortauthors{M. Haslbauer, I. Banik, P. Kroupa, N. Wittenburg \& B. Javanmardi} 

\author[0000-0002-5101-6366]{Moritz Haslbauer}
\affiliation{Helmholtz-Institut f\"ur Strahlen- und Kernphysik (HISKP), University of Bonn, Nussallee 14$-$16, D-53115 Bonn, Germany}
\affiliation{Max-Planck-Institut f\"ur Radioastronomie, Auf dem H\"ugel 69, D-53121 Bonn, Germany}
\email{mhaslbauer@astro.uni-bonn.de}

\author[0000-0002-4123-7325]{Indranil Banik}
\affiliation{Scottish Universities Physics Alliance, University of Saint Andrews, North Haugh, Saint Andrews, Fife, KY16 9SS, UK}
\affiliation{Helmholtz-Institut f\"ur Strahlen- und Kernphysik (HISKP), University of Bonn, Nussallee 14$-$16, D-53115 Bonn, Germany}
\email{ib45@st-andrews.ac.uk}

\author[0000-0002-7301-3377]{Pavel Kroupa}
\affiliation{Helmholtz-Institut f\"ur Strahlen und Kernphysik (HISKP), University of Bonn, Nussallee 14$-$16, D-53115 Bonn, Germany}
\affiliation{Charles University in Prague, Faculty of Mathematics and Physics, Astronomical Institute, V Hole\v{s}ovi\v{c}k\'ach 2,\\ CZ-180 00 Praha 8, Czech Republic}

\author[0000-0001-9332-0000]{Nils Wittenburg}
\affiliation{Helmholtz-Institut f\"ur Strahlen und Kernphysik (HISKP), University of Bonn, Nussallee 14$-$16, D-53115 Bonn, Germany} 
	
\author[0000-0002-9317-6114]{Behnam Javanmardi}
\affiliation{Institute for Genomic Statistics and Bioinformatics, University of Bonn, Venusberg-Campus 1, D-53127 Bonn, Germany}

\begin{abstract} 

Any viable cosmological framework has to match the observed proportion of early- and late-type galaxies. In this contribution, we focus on the distribution of galaxy morphological types in the standard model of cosmology (Lambda cold dark matter, $\Lambda$CDM). Using the latest state-of-the-art cosmological $\Lambda$CDM simulations known as Illustris, IllustrisTNG, and EAGLE, we calculate the intrinsic and sky-projected aspect ratio distribution of the stars in subhalos with stellar mass $M_* > 10^{10}\,M_\odot$ at redshift $z=0$. There is a significant deficit of intrinsically thin disk galaxies, which however comprise most of the locally observed galaxy population. Consequently, the sky-projected aspect ratio distribution produced by these $\Lambda$CDM simulations disagrees with the Galaxy And Mass Assembly (GAMA) survey and Sloan Digital Sky Survey at $\geq 12.52\sigma$ (TNG50-1) and $\geq 14.82\sigma$ (EAGLE50) confidence. The deficit of intrinsically thin galaxies could be due to a much less hierarchical merger-driven build-up of observed galaxies than is given by the $\Lambda$CDM framework. It might also arise from the implemented sub-grid models, or from the limited resolution of the above-mentioned hydrodynamical simulations. We estimate that an $8^5$ times better mass resolution realization than TNG50-1 would reduce the tension with GAMA to the $5.58\sigma$ level. Finally, we show that galaxies with fewer major mergers have a somewhat thinner aspect ratio distribution. Given also the high expected frequency of minor mergers in $\Lambda$CDM, the problem may be due to minor mergers. In this case, the angular momentum problem could be alleviated in Milgromian dynamics (MOND) because of a reduced merger frequency arising from the absence of dynamical friction between extended dark matter halos.

\end{abstract}


\section{Introduction}
\label{sec:Introdcution}

Observed galaxies show a wide spectrum of structural and dynamical properties. According to the morphological classification scheme, early-type galaxies typically have a smooth ellipsoidal shape whereas late-type galaxies have a flattened disk that often contains spiral features. A dynamical characterization divides galaxies into dispersion- and rotation-dominated systems. These classifications are not identical, e.g. most early-type galaxies in the ATLAS$^{\mathrm{3D}}$ sample are rotation-supported \citep{Emsellem_2011}. 

The vast majority of nearby galaxies with stellar mass $M_{*} > 10^{10}\,M_\odot$ are of late type \citep[e.g.][]{Kautsch_2006, DelgadoSerrano_2010, Kormendy_2010}. In particular, \citet{DelgadoSerrano_2010} analyzed local galaxies with an absolute magnitude $J < - 20.3$ ($M_{*} \ga 1.5 \times 10^{10} \, M_\odot$) from the Sloan Digital Sky Survey \citep[SDSS;][]{York_2000} and found that only $3\% \pm 1\%$ of galaxies are elliptical, $15\% \pm 4\%$ are lenticular, $72\% \pm 8\%$ are spiral, and $10\% \pm 3\%$ are peculiar. Interestingly, they also showed that the relative fraction of ellipticals and lenticulars has hardly evolved over the last 6~Gyr \citep[see table~3 and figure~5 of][]{DelgadoSerrano_2010}, which is consistent with their early and rapid formation \citep[][and references therein]{Kroupa_2020}. Moreover, the relative fraction of non-spheroidals ($70-80\%$; i.e. spirals, irregulars, and interacting galaxies) and early types ($20-30\%$; i.e. ellipticals and transition E/S0 galaxies) with an apparent $K_s$ magnitude brighter than 22 selected from the GOODS-MUSIC catalog \citep{Grazian_2006, Santini_2009} of the Great Observatory Origins Deep Survey \citep{Giavalisco_2004} remains constant over the redshift range $0.6 \leq z \leq 2.5$ \citep[see figure~8 of][]{Tamburri_2014}. 

The morphology of a galaxy is closely related to internal physical processes (e.g. rapid monolithic collapse of post-Big Bang gas clouds, star formation, feedback from supernovae and active galactic nuclei), its dynamical history (interactions and mergers with other galaxies), and its environment (e.g. tidal and ram pressure effects). Thus, the observed distribution of galaxy morphological types constrains models of galaxy formation and evolution. Indeed, \citet{Disney_2008} emphasised that the observed population of galaxies shows a significantly smaller variation of individual properties than expected in a hierarchical formation model where galaxies undergo mergers stochastically. Several simulations in the standard cosmological model known as Lambda Cold Dark Matter \citep[$\Lambda$CDM;][]{Efstathiou_1990, Ostriker_Steinhardt_1995} show an excessive loss of angular momentum \citep[e.g.][]{Katz_1991, Navarro_1991, Navarro_1994, Navarro_2000, vandenBosch_2001, Piontek_2011, Scannapieco_2012}. This hampers the formation of bulgeless disk galaxies. Due to dynamical friction on the extended dark matter halos \citep{Kroupa_2015}, galactic mergers are common in $\Lambda$CDM. Indeed, $N$-body simulations yield that $\approx 95\%$ of galaxies with dark matter halo mass $M_{\mathrm{halo}} \approx 10^{12} \, h^{-1} \, M_\odot$ accreted at least one galaxy with $M_{\mathrm{halo}} > 5 \times 10^{10} \, h^{-1}\,M_\odot$ within the last 10~Gyr, where $h$ is the present Hubble constant $H_0$ in units of $100\,km\,s^{-1}\,Mpc^{-1}$ \citep{Stewart_2008}. In the Millennium-II simulation \citep{Springel_2005_Millennium}, 69\% of galaxies with a similar halo mass have had a major merger since $z = 3$ \citep{Fakhouri_2010}. Galaxy mergers thicken the stellar disk and grow the bulge component, making it difficult for $\Lambda$CDM to account for the observed large population of bulgeless disk galaxies \citep{Graham_2008, Kormendy_2010}. \citet{Trayford_2017} showed that disk galaxies in the Evolution and Assembly of GaLaxies and their Environments (EAGLE) simulation are thicker than those observed (see their figure~3), but concluded that this is because of the underyling sub-grid physics (we discuss this further in Section~\ref{sec:Discussion}).

Although the formation of such galaxies is generally known to be a challenge for the $\Lambda$CDM paradigm, some works claimed that the angular momentum problem has been resolved in the latest self-consistent $\Lambda$CDM simulations. For example, \citet{Vogelsberger_2014} argued that the loss of angular momentum was caused by numerical and physical modeling limitations rather than a failure of the $\Lambda$CDM paradigm, because the Illustris-1 simulation produces a mix of disk galaxies and ellipticals (but see our Section~\ref{subsec:Sky-projected aspect ratio distribution and comparison with observations}, which comes to a different conclusion). Indeed, it is possible to form a Milky Way-like galaxy with a small bulge in the $\Lambda$CDM framework, but only under very special conditions of a quiescent merger history and rapid star formation, which removes low angular momentum gas from the inner part of the galaxy \citep{Guedes_2011}. However, in the observed universe, such late-type galaxies are frequent, with $\approx 50\%$ of them having no classical merger-built bulge \citep{Kormendy_2010}. The same problem was identified by \citet{Graham_2008} two years earlier, who found that most real lenticular galaxies have bulge/total fractions $\la 1/3$. A recent attempt to quantify the tension \citep{RodriguezGomez_2019} showed that the median of the sky-projected ellipticity distribution of galaxies in the ``Illustris The Next Generation'' simulation \citep[IllustrisTNG;][]{Pillepich_2018, Nelson_2019} lies within the $16^{\rm{th}}-84^{\rm{th}}$ percentile range of the Panoramic Survey Telescope And Rapid Response System (Pan-STARRS) observational sample \citep{Chambers_2016}. However, this is only a very crude test. In order to rigorously assess the angular momentum problem in a cosmological framework, one has to consider the overall distribution of galaxy morphology, as addressed by this contribution.

The present-day morphological distribution of galaxies has already been studied in the Illustris-1 simulation using mock photometry. Based on dust-free synthetic images of simulated galaxies with $M_{*} > 10^{10} \,M_\odot$ at $z = 0$, \citet{BOTTRELL_2017a} derived photometric bulge/total fractions $\left(B/T\right)_{\mathrm{phot}}$ by performing a 2D parametric surface brightness decomposition with a fixed S\'{e}rsic index of $n_{\mathrm{d}} = 1$ for the disk component and $n_{\mathrm{b}} = 4$ for the bulge (see their section~3.2). This was done in the SDSS $g$ and $r$ bands at four different camera angles. In a subsequent study \citep{BOTTRELL_2017b}, those authors applied the same decomposition analysis to observed galaxies from the SDSS. By comparing the simulated and observed galaxy samples in the space of $M_*$ and $\left( B/T \right)_{\mathrm{phot}}$, they surprisingly found a significant deficit of bulge-dominated subhalos in the Illustris-1 simulation at $10^{10} \leq M_{*}/ M_\odot \leq 10^{11}$ (see also their figures~4 and 6). This would imply that the angular momentum problem has been resolved in $\Lambda$CDM cosmology despite mergers being very common. However, we will argue in Section~\ref{subsec:Photometric bulge/total ratio of subhalos in the Illustris simulation} that their derived $\left( B/T \right)_{\mathrm{phot}}$ is not an appropriate measure to quantify the morphology of a galaxy in the Illustris-1 simulation. Instead of a deficit of bulge-dominated subhalos, the simulation in fact overproduces these and lacks disk-dominated galaxies, contrary to the claims of \citet{BOTTRELL_2017b}.

In this contribution, we statistically compare the observed sky-projected aspect ratio ($q_\mathrm{sky}$) distribution with that provided by the $\Lambda$CDM framework based on the latest state-of-the-art hydrodynamical cosmological $\Lambda$CDM simulations from the projects known as EAGLE \citep{Schaye_2015,McAlpine_2016}, Illustris \citep{Vogelsberger_2014, Nelson_2015}, and IllustrisTNG \citep{Pillepich_2018, Nelson_2019}. Our analysis focuses on the stellar distribution in a galaxy without regards to individual structural components like its thin or thick disk. The main aim of our work is to test whether state-of-the-art $\Lambda$CDM simulations form galaxies with a realistic distribution of morphologies.

The layout of this paper is as follows: Section~\ref{subsec:Cosmological LCDM simulations} describes the here assessed hydrodynamical cosmological $\Lambda$CDM simulations. The methods to calculate the intrinsic and sky-projected aspect ratio of a galaxy are given in Section~\ref{subsec:Quantifying the shape of a galaxy}. In Section~\ref{subsec:Observational galaxy samples}, we introduce the observational galaxy samples from which we extract the aspect ratio distribution. The statistical method to quantify the tension between the simulated and observed galaxy populations is explained in Section~\ref{sec:Quantifying the tension between simulations and observations}. The present-day intrinsic and sky-projected aspect ratio distributions yielded by the $\Lambda$CDM framework are presented and the latter are compared with observations in Section~\ref{sec:Results}. In Section~\ref{sec:Discussion}, we test the numerical convergence of the TNG50 and EAGLE runs, seek to understand the mismatch between the photometric parameters \citep{BOTTRELL_2017a,BOTTRELL_2017b} and intrinsic aspect ratios in the Illustris-1 simulation, and investigate the effect of different merger histories on galaxy shapes. We also qualitatively compare the $q_\mathrm{sky}$ distribution of SDSS spirals with that of disk galaxies formed in hydrodynamical Milgromian dynamics \citep[MOND;][]{Milgrom_1983} simulations. Our conclusions are given in Section~\ref{sec:Conclusion}.

\section{Cosmological \texorpdfstring{$\Lambda$CDM}{LCDM} simulations}
\label{subsec:Cosmological LCDM simulations}

\begin{table*}
  \centering
  \resizebox{\linewidth}{!}{
		\begin{tabular}{lllllllll}
			\hline
    Simulation & $L$ & $N$ & $H_{0}$ & $\Omega_{\mathrm{b},0}$ & $\Omega_{\mathrm{m},0}$ & $\Omega_{\mathrm{\Lambda},0}$ & $m_{\mathrm{b}}$ & $m_{\mathrm{dm}}$ \\ 
     & $(\rm{cMpc})$ & $-$ & $(\rm{km\, s^{-1}\, Mpc^{-1}})$ & $-$ & $-$ & $-$ & $[M_\odot]$ & $[M_\odot]$ \\  \hline
    EAGLE100  & $100$ & $2 \times 1504^{3}$ & $67.77$ & $0.04825$ & $0.307$ & $0.693$ & $1.81 \times 10^{6}$ & $9.70 \times 10^{6}$ \\
    EAGLE50   & $50$ & $2 \times 752^{3}$ & $67.77$ & $0.04825$ & $0.307$ & $0.693$ & $1.81 \times 10^{6}$ & $9.70 \times 10^{6}$ \\ 
    EAGLE25   & $25$ & $2 \times 752^{3}$ & $67.77$ & $0.04825$ & $0.307$ & $0.693$ & $2.26 \times 10^{5}$ & $1.21 \times 10^{6}$ \\
    Illustris-1 & $106.5$ & $2 \times 1820^{3}$ & $70.4$ & $0.0456$ & $0.2726$ & $0.7274$ & $1.3 \times 10^{6}$ & $6.3 \times 10^{6}$ \\
    TNG50-1    & $51.7$ & $2 \times 2160^{3}$ & $67.74$ & $0.0486$ & $0.3089$& $0.6911$ & $8.5 \times 10^{4}$ & $4.5 \times 10^{5}$ \\
    TNG50-2    & $51.7$ & $2 \times 1080^{3}$ & $67.74$ & $0.0486$ & $0.3089$ & $0.6911$ & $6.8 \times 10^{5}$ & $3.6 \times 10^{6}$ \\
    TNG50-3    & $51.7$ & $2 \times 540^{3}$ & $67.74$ & $0.0486$ & $0.3089$ & $0.6911$ & $5.4 \times 10^{6}$ & $2.9 \times 10^{7}$ \\
    TNG50-4    & $51.7$ & $2 \times 270^{3}$ & $67.74$ & $0.0486$ & $0.3089$ & $0.6911$ & $4.3 \times 10^{7}$ & $2.3 \times 10^{8}$ \\
    TNG100-1    & $110.7$ & $2 \times 1820^{3}$ & $67.74$ & $0.0486$ & $0.3089$ & $0.6911$ & $1.4 \times 10^{6}$ & $7.5 \times 10^{6}$ \\ \hline 
		\end{tabular}}
	\caption{Numerical and cosmological parameters of the here analyzed $\Lambda$CDM simulations. From left to right: simulation name; co-moving box size (cubic side length); number of dark matter particles plus the initial number of gas cells/particles (hence the factor of 2); present-day Hubble constant; present baryonic density in units of the cosmic critical density; same for the total matter density; the dark energy density; the baryonic mass resolution; and the dark matter mass resolution. The highest resolution realization TNG50-1 has a Plummer-equivalent gravitational softening length for the collisionless component of 288~pc and a minimum adaptive gas gravitational softening length of 72~pc at redshift $z=0$ \protect\citep[for more information, see table~1 of][]{Pillepich_2019}. EAGLE25 refers to the EAGLE Recal-L0025N0752 simulation. Additional parameters for the EAGLE set can be found in table~1 of \protect\citet{McAlpine_2016}, for Illustris-1 in table~1 of \protect\citet{Nelson_2015}, and for TNG100-1 in table~1 of \protect\citet{Nelson_2018}.}
  \label{tab:simulations_parameters}
\end{table*}

We investigate the aspect ratio distribution provided by different simulation runs of the projects known as EAGLE \citep{Schaye_2015,McAlpine_2016}, Illustris \citep{Vogelsberger_2014,Nelson_2015}, and TNG \citep{Pillepich_2018,Nelson_2019}.\footnote{IllustrisTNG (abbreviated as TNG hereafter) is a further development of the Illustris project with an improved galaxy formation and evolution model \citep{Nelson_2019}.} These state-of-the-art simulations self-consistently evolve dark matter and baryons from shortly after the Big Bang up to the present time in a $\Lambda$CDM cosmology consistent with the WMAP-9 \citep{Hinshaw_2013_Illustris}, Planck 2015 \citep{Planck_2016_IllustrisTNG}, and Planck 2013 \citep{Planck_2014_EAGLE} measurements of the cosmic microwave background for Illustris, TNG, and EAGLE, respectively. These simulations differ in the implemented baryonic feedback models and underlying grid solvers. Both the Illustris and TNG simulations were performed with the moving-mesh code \textsc{arepo} \citep{Springel_2010}, whereas the EAGLE simulations employed the \textsc{GADGET-3} \citep{Springel_2005} smoothed particle hydrodynamics (SPH) code. Here, we use the TNG50-1, TNG100-1, Illustris-1, EAGLE Ref-L050N0752 (hereafter EAGLE50), and EAGLE Ref-L100N1504 (EAGLE100) simulations, with TNG50-1 having the highest resolution. In Section~\ref{subsec:Comparison of different TNG50 and EAGLE resolution realizations}, we also employ the lower-resolution realizations TNG50-2, TNG50-3, and TNG50-4 of the TNG50 sets and the higher-resolution realization EAGLE Recal-L0025N0752 (hereafter EAGLE25) in order to study the effect of resolution on the shapes of simulated galaxies. The numerical and cosmological parameters of these simulations are listed in Table~\ref{tab:simulations_parameters}.\footnote{In the Illustris and TNG simulations, the initial speed of a baryonic wind particle has an unphysical dependence on the local one-dimensional dark matter velocity dispersion \citep[see equation~1 in][]{Pillepich_2018}.}

\section{Quantifying the shape of a galaxy}
\label{subsec:Quantifying the shape of a galaxy}

The shape of a galaxy can be quantified by the 3D intrinsic aspect ratio of its mass distribution as defined by $q_{\mathrm{int}} \equiv \lambda_{1}/\sqrt{\lambda_{2} \lambda_{3}}$, where $\lambda_{1}$, $\lambda_{2}$, and $\lambda_{3}$ (sorted so $\lambda_{1}~<~\lambda_{2}~<~\lambda_{3}$) are the square roots of the eigenvalues of the mass distribution tensor (MDT, sometimes also called the moment of inertia tensor) divided by the total mass \citep[see e.g. equation~D.39 in appendix~D of][]{Binney_2008}. In this contribution, we consider only the stellar MDT because we are interested in the appearance of a galaxy in optical images. This analysis does not distinguish different structural components of the galaxy like its thin or thick disk. A completely intrinsically thin disk galaxy has $q_{\mathrm{int}} = 0$, whereas a perfectly spherical galaxy has $q_{\mathrm{int}} = 1$. Spiral galaxies account for the bulk of galaxies in the locally observed Universe \citep{DelgadoSerrano_2010}, with typical $q_{\mathrm{int}} \approx 0.2$ (see e.g. figure~1 of \citealt{Mosenkov_2010} and figure~1 of \citealt{Hoffmann_2020}). About 57\% (79\%) of all galaxies in the Sydney–AAO Multi-object Integral field spectrograph (SAMI) Galaxy Survey \citep{Croom_2012, Bryant_2015} have $q_{\mathrm{int}} < 0.4$ ($<0.6$) \citep[see figure~15 of][]{Oh_2020}. The Galactic thin disk has an exponential scale length of $l = 2.6\pm0.5$~kpc and an exponential scale height of $h = 220-450$~pc \citep{BlandHawthorn_2016}, which results in an aspect ratio of $h/l \approx 0.07-0.21$. The Andromeda galaxy (M31) has a thin disk with $1 - \epsilon \equiv b/a \equiv q_\mathrm{sky} = 0.27\pm0.03$, where $\epsilon$ and $b/a$ are the sky-projected ellipticity and axis ratio, respectively \citep{Courteau_2011}. Because M31 is not viewed exactly edge-on, it must be intrinsically thinner than it appears on the sky \citep[section~2.1 of][]{Banik_2018_anisotropy}. 

We extract the eigenvalues of the stellar MDT from supplementary data catalogs provided by the Illustris, TNG, and EAGLE teams \citep{Thob_2019}.\footnote{The eigenvalues of the stellar MDT for Illustris and TNG subhalos can be downloaded from \url{https://www.tng-project.org/data/docs/specifications/\#sec5c} [21.07.2020].} The MDTs of subhalos in the Illustris and TNG simulations are calculated from the stellar particles within twice the stellar half-mass radius $r_{0.5,*}$ \citep{Genel_2015}. This slightly differs from the EAGLE simulations in which an iterative form of the reduced MDT is used, with the initial selection being all stellar particles inside a spherical aperture of physical radius 30~kpc \citep[section~2.3 of][]{Thob_2019}. As demonstrated in Section~\ref{subsec:disk galaxies in Milgromian dynamics}, this method provides a secure division into spirals and ellipticals, whereas other bulge-disk decomposition methods face problems when applied to $\Lambda$CDM simulations (Section~\ref{subsec:Photometric bulge/total ratio of subhalos in the Illustris simulation}).

\subsection{Projecting an Ellipsoid onto the Sky}
\label{subsec:Projecting an ellipsoid onto the sky}

In order to compare simulations with observations, we have to determine $q_\mathrm{sky}$ for a galaxy with $\lambda_i$, where $i = 1-3$ (Section~\ref{subsec:Quantifying the shape of a galaxy}). Only the ratios of the $\lambda_i$ are relevant for our analysis, but it will be helpful to think of them as actual lengths. We approximate that the galaxy is an ellipsoid and find what this looks like when viewed by a distant observer. We work in Cartesian coordinates in a reference frame centered on the galaxy and aligned with the eigenvectors of its inertia tensor. Thus, the `edge' of the galaxy is given by
\begin{eqnarray}
	\frac{x^2}{{\lambda_1}^2} \, + \, \frac{y^2}{{\lambda_2}^2} \, + \, \frac{z^2}{{\lambda_3}^2} ~=~ 1 \, .
	\label{Surface_equation}
\end{eqnarray}

Suppose a distant observer is located toward the direction $\widehat{\bm{n}}$, where we use hats to denote unit vectors. Our approach is to find the extent of the image along the direction $\widehat{\bm{n}}_3$, which lies entirely within the sky plane (i.e. $\widehat{\bm{n}} \cdot \widehat{\bm{n}}_3 = 0$). Our main goal is to find the position vector $\bm{r}$ of the point corresponding to the edge of the galaxy image along the direction $\widehat{\bm{n}}_3$. For this purpose, it will be useful to define the unit vector within the sky plane in the orthogonal direction, which we call $\widehat{\bm{n}}_2$ (orthogonal to both $\widehat{\bm{n}}$ and $\widehat{\bm{n}}_3$).

We know that $\bm{r} \cdot \widehat{\bm{n}}_2 = 0$, or else the point would not appear to be in the direction $\widehat{\bm{n}}_3$ from the galaxy's center. To get an additional constraint, we note that the tangent plane to the galaxy at the point $\bm{r}$ must not contain $\widehat{\bm{n}}_3$, or else it would be possible to move along the galaxy's boundary and reach a larger apparent separation along $\widehat{\bm{n}}_3$. The plane normal must therefore be $\pm \widehat{\bm{n}}_3$, but for our analysis, it is sufficient to know that the plane normal is orthogonal to $\widehat{\bm{n}}$.

These two constraints are sufficient to determine the direction of $\bm{r}$. Its magnitude is found through Equation~\ref{Surface_equation}. Thus, the extent of the image along $\widehat{\bm{n}}_3$ is $d$, which we find through the following procedure involving the intermediate vectors $\bm{q}$ and $\bm{v}$:
\begin{eqnarray}
	\bm{q}_i ~&\equiv&~ \frac{\widehat{\bm{n}}_i}{{\lambda_i}^2} \, , \\
	\bm{v} ~&\equiv&~ \widehat{\bm{n}}_2 \times \bm{q} \, , \\
	d ~&=&~ \frac{\bm{v} \cdot \widehat{\bm{n}}_3}{\sqrt{\sum_{i = 1}^3 \left( \frac{\bm{v}_i}{\lambda_i} \right)^2}} \, .
\end{eqnarray}
We repeat this for a low-resolution grid of $\widehat{\bm{n}}_3$, whose direction we parameterize using the so-called position angle. Starting from the $\widehat{\bm{n}}_3$ in this grid which gives the lowest $d$, we apply the gradient descent algorithm \citep{Fletcher_1963} to find the minimum value of $d$ to high precision. We then rotate $\widehat{\bm{n}}_3$ through a right angle, and start a gradient ascent stage to search for the maximum extent of the image. The ratio of these $d$ values is the sky-projected aspect ratio $q_\mathrm{sky} \leq 1$, which forms the heart of our analysis.

To build up the $q_\mathrm{sky}$ distribution, we repeat this procedure for a 2D grid of viewing angles $\widehat{\bm{n}}$, with each result weighted according to the solid angle it represents. The observer is thus assumed to be in a random direction relative to the galaxy.

\section{Observational galaxy samples} \label{subsec:Observational galaxy samples}

The sky-projected aspect ratio distributions of the following observational galaxy samples are statistically compared with the simulations discussed in Section~\ref{subsec:Cosmological LCDM simulations}.

\subsection{GAMA Survey}
\label{subsec:GAMA Survey}

The Galaxy And Mass Assembly survey \citep[GAMA;][]{Driver_2009, Driver_2011} is a multiwavelength photometric and spectroscopic redshift survey. An overview of the survey regions and their corresponding magnitude limits is given in table~1 of \citet{Baldry_2018}. Here, we download the stellar masses, redshifts, and ellipticities by submitting an \textsc{sql} query to the GAMA DR3 database \citep{Baldry_2018}.\footnote{\url{http://www.gama-survey.org/dr3/query/} [11.11.2021]} In detail, we extract the \textsc{galfit} \citep{Peng_2002, Peng_2010} $r$-band ellipticity ($1 - b/a$) from the SersicPhotometry (v09) - SersicCatSDSS catalog \citep{Kelvin_2012}. We use the stellar mass labeled as `logmstar' from the StellarMasses (v20) catalog \citep{Taylor_2011}, which is derived from matched aperture photometry in the $r$ band, missing therewith flux beyond the AUTO aperture.\footnote{For further information on the stellar masses, see also {\url{http://www.gama-survey.org/dr3/schema/dmu.php?id=9} [11.11.2021]}}
Consequently, the stellar masses $M_{*,\mathrm{AUTO}}$ within the photometric aperture (`logmstar') are corrected by 
\begin{eqnarray}
    \log_{10}\bigg(\frac{M_{*}}{M_{\odot}}\bigg)=\log_{10}\bigg(\frac{M_{*,\mathrm{AUTO}}}{M_{\odot}}\bigg) + \log_{10}\bigg(\frac{f_{\text{S\'{e}rsic}}}{f_{\mathrm{AUTO}}} \bigg) \, , \nonumber
    \label{eq:GAMA_stellar_masses}
\end{eqnarray}
where $f_{\text{S\'{e}rsic}}/f_{\mathrm{AUTO}}$ is the so-called `fluxscale' parameter, i.e. the linear ratio between the total $r$-band flux from a S\'{e}rsic profile fit cut at 10 effective radii and the $r$-band AUTO aperture flux \citep[see also, e.g.][]{Taylor_2011, Kelvin_2014b, Lange_2016, VazquezMata_2020}. The stellar masses are calculated by assuming concordance cosmology with the cosmological parameters being $\Omega_{\mathrm{m},0} = 0.3$, $\Omega_{\mathrm{\Lambda},0} = 0.7$, and $h = 0.7$ \citep{Taylor_2011}.

For our analysis, we only consider galaxies with a fluxscale correction of $0.5 < \rm{fluxscale} < 2$ and a spectral energy distribution (SED) fit with a posterior predictive $P$-value $PPP > 0$, which removes failed SED fits. In addition, we exclude objects with a heliocentric redshift $z < 0.005$ to remove stars.\footnote{Private communication with Edward Taylor.} Also requiring $M_{*} > 10^{10}\,M_\odot$ and $z < 0.1$ yields a final sample of $5304$ galaxies that pass the above quality cuts.

\subsection{GAMA and Cluster Input Catalogs for the SAMI Galaxy Survey}
\label{subsec:SAMI Galaxy Survey}

The SAMI Galaxy Survey Data Release 3 (DR3) includes 3068 galaxies in the redshift range $0.004 < z < 0.095$. The ellipticity distribution of a subsample of the SAMI Galaxy Survey consisting of 826 galaxies is shown in the third panel of figure~6 in \citet{Oh_2020}. 

In order to increase the sample size, we analyze the aspect ratio distribution of galaxies listed in the input catalogs of the three equatorial regions of the GAMA survey \citep[i.e. G09, G12, and G15;][]{Bryant_2015} and the eight cluster regions \citep[i.e. APMCC0917, A168, A4038, EDCC442, A3880, A2399, A119, and A85;][]{Owers_2017} for the SAMI DR3 Galaxy Survey. For this, we download the stellar masses, redshifts, and projected ellipticities by using an \textsc{sql}/Astronomical Data Query Language (\textsc{adql}) query to the Data Central database servers\footnote{\url{https://datacentral.org.au/services/query/} [04.11.2021]} of the InputCatGAMADR3 and InputCatClustersDR3 catalogs \citep[for more details, see also table~1 of][]{Croom_2021}. The ellipticities are derived from S\'{e}rsic fits in the $r$ band \citep{Bryant_2015,Owers_2019}.

Requiring $M_{*} > 10^{10} \, M_{\odot}$ gives a sample of $4252$ galaxies, of which $3238$ are from GAMA and $1014$ are from the cluster regions. As we show in Section~\ref{sec:Quantifying the tension between simulations and observations}, the aspect ratio distributions of the GAMA survey and the here described sample disagree only at the $0.013\sigma$ confidence level. Although the ellipticity values of the same galaxy listed in the InputCatGAMADR3 and the GAMA survey are slightly different, these catalogs are not fully independent of each other. Because of these reasons and since the GAMA survey contains more galaxies (5304), we do not use the input catalogs of the GAMA and cluster regions for SAMI DR3 in our statistical comparison with simulations (Section~\ref{sec:Quantifying the tension between simulations and observations}).

\subsection{SDSS}
\label{subsec:SDSS}

The SDSS is a flux-limited galaxy survey including galaxies with a Petrosian $r$ magnitude brighter than $17.77$ \citep{York_2000}. Here, we download the stellar masses, redshifts, and photometric parameters from its DR16\footnote{\url{https://www.sdss.org/dr16/}} \citep{Ahumada_2020} using an \textsc{sql} query to the SDSS database server.\footnote{\url{http://skyserver.sdss.org/dr16/en/home.aspx} [11.11.2021]}

We selected galaxies listed in the PhotoPrimary catalog, which also contains the aspect ratios of galaxies derived from an exponential and a de Vaucouleurs profile \citep{deVaucouleurs_1948} for the SDSS $r$, $i$, $u$, $z$, and $g$ filters. Throughout this analysis, we use the aspect ratio parameters based on the SDSS $r$-band magnitude. In order to distinguish between spiral and elliptical galaxies, we access the fracDeV parameter, the weight of the de Vaucouleurs component in a linear combination of the exponential and de Vaucouleurs profiles \citep[for a more detailed description, see section~3.1 of][]{Abazajian_2004}. Following \citet{Padilla_2008}, we use the aspect ratio derived from the exponential fit if $\mathrm{fracDeV} < 0.8$ and from the de Vaucouleurs fit if $\mathrm{fracDeV} \geq 0.8$.

The here used stellar masses from the galSpecExtra table correspond to the median of the estimated logarithmic stellar mass probability density function using model photometry \citep{Kauffmann_2003, Salim_2007}. The redshifts are extracted from the SpecObj table.\footnote{Although the SpecObj catalog does not contain duplicate observations, there is the rare situation that the catalog lists more than one redshift measurement for the same object. In this case, the R.A. and decl. values of the fibers are different during the spectroscopic observation. We take this into account by checking if our downloaded catalog lists more than one redshift measurement for the same object ID. If so, we remove the measurement with the lower signal-to-noise ratio. Of our $232,315$ selected galaxies, a duplicate happened only once for the object with ID 1237663782590021834.}


Selecting galaxies with $z < 0.1$ and $M_{*} > 10^{10}\,M_{\odot}$ gives a sample of $232,315$ galaxies. The so-obtained and here analyzed $q_{\mathrm{sky}}$ distribution is consistent with \citet{Padilla_2008}, as shown in Section~\ref{sec:Quantifying the tension between simulations and observations}. They obtained a volume-limited sample by weighting each galaxy in the SDSS DR6 \citep{AdelmanMcCarthy_2008} by $1/V_{\mathrm{max}}$, where $V_{\mathrm{max}}$ is the volume corresponding to the maximum distance at which we can observe a galaxy with its absolute magnitude \citep[see section~3.1 of][]{Padilla_2008}. In total, their sample contains $303,290$ `spirals' ($\mathrm{fracDeV} < 0.8$) and $282,203$ `ellipticals' ($\mathrm{fracDeV} \geq 0.8$). Here, we extract the $1/V_{\mathrm{max}}$ weighted $q_{\mathrm{sky}}$ distributions of the spiral and elliptical samples from their figure~1 (bottom panels), and combine these in order to obtain the $q_{\mathrm{sky}}$ distribution of the total galaxy sample.

\subsection{The Catalog of Neighboring Galaxies} \label{subsubsec:The Catalog of Neighboring Galaxies}

\begin{figure*}
	\begin{center}
		\includegraphics[width=8.5cm]{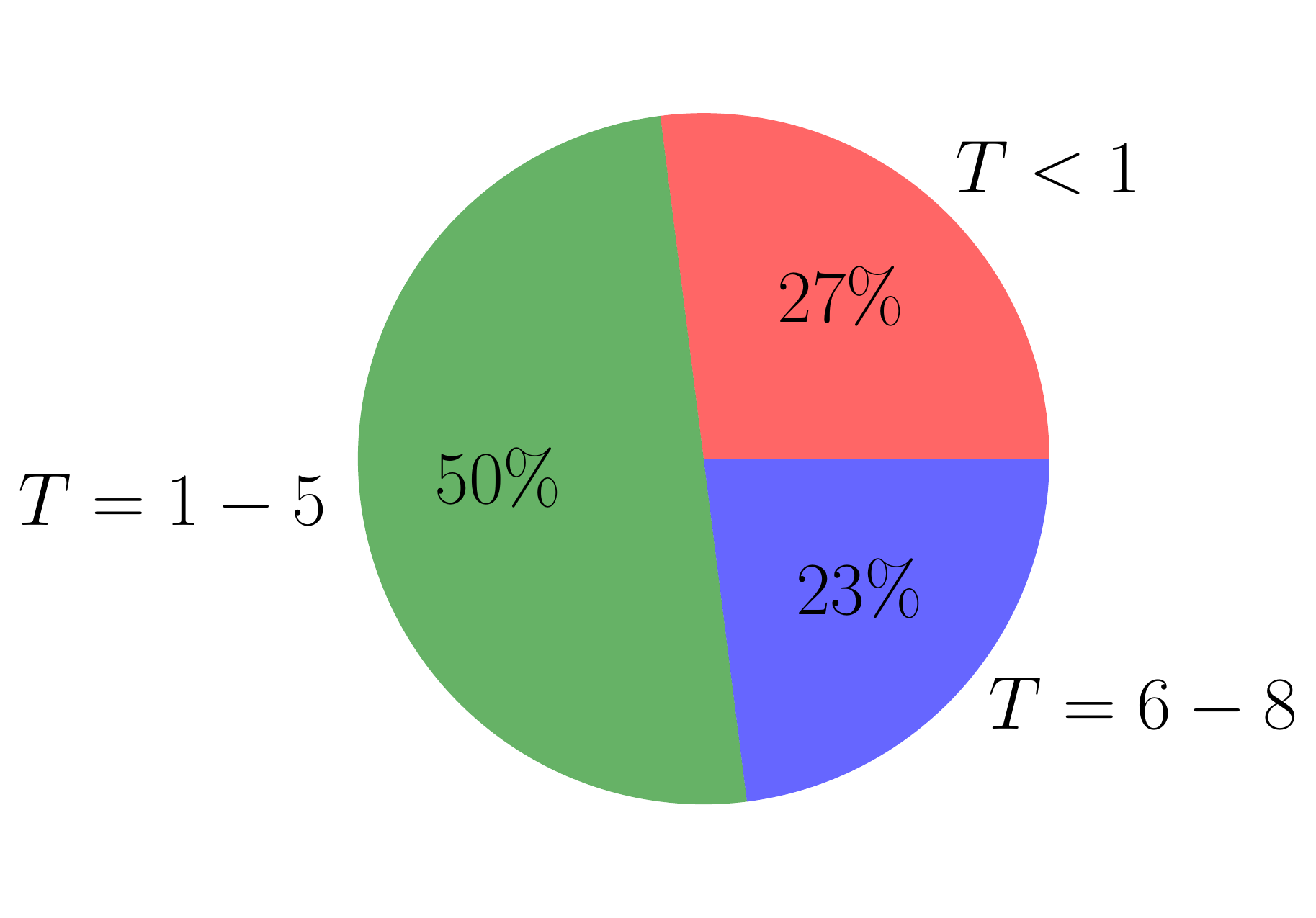}
		\includegraphics[width=8.5cm]{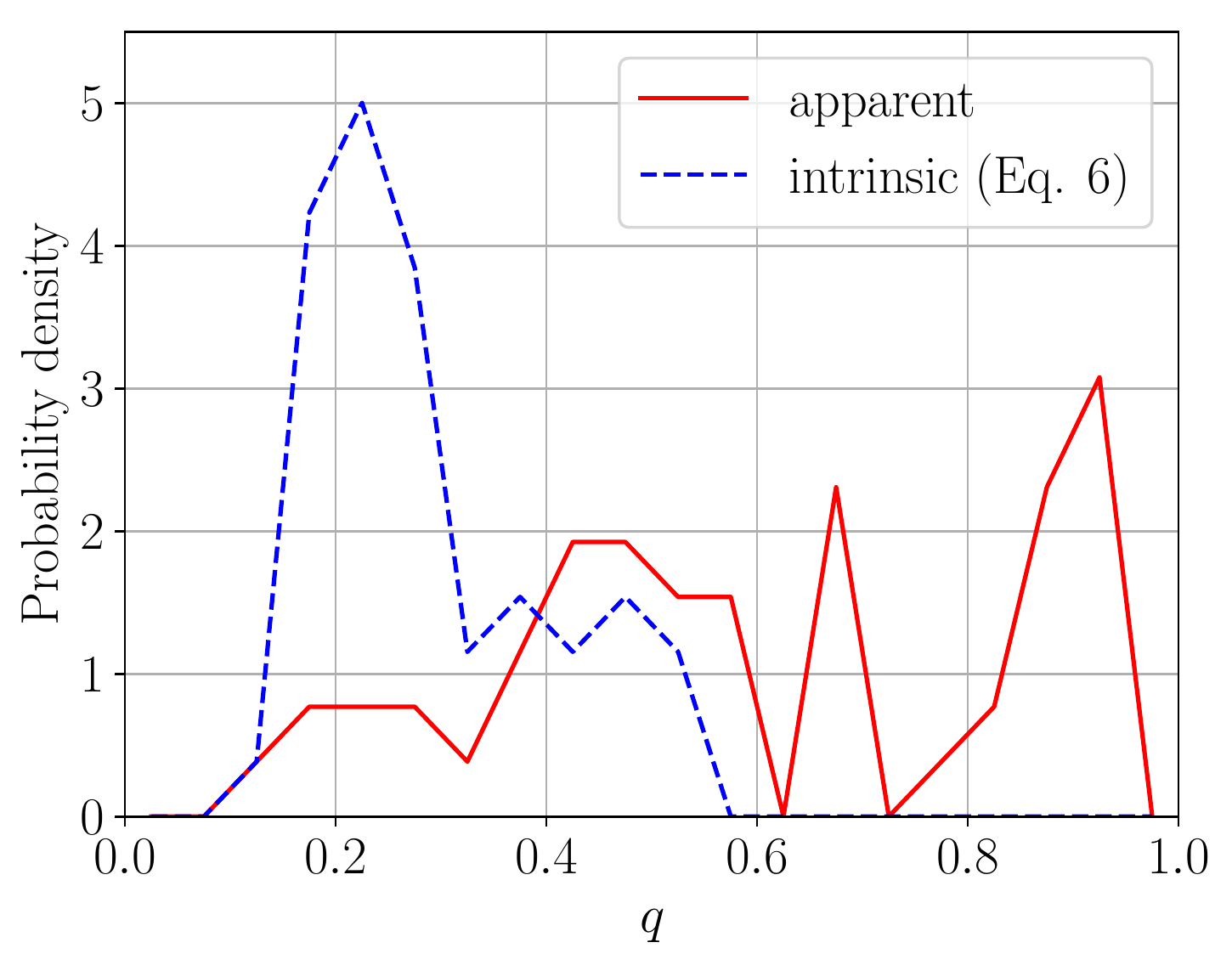}
    \end{center}
	\caption{\emph{Left:} distribution of the morphological $T$-types of galaxies with $M_{*} > 10^{10}\,M_\odot$ in the Local Volume (LV). The red slice refers to E, S0, dSph, and S0a galaxies ($T < 1$), the green slice to early-type spirals (Sa, Sab, Sb, Sbc, Sc; $T = 1 - 5$), and the blue slice to late-type spirals (Scd, Sd, Sdm, Sm; $T=6-8$). \emph{Right:} the sky-projected (solid red) and intrinsic (dashed blue) aspect ratio distribution of galaxies with $M_{*} > 10^{10}\,M_\odot$ in the LV. The bin width is $\Delta q_{\mathrm{sky}} = \Delta \widetilde{q}_{\mathrm{int}} = 0.05$. The intrinsic aspect ratios are derived using Equation~\ref{eq:intrinsic_ttype}, which is a best guess based on the morphological $T$-type (see the text). It is statistically compared with the TNG50-1 simulation run in Section~\ref{subsec:Intrinsic aspect ratio distribution}, but the model dependence of the $\widetilde{q}_{\mathrm{int}}$ values means this is not our main result.}
	\label{fig:skyprojected_intrinsic_aspect_ratio_distribution}
\end{figure*}

As a consistency check, we also investigate the $q_{\mathrm{sky}}$ distribution of galaxies within the Local Volume (LV), a sphere of radius $11\,\rm{Mpc}$ centered on the Sun. The stellar masses are calculated from $K$-band luminosities using a mass-to-light ratio of 0.6 \citep[e.g.][]{McGaugh_2014}. We select 52 galaxies with $10.0 < \log_{10}(M_{*}/M_{\odot}) \leq 11.65$ from the Updated Nearby Galaxy Catalog \citep{Karachentsev_2013}, a renewed version of The Catalog of Neighboring Galaxies\footnote{\url{https://www.sao.ru/lv/lvgdb/tables.php} [22.11.2021]} \citep{Karachentsev_2004}. These galaxies cover a wide range of $q_{\mathrm{sky}}$ ($0.13-0.94$), as shown in the right panel of Figure~\ref{fig:skyprojected_intrinsic_aspect_ratio_distribution}. Due to the small sample size and high resulting Poisson uncertainties, we do not calculate its tension with the simulated $q_{\mathrm{sky}}$ distributions.

These are nearby galaxies with well-established morphological types, allowing direct well-resolved images to inform us of which galaxies are thin disks. Therefore, we show the distribution of the morphological $T$-types of our LV galaxy sample in the left panel of Figure~\ref{fig:skyprojected_intrinsic_aspect_ratio_distribution}. The $T$-types of these galaxies are listed in the Updated Nearby Galaxy Catalog and discussed and analyzed in detail in \citet{Karachentsev_2018}. Of the 52 galaxies, 50\% are early-type spirals (Sa, Sab, Sb, Sbc, Sc; morphological type $T = 1-5$), 23\% are late-type spirals (Scd, Sd, Sdm, Sm; $T = 6-8$), and the remaining 27\% are E, S0, dSph, or S0a galaxies ($T < 1$). Thus, the morphological classification scheme used by \citet{Karachentsev_2013, Karachentsev_2018} is slightly different to the de Vaucouleurs system.

It is not in general possible to disentangle the contributions to $q_{\mathrm{sky}}$ from inclination and intrinsic thickness. The inclination $i$ between disk and sky planes listed in the Updated Nearby Galaxy Catalog is derived with equation~5 of \citet{Karachentsev_2013}: 
\begin{eqnarray}
    \sin^2 i ~=~\frac{1 - {q_{\mathrm{sky}}}^2}{1 - {{\widetilde{q}_{\mathrm{int}}}{}}^2} \, ,
    \label{eq:inclination}
\end{eqnarray}
where the assumed intrinsic aspect ratio $\widetilde{q}_{\mathrm{int}}$ is assigned a value depending on the morphological $T$-type ($T$) of the considered galaxy as given by their equation~6:\footnote{There seems to be a typo in equation~6 of \citet{Karachentsev_2013}. The relation between the assumed intrinsic aspect ratio and the $T$-type parameter for galaxies with $T \leq 8$ should be $\log_{10}\left( a/b \right)_{0} = -\log_{10} \widetilde{q}_{\mathrm{int}} = 0.43 + 0.053 \, T$ \citep[see also equation~14 of][]{Paturel_1997} rather than $0.43 + 0.53 \, T$, which would make late-type galaxies far too thin.}
\begin{eqnarray}
    \log_{10} \widetilde{q}_{\mathrm{int}} ~=~\left \{\begin{array}{ll}
    -0.43 - 0.053 \, T \quad \left(T \leq 8\right) \, ,\\
    -0.38 \qquad \qquad \left( T = 9, 10 \right) \, ,
    \end{array}
    \right. \label{eq:intrinsic_ttype}
\end{eqnarray}
where according to \citet{Karachentsev_2018}, $T = 9$ and $T = 10$ refer to Im/BCD and Ir galaxies, respectively. Thus, the so-obtained intrinsic aspect ratios are a best guess based on the observed morphology. As an example, the M31 galaxy with $T = 3$ has with this approach $\widetilde{q}_{\mathrm{int}} = 0.26$ ($q_{\mathrm{sky}} = 0.33$).

Applying Equation~\ref{eq:intrinsic_ttype} to the 52 selected LV galaxies shows that their intrinsic aspect ratio distribution covers the range $\widetilde{q}_{\mathrm{int}} = 0.07 - 0.55$ with a global peak at $\widetilde{q}_{\mathrm{int}} \approx 0.23$, as presented in the right panel of Figure~\ref{fig:skyprojected_intrinsic_aspect_ratio_distribution}. Converting $q_{\mathrm{sky}}$ to $q_{\mathrm{int}}$ depends on the relation between $\widetilde{q}_{\mathrm{int}}$ and the morphological $T$-type, so the intrinsic aspect ratio distribution of our LV sample is mainly considered for illustrative purposes. However, it helps to demonstrate that the LV galaxies seem to be mostly thin disks. According to this, if the LV were to be representative of the universe, then $81\% \pm 13\%$ (42/52) of all galaxies heavier than $M_{*} = 10^{10} \, M_\odot$ are thin disk galaxies with $\widetilde{q}_{\mathrm{int}} < 0.4$. For completeness, we quantify the tension between the here obtained intrinsic aspect ratio distribution of LV galaxies and the $\Lambda$CDM models in Section~\ref{subsec:Intrinsic aspect ratio distribution}.

In order to avoid any bias on the shapes of galaxies introduced by model assumptions when converting $q_{\mathrm{sky}}$ to $q_{\mathrm{int}}$, we only consider the former when statistically comparing observations to simulations in our main analysis. As explained in Section~\ref{subsec:Projecting an ellipsoid onto the sky}, this requires us to generate the $q_{\mathrm{sky}}$ distribution given the intrinsic shapes of simulated galaxies. We can then directly compare the resulting distribution with the observed $q_{\mathrm{sky}}$ distribution (Section~\ref{subsec:Sky-projected aspect ratio distribution and comparison with observations}).

\section{Quantifying the Tension between Simulations and Observations}
\label{sec:Quantifying the tension between simulations and observations}

\begin{figure*}
	\begin{center}
		\includegraphics[width=8.5cm]{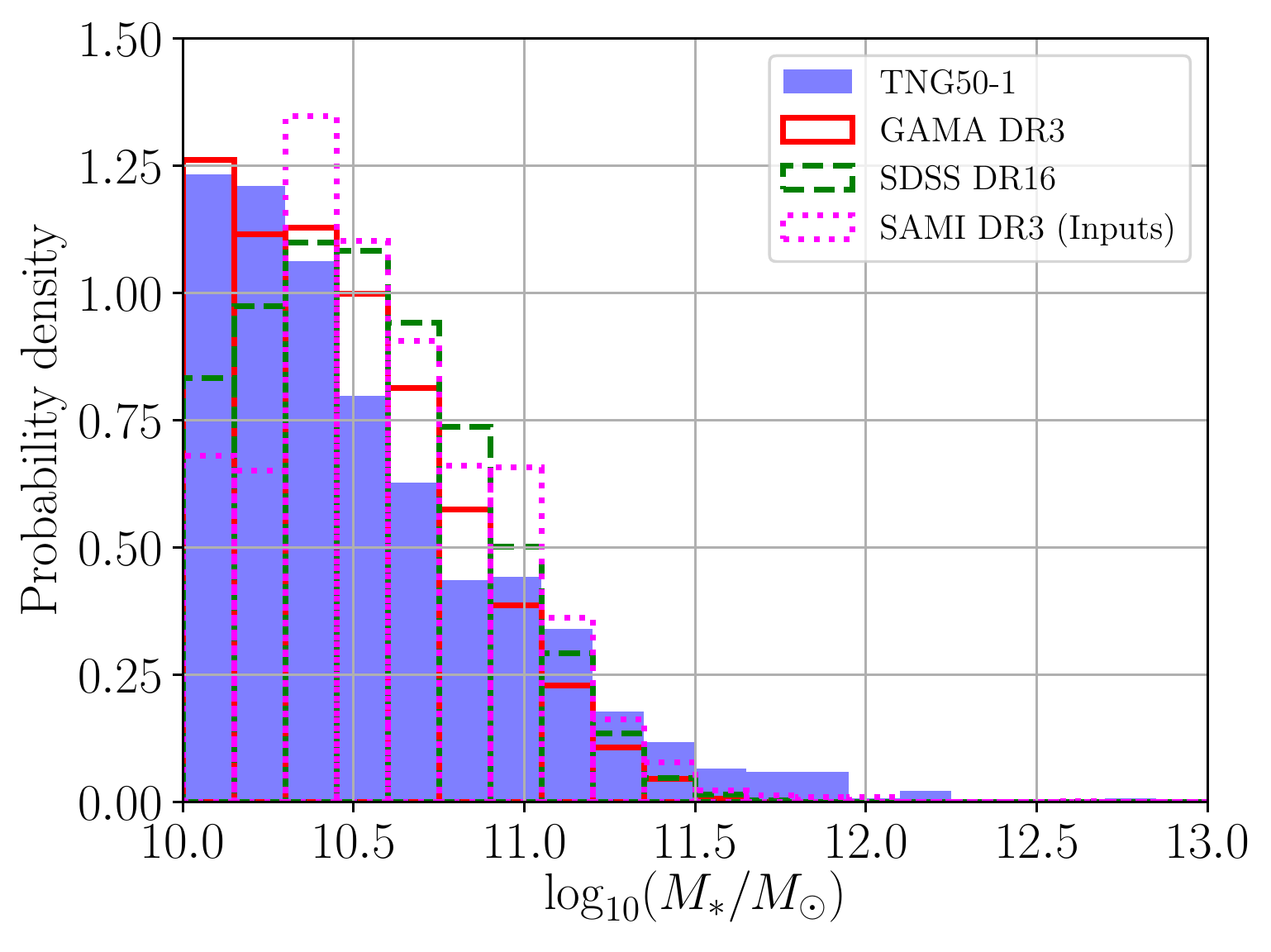}
		\includegraphics[width=8.5cm]{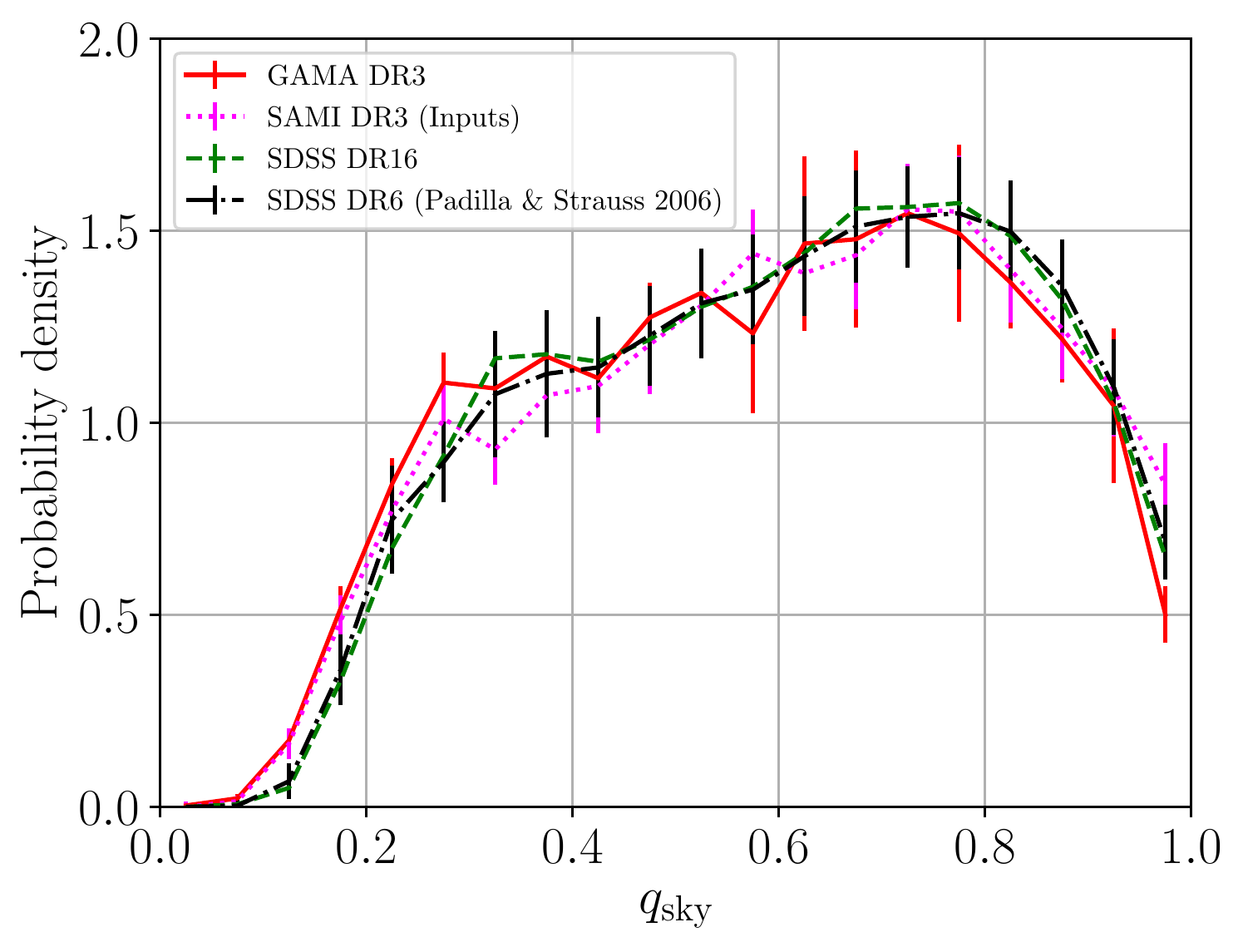}
    \end{center}
	\caption{\emph{Left:} stellar mass distribution of the TNG50-1 simulation run (filled blue) and on the observational side GAMA DR3 (open solid red, Section~\ref{subsec:GAMA Survey}), the input catalogs of the GAMA and cluster regions for SAMI DR3 (open dotted magenta, Section~\ref{subsec:SAMI Galaxy Survey}), and SDSS DR16 (open dashed green, Section~\ref{subsec:SDSS}). We use a bin width of $\Delta \log_{10}(M_{*}/M_{\odot}) = 0.15$. \emph{Right:} sky-projected aspect ratio distribution of GAMA DR3 (solid red), the input catalogs of the GAMA and cluster regions for the SAMI DR3 (dotted magenta), and SDSS DR16 (dashed green) after weighting the galaxies in order to match the stellar mass distribution of the TNG50-1 run. Invisible error bars (Equation~\ref{eq:chi_sq_sigma_model}) are smaller than the data point symbol. The black dotted-dashed line refers to the volume-weighted SDSS DR6 sample \citep{Padilla_2008} with $N_{\mathrm{spirals}} = 303,290$ and $N_{\mathrm{ellipticals}} = 282,203$ (see the bottom panels of their figure~1). The error bars of the spiral and elliptical aspect ratio distributions have been obtained from the jackknife technique and were here weighted by the above-mentioned proportion of spirals and ellipticals. The bin width is $\Delta q_{\mathrm{sky}} = 0.05$.}
	\label{fig:stellarmass_aspectratio}
\end{figure*}

The stellar mass distributions of the GAMA survey and SDSS are compared with that of the TNG50-1 run in the left panel of Figure~\ref{fig:stellarmass_aspectratio}. Galaxies with $10.30 \la \log_{10} \left( M_{*}/M_{\odot} \right) \la 11.05$ are more abundant in the observational samples than in the simulation, while the opposite is true at higher mass. This is most likely because the used stellar masses in the Subfind catalog of the TNG50-1 simulation refer to the mass of all stellar particles bound to each considered subhalo, not to the stellar mass within a certain aperture size.

The different $M_*$ distributions would bias the comparison between the observed and simulated $q_{\mathrm{sky}}$ distributions. Thus, we apply an $M_{*}$-weighting scheme to each observed galaxy in order to match the simulated $M_*$ distribution, as explained in the following. The total $\chi^{2}$ between the simulated and observed sky-projected aspect ratio distributions is
\begin{eqnarray}
    \label{eq:chi_sq_model_observations}
    \chi^{2} ~&=&~ \sum_{i=1}^{N_{\mathrm{bins}}}  \frac{\left( \frac{w_{\mathrm{tot},i}}{w_{\mathrm{tot}}} - \frac{N_{\mathrm{model},i}}{N_{\mathrm{model,tot}}} \right)^{2}}{\sigma_{\mathrm{obs},i}^{2} + \sigma_{\mathrm{model},i}^{2}} \, , \\   
    \sigma_{\mathrm{obs},i} ~&=&~ \left \{\begin{array}{ll}
    \frac{w_{\mathrm{max},i}}{w_{\mathrm{tot}}} \sqrt{ \frac{w_{\mathrm{tot},i}}{w_{\mathrm{max},i}}+1} \quad \left(w_{\mathrm{obs},i} \neq 0 \right) \, ,\\
    \label{eq:chi_sq_sigma_model} 
    \frac{w_{\mathrm{max}}}{w_{\mathrm{tot}}} \quad \left( w_{\mathrm{obs},i} = 0 \right) \, ,
    \end{array}
    \right. \\
    \sigma_{\mathrm{model},i} ~&=&~ \frac{\sqrt{N_{\mathrm{model},i}+1}}{N_{\mathrm{model,tot}}} \, , \, 
\end{eqnarray}
where $N_{\mathrm{model},i}$ is the number of simulated galaxies in bin $i$, $N_{\mathrm{bins}} = 20$ is the number of bins in $q_{\mathrm{sky}}$, and $N_{\mathrm{model,tot}}$ is the total number of simulated galaxies. In order to avoid the bias caused by the different $M_*$ distributions of the simulations and observations, we group simulated and observed galaxies in $M_*$ bins of width $0.15$~dex over the range $10.0 < \log_{10}(M_{*}/M_{\odot}) \leq 11.65$. Each observed galaxy is weighted by $w_{\mathrm{obs}} = N_{\mathrm{sim}}/N_{\mathrm{obs}}$, where $N_{\mathrm{sim}}$ ($N_{\mathrm{obs}}$) is the number of simulated (observed) galaxies in the $M_{*}$ bin of the considered galaxy. The lower limit on $M_{*}$ is set to guarantee that only well-resolved simulated galaxies are analyzed. The maximum limit is applied because the GAMA DR3 sample runs out of galaxies at higher stellar mass, leading to an undefined $w_{\mathrm{obs}}$. These criteria give final GAMA, SAMI DR3 inputs, and SDSS sample sizes of $5304$, $4229$, and $232,128$, respectively. 
$w_{\mathrm{tot},i}$ is then the weighted number of galaxies in $q_{\mathrm{sky}}$ bin $i$, while $w_{\mathrm{max},i}$ is the weight of the galaxy in this $q_{\mathrm{sky}}$ bin with the maximum weight, $w_{\mathrm{max}}$ is the maximum weight of all considered observed galaxies (regardless of $q_{\mathrm{sky}}$), and $w_{\mathrm{tot}}$ is the total weight of all observed galaxies, which by definition must match the number of simulated galaxies used for the comparison. This approach is invariant to a uniform scaling of the weights. $\sigma_{\mathrm{obs},i}$ and $\sigma_{\mathrm{model},i}$ are Poisson uncertainties. The use of Poisson statistics is valid because we choose a bin width of $\Delta q_{\mathrm{sky}} = 0.05$, so each bin contains only a small fraction of the full sample.

This $M_{*}$-weighting scheme is not applied when the aspect ratio distributions are being compared between simulations. The total $\chi^{2}$ between any two models is
\begin{eqnarray}
    \chi^{2} ~&=&~ \sum_{i=1}^{N_{\mathrm{bins}}}  \frac{\left( \frac{N_{\mathrm{model 1},i}}{N_{\mathrm{model 1,tot}}} - \frac{N_{\mathrm{model 2},i}}{N_{\mathrm{model 2,tot}}} \right)^{2}}{\sigma_{\mathrm{model 1},i}^{2} + \sigma_{\mathrm{model 2},i}^{2}} \, . 
    \label{eq:chi_sq_models}
\end{eqnarray}
The uncertainties $\sigma_{\mathrm{model 1}}$ and $\sigma_{\mathrm{model 2}}$ can be thought of as given by Equation~\ref{eq:chi_sq_sigma_model} with all galaxies equally weighted.

The total $\chi^2$ between any model and observations (Equation~\ref{eq:chi_sq_model_observations}) or between two models (Equation~\ref{eq:chi_sq_models}) is converted to a probability or $P$-value of a more extreme outlier using the $\chi^2$ distribution for the appropriate number of degrees of freedom, which we call $n$. For convenience, we then express the $P$-value as an equivalent number of standard deviations for a single Gaussian variable. We denote this $x$, which we find by iteratively solving
\begin{eqnarray}
    1 - \frac{1}{\sqrt{2 \mathrm{\pi}}} \int_{-x}^x \exp \left( -\frac{x^2}{2} \right) \, dx ~\equiv~ P \, .
\label{P_chi}
\end{eqnarray}
A protocol to convert particularly large $\chi^{2}$ values is given in Appendix~\ref{subsec:Statistical significance of extreme events}, which provides a way to solve this equation despite numerical difficulties that arise for very high $\chi^{2}$.

The right panel of Figure~\ref{fig:stellarmass_aspectratio} shows the $q_{\mathrm{sky}}$ distributions of the GAMA Galaxy Survey and SDSS, weighted based on the TNG50-1 run for galaxies with $10.0 < \log_{10}(M_{*}/M_{\odot}) \leq 11.65$. The GAMA DR3 and SDSS DR16 results disagree at $3.25\sigma$ confidence, indicating good agreement considering the large sample sizes. The aspect ratio distributions of GAMA DR3 and the input catalogs of the GAMA and cluster regions for the SAMI DR3 differ only at $0.013\sigma$ confidence. Because of this and the larger sample size of the GAMA survey, we only use the GAMA survey and SDSS for the following analysis.

\section{Results}
\label{sec:Results}

In this section, we present both the intrinsic and the sky-projected aspect ratio distribution at $z = 0$ as produced by the $\Lambda$CDM paradigm. We then statistically compare the simulated $q_{\mathrm{sky}}$ distribution with local observations from the GAMA survey and SDSS.

\subsection{Intrinsic Aspect Ratio Distribution}
\label{subsec:Intrinsic aspect ratio distribution}

\begin{figure*}
	\begin{center}
		\includegraphics[width=8.5cm]{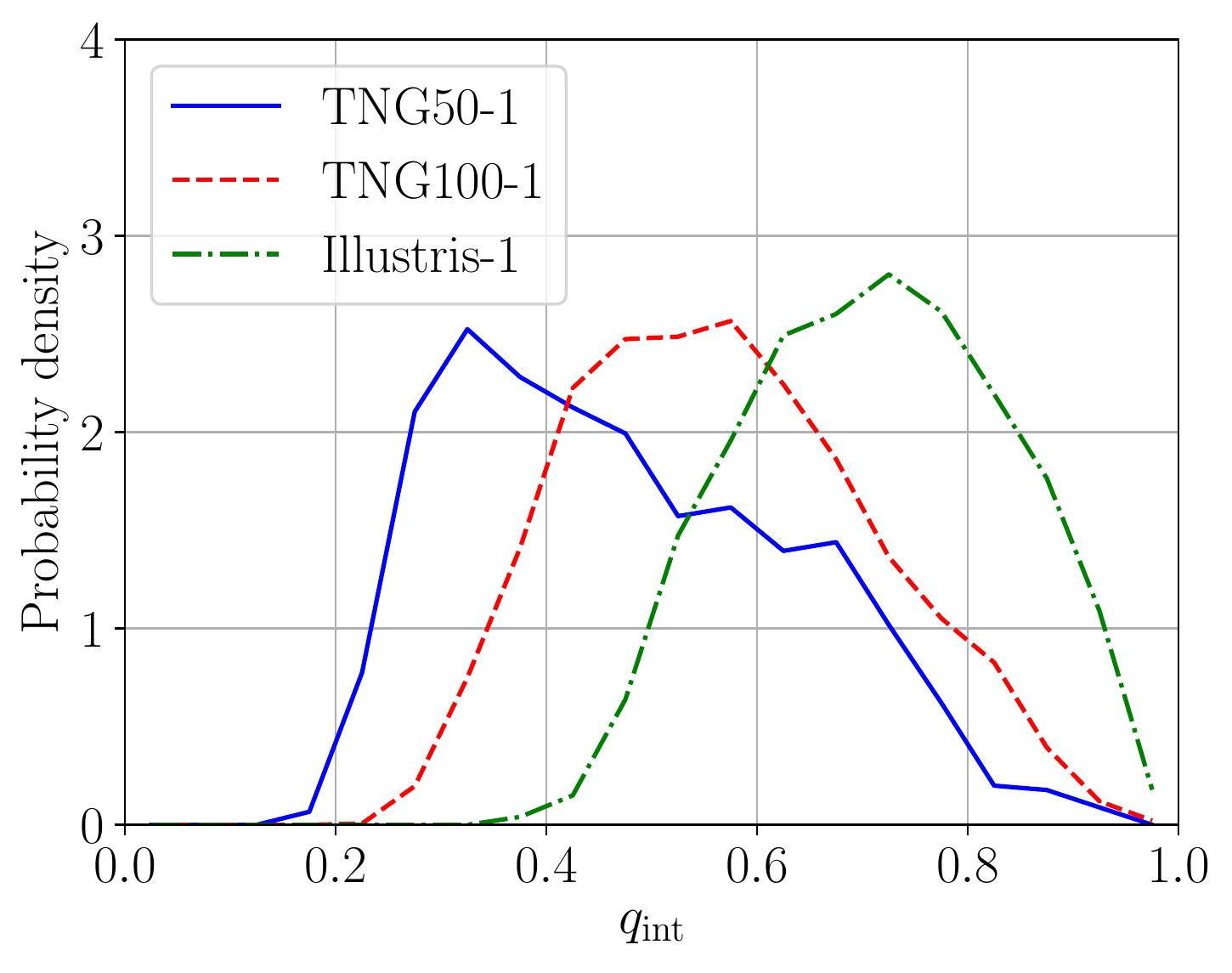}
		\includegraphics[width=8.5cm]{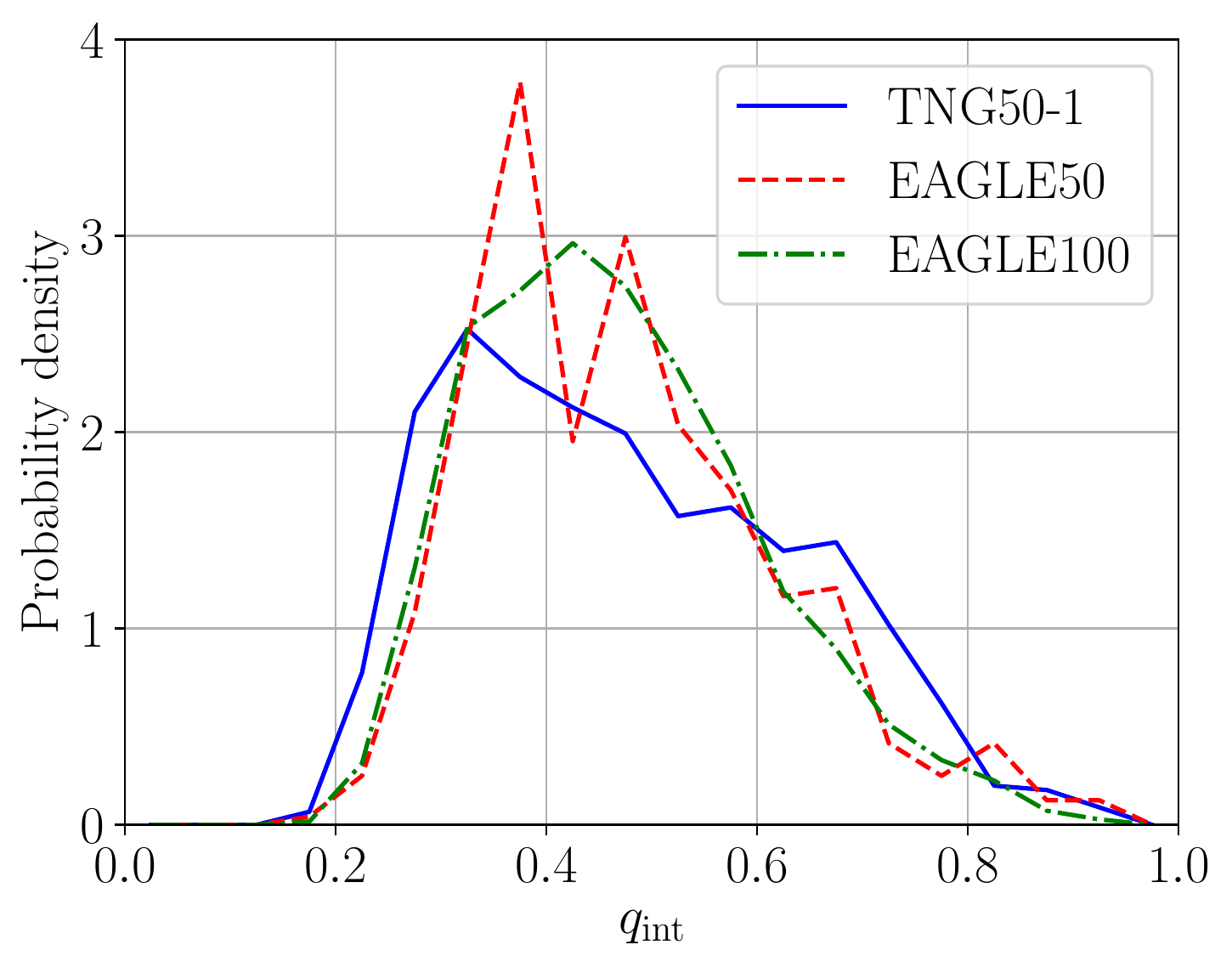}
	\end{center}
	\caption{Distribution of the intrinsic aspect ratio $q_{\mathrm{int}}$ for galaxies with $M_{*}~>~10^{10}\,M_\odot$ in the Illustris and TNG (\emph{left}) and EAGLE (\emph{right}) simulations, with the TNG50-1 result shown in both panels as a solid blue line for clarity. The bin width is $\Delta q_{\mathrm{int}} = 0.05$ throughout this work.}
	\label{fig:3D_aspect_ratio}
\end{figure*}


Figure~\ref{fig:3D_aspect_ratio} shows the $q_{\mathrm{int}}$ distribution of galaxies in central and noncentral subhalos with $M_{*} > 10^{10}\,M_\odot$ in the TNG50-1, TNG100-1, Illustris-1, EAGLE50, and EAGLE100 simulations. These all show a unimodal distribution with different peak positions. We show later that increasing the resolution broadens the peak and shifts it to smaller $q_{\mathrm{int}}$ (Section~\ref{subsec:Comparison of different TNG50 and EAGLE resolution realizations}). Galaxies in the EAGLE and TNG50-1 runs are typically thinner than in the Illustris-1 and TNG100-1 runs. The TNG50-1 run contains 133 galaxies with $q_{\mathrm{int}} < 0.3$ and $M_{*} > 10^{10}\,M_\odot$, with the thinnest galaxy of this sample having $q_{\mathrm{int}} \approx 0.19$. Clearly, the resolution and the adopted temperature floor of about $10^4$~K (10~kK; see section~2.1 of \citealt{Trayford_2017} for EAGLE and section~3.4 of \citealt{Nelson_2019} for TNG) allow for the formation of massive thin disk galaxies. This is consistent with the work of \citet{Sellwood_2019}, who managed to maintain a thin disk in a hydrodynamical CDM-based simulation of M33 with a slightly higher temperature floor of 12~kK (see their section 3.2).

The EAGLE runs have a very similar $q_{\mathrm{int}}$ distribution $-$ the peak position is similar to TNG50, and EAGLE forms galaxies as thin as $q_{\mathrm{int}} = 0.18$ (the thinnest galaxies in each run are $q_{\mathrm{int}} = 0.20$ for EAGLE25, $q_{\mathrm{int}} = 0.19$ for EAGLE50, and $q_{\mathrm{int}} = 0.18$ for EAGLE100). This is in agreement with TNG50-1 (right panel of Figure~\ref{fig:3D_aspect_ratio}). By using different resolution realizations of the TNG50 and EAGLE simulations, we test the numerical convergence of the aspect ratio distribution in Section~\ref{subsec:Comparison of different TNG50 and EAGLE resolution realizations}. While there is no evidence that the TNG50 simulation has numerically converged, the aspect ratio distributions of different EAGLE runs closely agree with each other, possibly because they have the same resolution. The similarity between EAGLE runs is consistent with \citet{Lagos_2018}, who showed that higher-resolution realizations of the EAGLE project yield galaxies with a similar ellipticity distribution (see their figures~A2 and A3). Moreover, the $q_{\mathrm{int}}$ distributions of EAGLE and TNG50-1 are very similar despite differences in resolution and other details of the models.

In the LV, $81\% \pm13 \%$ of galaxies with $M_{*} > 10^{10}\,M_\odot$ have $\widetilde{q}_{\mathrm{int}} < 0.4$ (Section~\ref{subsubsec:The Catalog of Neighboring Galaxies}). In contrast, the fractions of galaxies with such masses that have $q_{\mathrm{int}} < 0.4$ in the highest-resolution simulations analyzed here (TNG50-1, EAGLE25) are only $39\% \pm 2\%$ and $46\% \pm 8\%$, respectively. The stated Poisson uncertainties are estimated as $\sigma = \sqrt{N_{\mathrm{<0.4}} + 1}/N_{\mathrm{tot}}$, where $N_{\mathrm{<0.4}}$ and $N_{\mathrm{tot}}$ are the number of galaxies with $q_{\mathrm{int}} < 0.4$ and the total number of galaxies, respectively. The overall intrinsic aspect ratio distribution of the LV galaxies (blue dashed line in the left panel of Figure~\ref{fig:skyprojected_intrinsic_aspect_ratio_distribution}) is in $5.42\sigma$ tension with the TNG50-1 run. We emphasize again that the intrinsic aspect ratios of LV galaxies are based on the morphological $T$-type (Equation~\ref{eq:intrinsic_ttype}). Therefore, we provide in the following section another test of the $\Lambda$CDM framework in which we compare the sky-projected aspect ratio distributions of the $\Lambda$CDM simulations with the GAMA survey and SDSS.

\subsection{Sky-projected Aspect Ratio Distribution and Comparison with Observations}
\label{subsec:Sky-projected aspect ratio distribution and comparison with observations}

\begin{figure*}
	\begin{center}
		\includegraphics[width=8.5cm]{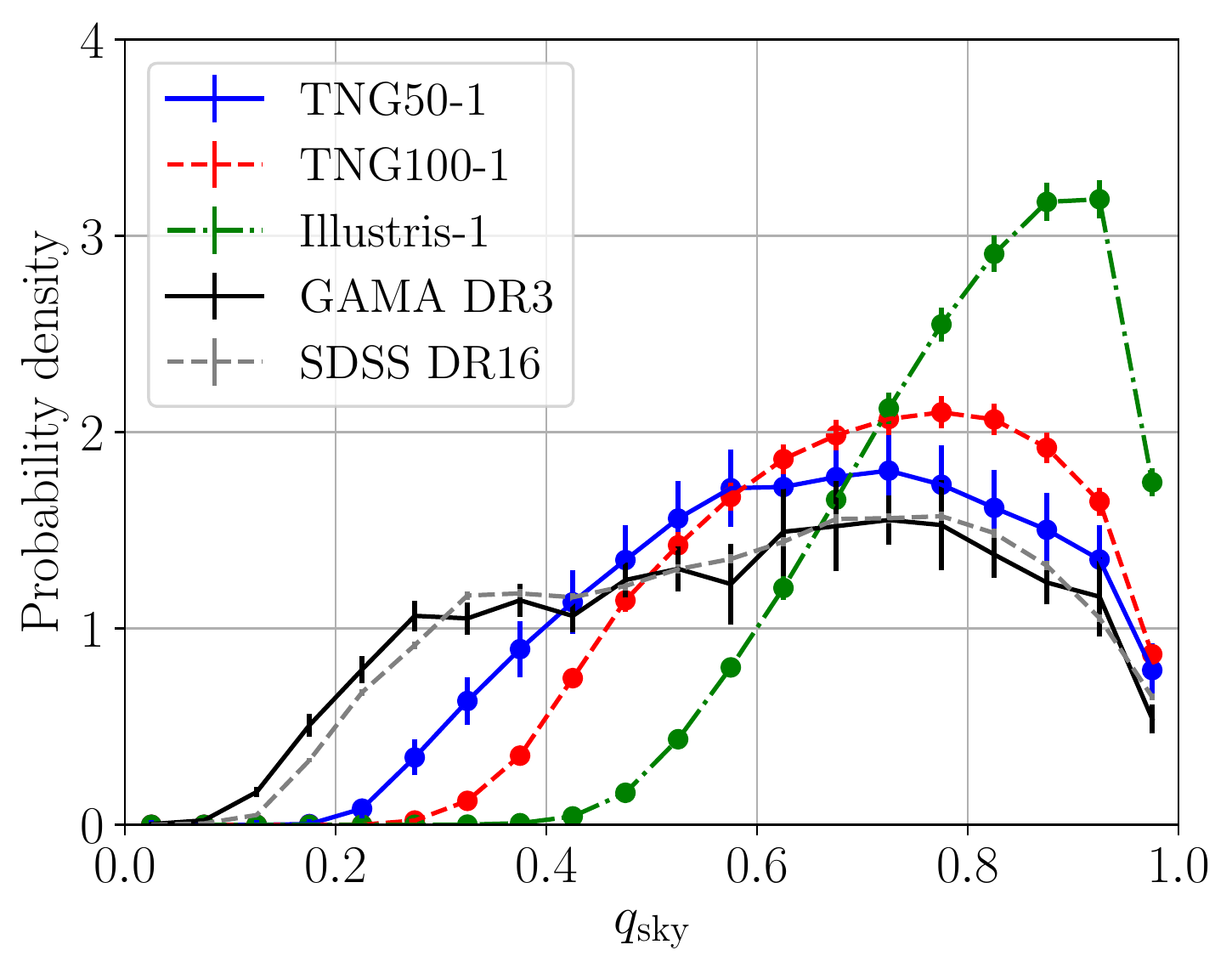}
		\includegraphics[width=8.5cm]{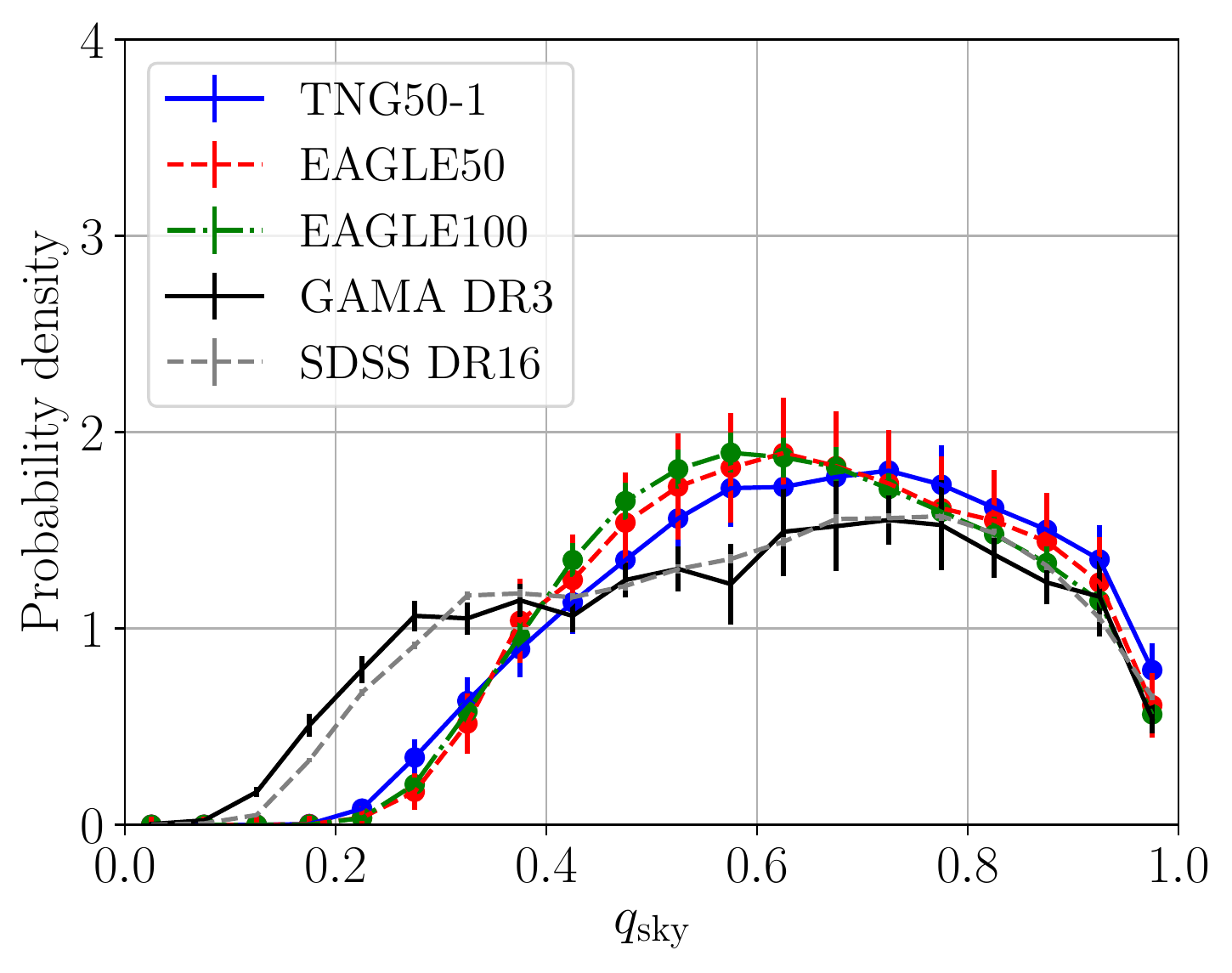}
	\end{center}
	\caption{Comparison between the observed distribution of $q_{\mathrm{sky}}$ and that produced by different cosmological $\Lambda$CDM simulations for galaxies with $10.0 < \log_{10}(M_{*}/M_\odot) \leq 11.65$. The observed $q_{\mathrm{sky}}$ distributions (black and gray points with error bars) have been weighted based on the stellar mass distribution of the TNG50-1 run (shown in Figure~\ref{fig:stellarmass_aspectratio}). Invisible error bars (Equation~\ref{eq:chi_sq_sigma_model}) are smaller than the data point symbol. The total $\chi^{2}$ values between the observed and simulated distributions (Equation~\ref{eq:chi_sq_model_observations}) and the corresponding levels of tension are reported in Table~\ref{tab:results}. The TNG50-1, TNG100-1, Illustris-1, EAGLE50, and EAGLE100 computations shown here use 882, 6424, 6842, 480, and 3613 subhalos, respectively.} 
	\label{fig:skyprojected_aspect_ratio}
\end{figure*}

In order to compare these simulation results with observations from the GAMA survey and SDSS (Section~\ref{subsec:Observational galaxy samples}), the simulated galaxies are projected onto the sky from a grid of viewing angles to find the distribution of $q_{\mathrm{sky}}$ (Section~\ref{subsec:Projecting an ellipsoid onto the sky}). The resulting $q_{\mathrm{sky}}$ distribution of each simulation is compared with the GAMA survey and SDSS in Figure~\ref{fig:skyprojected_aspect_ratio}. This reveals that the simulations significantly underproduce the fraction of galaxies with a low $q_{\mathrm{sky}}$. In particular, only about $10\%$ of simulated galaxies with $10.0 < \log_{10}(M_{*}/M_{\odot}) \leq 11.65$ have $q_{\mathrm{sky}} < 0.4$, while these galaxies make up about $22-25\%$ of locally observed galaxies (Table~\ref{tab:results_fraction}).

\begin{table}
  \centering
    \resizebox{\linewidth}{!}{
	\begin{tabular}{ll|ll}
	\hline
	\multicolumn{2}{c}{Simulations} & \multicolumn{2}{c}{Observations} \\
    \hline
    EAGLE100 & $0.089 \pm 0.005$ & Local 11 Mpc$^{*}$ & $0.212 \pm 0.067$\\
    EAGLE50 & $0.088 \pm 0.014$ & GAMA DR3$^{*}$ & $0.246 \pm 0.007$\\
    EAGLE25 & $0.106 \pm 0.038$  & GAMA DR3$^{**}$ & $0.238 \pm 0.009$\\
    Illustris-1 & $0.0005 \pm 0.0003$ & SDSS DR16$^{*}$ & $0.220 \pm 0.001$ \\
    TNG100-1 & $0.025 \pm 0.002$ & SDSS DR16$^{**}$ & $0.216 \pm 0.002$ \\ 
    TNG50-1 & $0.098 \pm 0.011$  & & \\
    \hline 
	\end{tabular}}
	\caption{Fraction of galaxies with $q_{\mathrm{sky}} < 0.4$ in the stellar mass range $10.0 < \log_{10}(M_{*}/M_\odot) \leq 11.65$ in different simulation runs (left columns) and observational surveys (right columns). The stated Poisson uncertainties are estimated as $\sigma = \sqrt{N_{\mathrm{<0.4}} + 1}/N_{\mathrm{tot}}$, where $N_{\mathrm{<0.4}}$ and $N_{\mathrm{tot}}$ are the number of galaxies with $q_{\mathrm{sky}} < 0.4$ and the total number of galaxies, respectively. The Poisson uncertainties for $M_{*}$-weighted distributions are given by Equation~\ref{eq:chi_sq_sigma_model}, where $w_{\mathrm{tot,i}}$ is here the sum of weights of galaxies with $q_{\mathrm{sky}} < 0.4$ and $w_{\mathrm{max,i}}$ is the maximum weight of such galaxies. \newline $^{*}$ No $M_{*}$-weighting applied. \newline $^{**}$ $M_{*}$-weighting applied based on the TNG50-1 stellar mass distribution.}
  \label{tab:results_fraction}
\end{table}

The tension between the simulated and observed $q_{\mathrm{sky}}$ distributions is quantified using a standard $\chi^{2}$ statistic (Section~\ref{sec:Quantifying the tension between simulations and observations}). All simulations significantly disagree with local observations. The smallest tension is $12.52\sigma$, which arises from comparing the GAMA survey with TNG50-1. The smallest tension for the SDSS is $14.82\sigma$, corresponding to a comparison with EAGLE50 (we do not consider results from the EAGLE25 run as it only has $81$ galaxies within the analyzed $M_*$ range, but see Section~\ref{subsec:Comparison of different TNG50 and EAGLE resolution realizations}). The total $\chi^{2}$ values and the corresponding levels of tension are listed in Table~\ref{tab:results} for different simulations.

\begin{table}
  \centering
	\begin{tabular}{lll}
	\hline
    Simulation & GAMA DR3 & SDSS DR16 \\ \hline 
    EAGLE100  & $551.20$ ($21.68\sigma$) & $2679.88$ ($50.68\sigma$) \\
    EAGLE50  & $252.89$ ($13.65\sigma$) & $288.51$ ($14.82\sigma$) \\
    Illustris-1 & $1768.16$ ($40.80\sigma$)  & $14689.73$ ($120.61\sigma$) \\
    TNG100-1 & $964.54$ ($29.54\sigma$) & $7421.62$ ($85.39\sigma$) \\ 
    TNG50-1 & $220.69$ ($12.52\sigma$) & $358.36$ ($16.89\sigma$) \\ \hline
    \end{tabular}
	\caption{Statistical comparison of the observed sky-projected aspect ratio distribution from GAMA DR3 and SDSS DR16 with the results of different cosmological $\Lambda$CDM simulations. The numbers show the total $\chi^{2}$ calculated from 20 bins (Equation~\ref{eq:chi_sq_model_observations}). The bracketed numbers correspond to the level of tension for $20$ degrees of freedom. The method for calculating the statistical significance of such extreme events is presented in Appendix~\ref{subsec:Statistical significance of extreme events}.}
  \label{tab:results}
\end{table}

\section{Discussion}
\label{sec:Discussion}

Although state-of-the-art cosmological $\Lambda$CDM simulations produce a variety of galaxy types \citep{Vogelsberger_2014, Schaye_2015}, we showed that the overall morphological distribution produced by the $\Lambda$CDM framework significantly disagrees with local observations. The $q_{\mathrm{sky}}$ distribution is similar between different observational samples, with the GAMA survey and SDSS being in $3.25\sigma$ tension with each other despite the large sample size of both (Figure~\ref{fig:stellarmass_aspectratio}). Here the important result has been documented that the different $\Lambda$CDM simulations disagree with the observed galaxy population. Galaxies formed in cosmological $\Lambda$CDM simulations are typically intrinsically thick rather than intrinsically thin (Section~\ref{subsec:Sky-projected aspect ratio distribution and comparison with observations}), making it challenging to explain the observed common formation of galaxies like the Milky Way or M31. This contrasts with the conclusion of \citet{Vogelsberger_2014} based on the Illustris-1 simulation, who claimed that the angular momentum problem of $\Lambda$CDM has been resolved. Although the simulations can produce \emph{some} disk galaxies, the fraction of such galaxies is far too small compared with observations (Table~\ref{tab:results_fraction}). The lack of such thin disk galaxies in simulations underlies the significant discrepancy between the observed and simulated distributions of galaxy shapes.

Our results broadly agree with those of \citet{vandenSande_2019}, who showed that the intrinsic (see their figure~10) and sky-projected aspect ratio distributions (their figures~4 and 8) of the EAGLE \citep{Schaye_2015, Crain_2015, McAlpine_2016}, HYDRANGEA \citep{Bahe_2017, Barnes_2017}, HORIZON-AGN \citep{Dubois_2014}, and MAGNETICUM \citep{Hirschmann_2014, Dolag_2016, Dolag_2017} simulations all disagree with observational data from the ATLAS\textsuperscript{3D}, Calar Alto Legacy Integral Field Area Survey (CALIFA), and MASSIVE surveys, and also with the SAMI Galaxy Survey (MAGNETICUM produced too few very round galaxies). Similar results were obtained by \citet{Peebles_2020}, who recently showed that stellar particles in $\Lambda$CDM subhalos have kinematics different to stars in local galaxies. However, the very small sample size made a statistical comparison difficult.

\subsection{The Effect of Numerical Resolution}
\label{subsec:Comparison of different TNG50 and EAGLE resolution realizations}

The fact that zoom-in $\Lambda$CDM simulations can produce thin disk galaxies \citep[e.g.][]{Wetzel_2016} suggests that further increasing the numerical resolution may solve the discrepancy reported in Section~\ref{subsec:Sky-projected aspect ratio distribution and comparison with observations}. For example, \citet{Ludlow_2021} recently argued that the coarse-grained implementation of dark matter halos causes an artificial heating of the stellar particles, which increases the vertical velocity dispersion of the galaxies. This yields thicker disks than better resolved galaxies, which could be the reason for the here reported tension. According to their table~3, the vertical velocity dispersion $\sigma_z$ of TNG50-1 galaxies embedded in a dark matter halo of $M_{200} < 7.9 \times 10^{11} \, M_{\odot}$ is numerically increased by $\Delta \sigma_z > 0.1 \, V_{200}$, where $V_{200}$ is the virial velocity of the halo, and $M_{200}$ is its virial mass.

In this section, we analyze the effect of numerical resolution on the distribution of galaxy shapes using the TNG50 simulation. The flagship of the TNG project is called TNG50-1, which has the highest resolution among the here analyzed simulation runs. In addition, the TNG50 simulation suite includes the runs TNG50-2, TNG50-3, and TNG50-4, where a higher suffixed number indicates a lower-resolution realization as summarized in Table~\ref{tab:simulations_parameters}. The particle mass differs by a factor of eight between any TNG50 run and the next higher-resolution realization. 

\begin{figure*}
	\includegraphics[width=8.5cm]{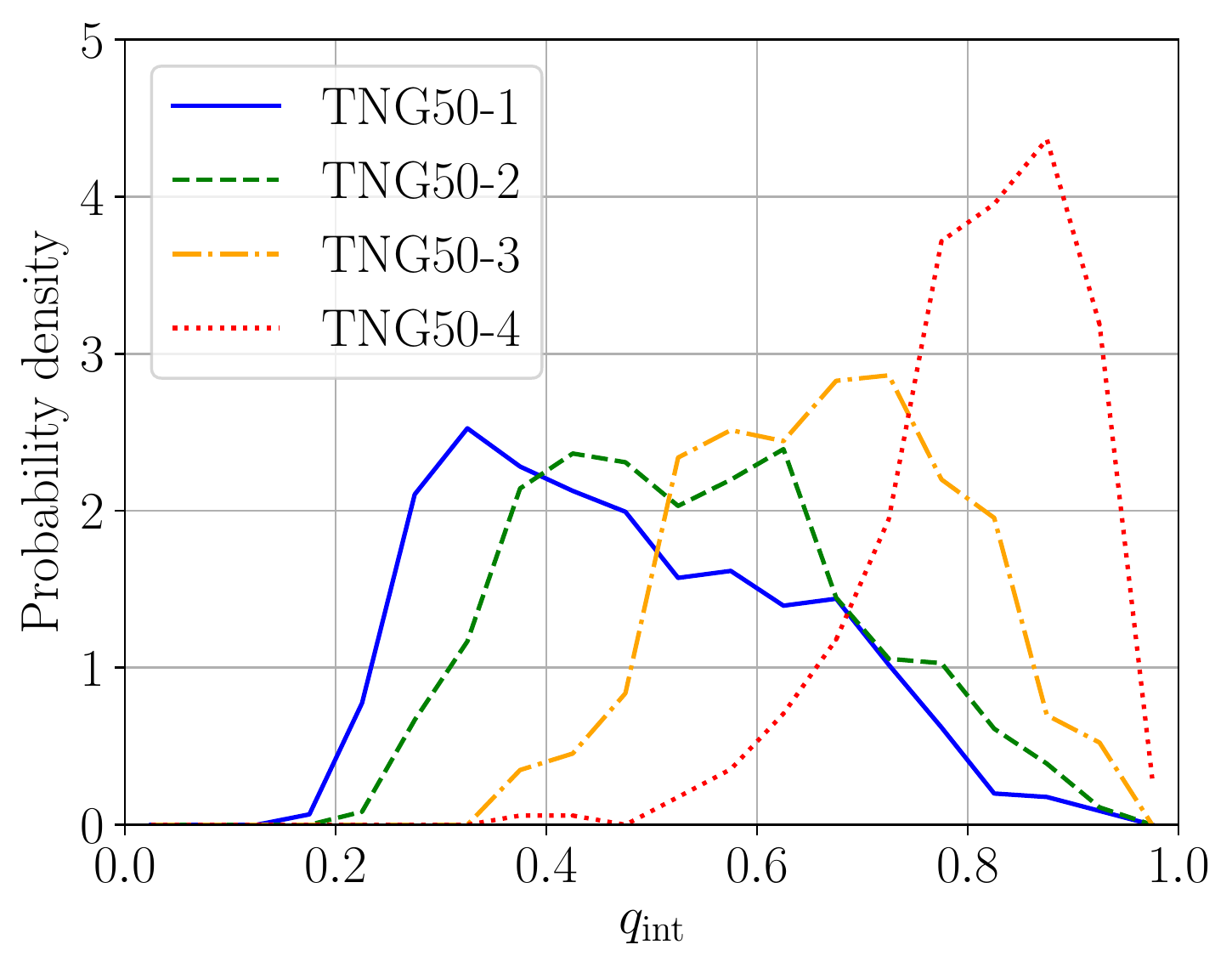}
	\includegraphics[width=8.5cm]{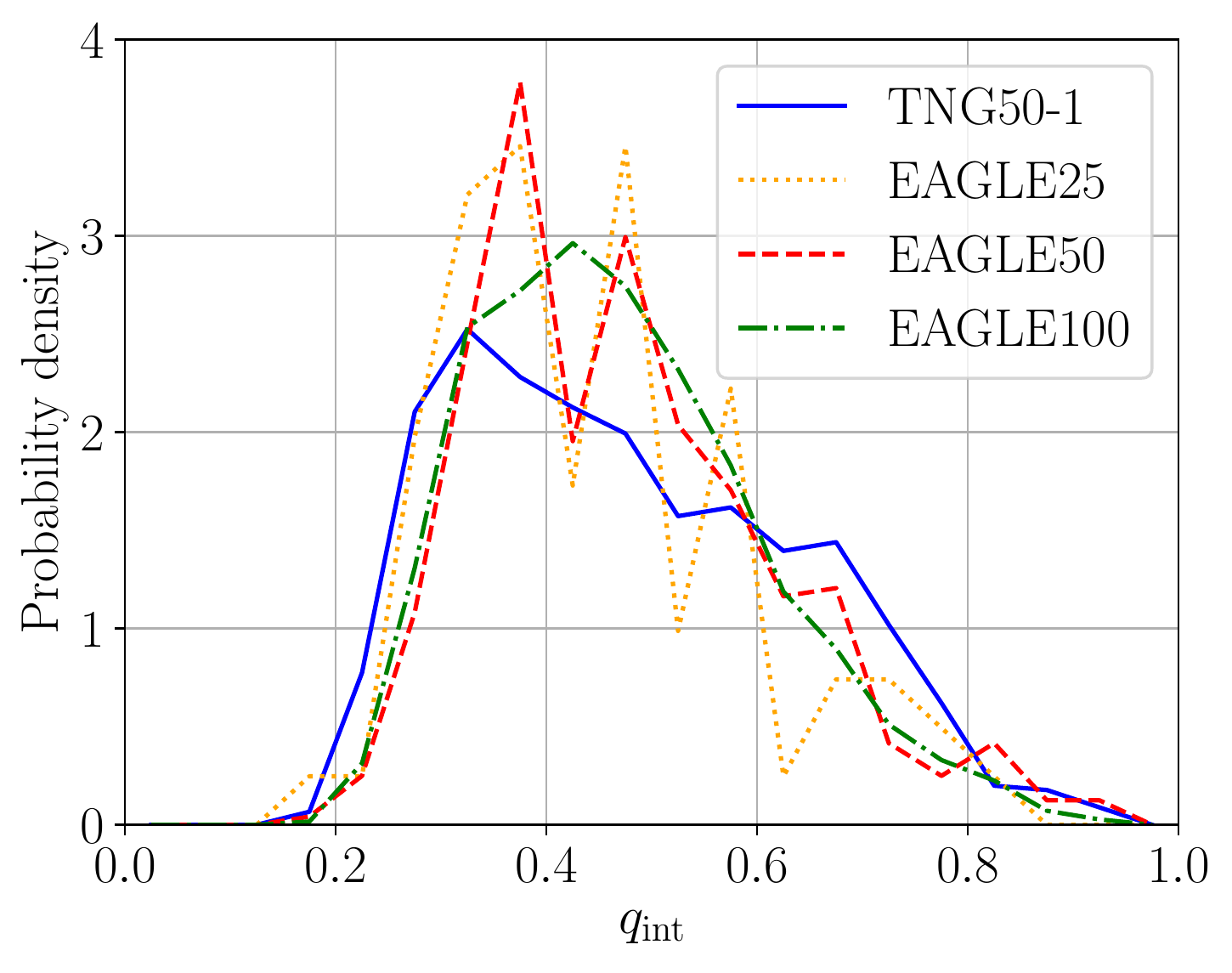}
    \caption{Intrinsic aspect ratio distribution of subhalos with $M_{*} > 10^{10}\,M_{\odot}$ for different resolution realizations of the simulation TNG50 (\emph{left}) and EAGLE (\emph{right}). A higher suffixed number for the TNG50 runs indicates a lower-resolution realization (i.e. TNG50-1 has the highest resolution). The simulated distributions are statistically compared in Table~\ref{tab:results_convergence}. EAGLE100 and EAGLE50 have the same resolution, while EAGLE25 has a higher resolution (Table~\ref{tab:simulations_parameters}).}
	\label{fig:TNG50_resolution}
\end{figure*}

The left panel of Figure~\ref{fig:TNG50_resolution} shows the intrinsic aspect ratio distribution for different resolution realizations of the TNG50 simulation. The formation of thin galaxies and the peak position of the $q_{\mathrm{int}}$ distribution strongly depend on the resolution: the mode shifts from $q_{\mathrm{int}} = 0.88$ for TNG50-4 to $q_{\mathrm{int}} = 0.33$ for TNG50-1. 

The convergence of different simulation runs is statistically quantified in Table~\ref{tab:results_convergence}. The $q_{\mathrm{int}}$ distributions of the TNG50-1 and TNG50-2 runs disagree with each other at $7.60\sigma$. Although this is still a high tension, it is much lower compared to that between TNG50-2(-3) and TNG50-3(-4), which differ at the $13.75\sigma$ ($14.17\sigma$) confidence level. Based on the $q_{\mathrm{sky}}$ distribution, the TNG50-1 and TNG50-2 runs disagree at only $0.83\sigma$.

The right panel of Figure~\ref{fig:TNG50_resolution} compares the $q_{\mathrm{int}}$ distribution of TNG50-1 with EAGLE100, EAGLE50, and the higher-resolution realization EAGLE25 (Recal-L0025N0752). The similarity of the EAGLE results (Table~\ref{tab:results_convergence}) implies that the EAGLE simulations have numerically converged, so the deficit of intrinsically thin galaxies cannot be explained by resolution effects $-$ as also concluded by \citet{Lagos_2018} and apparent in their figures A2 and A3. Although the $q_{\mathrm{int}}$ distribution of the EAGLE25 run is very similar to EAGLE50 or EAGLE100, the tension between the sky-projected aspect ratio distribution of EAGLE25 and GAMA DR3 (SDSS) is only $1.44\sigma$ ($0.37\sigma$). This is because EAGLE25 only has 81 galaxies in the stellar mass range $10.0 < \log_{10}(M_{*}/M_\odot) \leq 11.65$, which results in large Poisson uncertainties that decrease therewith the total~$\chi^{2}$.

Importantly, the $q_{\mathrm{sky}}$ distributions of galaxies in TNG50-1 and EAGLE50 (EAGLE100) are consistent with each other at $1.4 \times 10^{-3}\sigma$ ($0.58\sigma$) confidence (Table~\ref{tab:results_convergence}). Thus, we demonstrate for the first time that $\Lambda$CDM simulations with SPH-based (EAGLE) and adaptive grid (Illustris/TNG) methods produce the same sky-projected aspect ratio distribution. The baryonic algorithms are also quite different between these simulations.

\begin{table*}
  \centering
    \begin{tabular}{lll}
	\hline
    Simulations & $\chi_{\mathrm{int}}^2$ between them (tension) & $\chi_{\mathrm{sky}}^2$ between them (tension) \\
    \hline
    TNG50-3 vs. TNG50-4 & $268.26$ ($14.17\sigma$) & $91.80$ ($6.69\sigma$) \\
    TNG50-2 vs. TNG50-3 & $255.73$ ($13.75\sigma$) & $105.10$ ($7.45\sigma$) \\
    TNG50-1 vs. TNG50-2 & $107.85$ ($7.60\sigma$) & $20.82$ ($0.83\sigma$) \\
    TNG50-1 vs. EAGLE100 & $87.83$ ($6.35\sigma$) & $18.35$ ($0.58\sigma$) \\
    TNG50-1 vs. EAGLE50 & $56.78$ ($4.24\sigma$) & $6.04$ ($0.0014\sigma$) \\
    TNG50-1 vs. EAGLE25 & $19.70$ ($0.71\sigma$) & $2.98$ ($4.83 \times 10^{-6}\sigma$) \\
    EAGLE100 vs. EAGLE25 & $19.23$ ($0.66\sigma$) & $0.50$ ($2.80 \times 10^{-13}\sigma$) \\
    EAGLE50 vs. EAGLE25 & $12.87$ ($0.15\sigma$) & $0.95$ ($1.26 \times 10^{-10}\sigma$) \\
    \hline 
	\end{tabular}
	\caption{Testing the numerical convergence of different simulation runs (column 1) for subhalos with $M_{*} > 10^{10}\,M_\odot$ by showing the total $\chi^2$ (Equation~\ref{eq:chi_sq_models}) between their distributions of $q_{\mathrm{int}}$ (column 2) and $q_{\mathrm{sky}}$ (column 3), along with the corresponding level of tension (bracketed numbers). We use 20 bins in $q_{\mathrm{int}}$ and $q_{\mathrm{sky}}$, giving 20 degrees of freedom. EAGLE50 and EAGLE100 use the same resolution, so the similarity between their results ($\chi^2$ not shown) is not a strong test of numerical convergence. Note that the higher resolution realization EAGLE25 only has 81 galaxies with $M_{*} > 10^{10}\,M_{\odot}$, which reduces the total $\chi^{2}$ because of the higher Poisson uncertainties. The intrinsic aspect ratio distributions underlying these comparisons are plotted in Figure~\ref{fig:TNG50_resolution}.}
  \label{tab:results_convergence}
\end{table*}

In order to quantify the thickening of simulated galaxies due to limited numerical resolution \citep[e.g.][]{Ludlow_2021}, we apply as an ansatz a parametric correction to the intrinsic shape distribution. The short axis $\lambda_1$ of each galaxy is scaled down by the factor $x$ such that its $q_{\mathrm{int}}$ is corrected by
\begin{eqnarray}
   q_{\mathrm{int, corr}} ~=~ x q_{\mathrm{int}} \, , \quad x ~=~ \alpha + \left( 1 - \alpha \right) q_{\mathrm{int}} \, .
    \label{eq:alpha_prediction}
\end{eqnarray}
$\alpha$ is a variable ranging from $0-1$ in steps of $\Delta \alpha = 0.01$. Applying this correction to the $q_{\mathrm{int}}$ of simulated galaxies keeps a galaxy with a high $q_{\mathrm{int}}$ round, but makes a thin disk galaxy even thinner. $\alpha = 1$ does not change the $q_{\mathrm{int}}$ of a galaxy, while $\alpha = 0$ yields a corrected $q_{\mathrm{int, corr}} = {q_{\mathrm{int}}}^2$, therewith causing the most substantial thinning of the galaxy population in our parameterization.

\begin{figure*}
	\begin{center}
		\includegraphics[width=8.5cm]{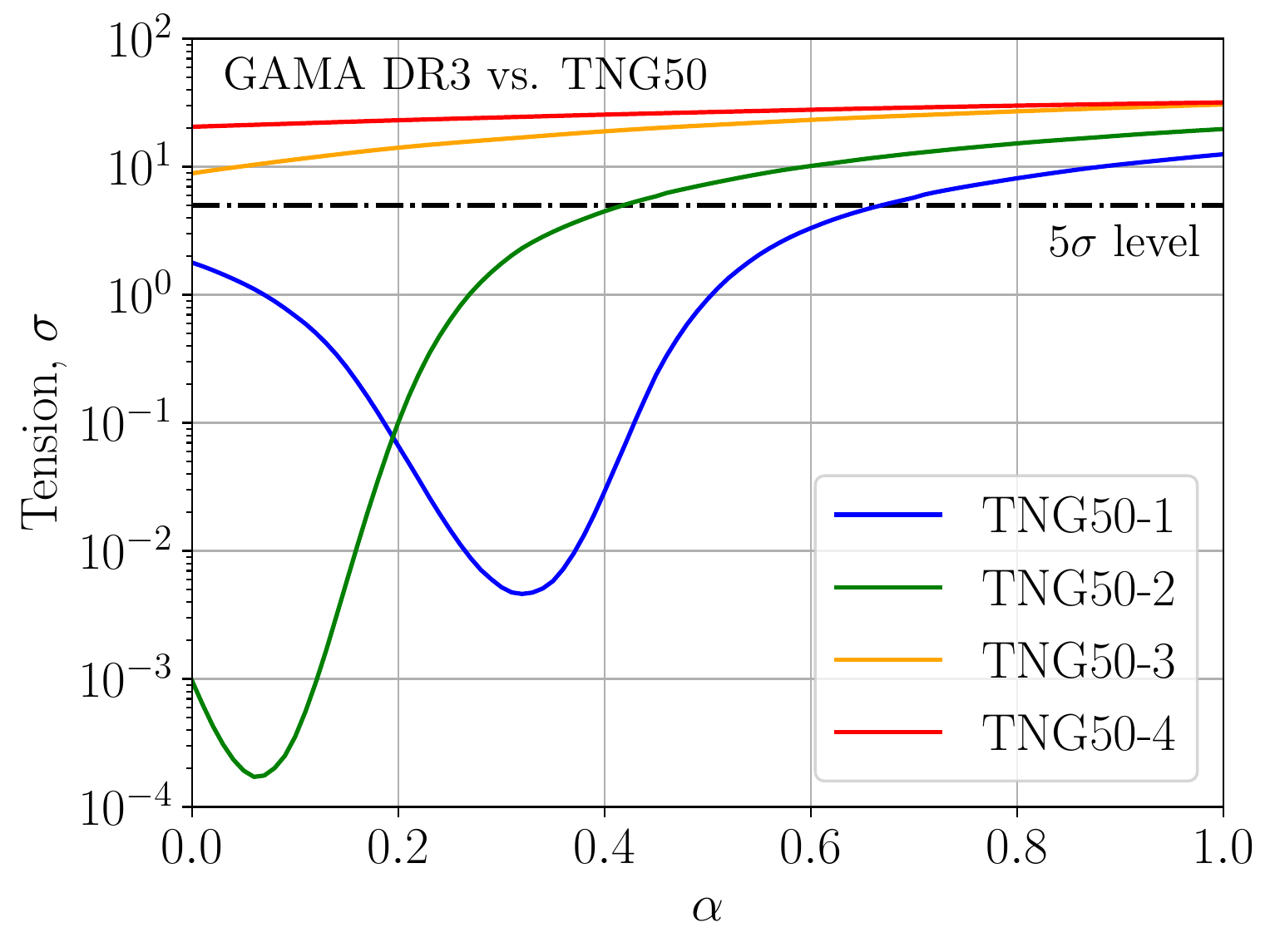}
		\includegraphics[width=8.5cm]{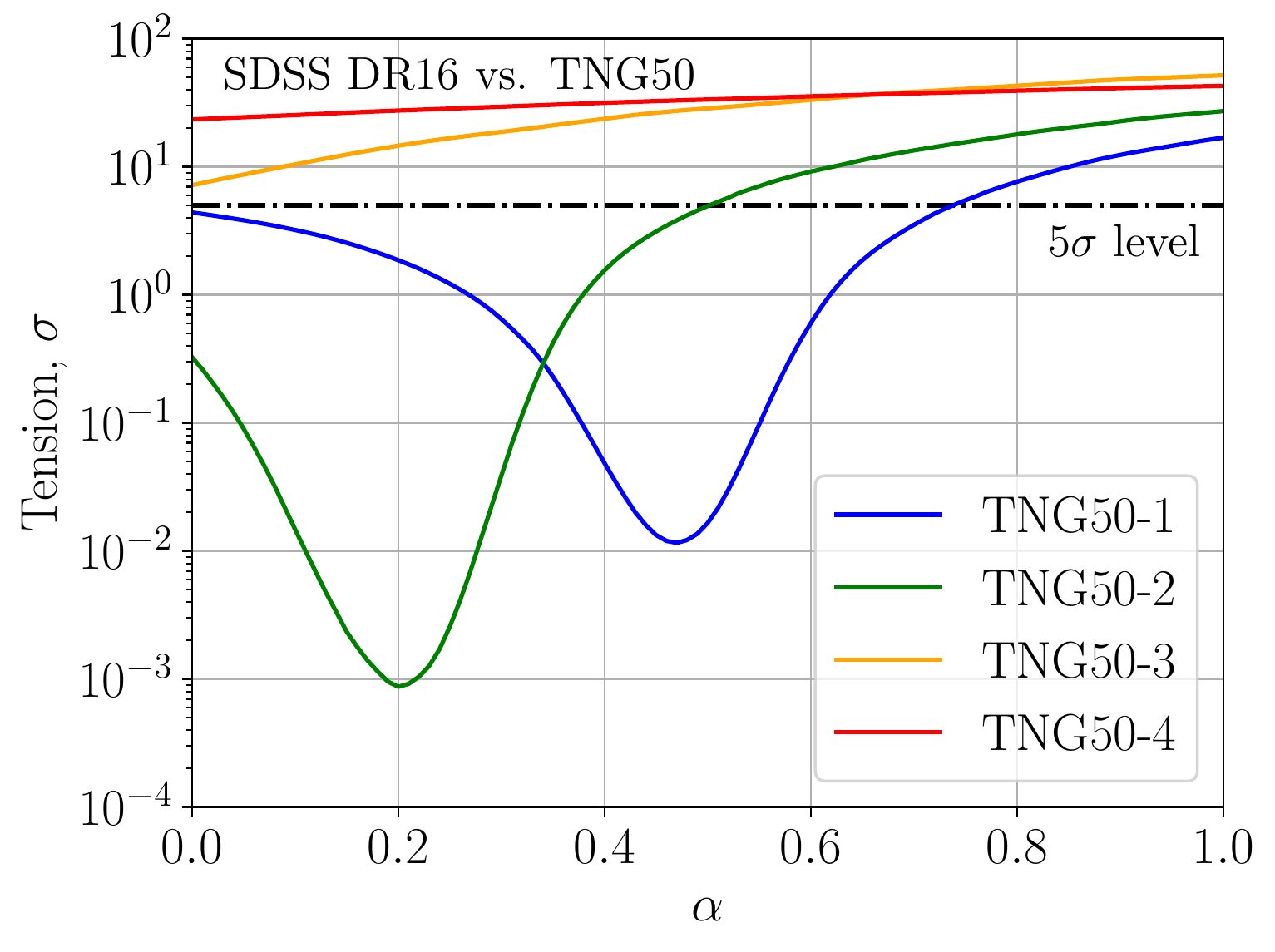}
	\end{center}
	\caption{Level of tension between the observed and simulated $q_{\mathrm{sky}}$ distribution of galaxies with $10.0 < \log_{10}(M_{*}/M_\odot) \leq 11.65$ in dependence of $\alpha$ (Equation~\ref{eq:alpha_prediction}) for different resolution realizations of the TNG50 simulation, which we show with different colored lines (see the legend). The left panel considers GAMA as the observational sample, while the right panel uses SDSS. The tension between TNG50-1 and GAMA DR3 (SDSS DR16) reaches the $5\sigma$ confidence level (dashed horizontal line) if $\alpha = 0.668$ ($\alpha = 0.738$), with the tension becoming minimal if instead $\alpha = 0.32$ ($\alpha = 0.47$), the tension in this case being only $4.6\times 10^{-3}\sigma$ ($1.2\times10^{-2}\sigma$).} 
	\label{fig:TNG50_GAMA_SDSS_alpha}
\end{figure*}

\begin{figure*}
	\begin{center}
		\includegraphics[width=8.5cm]{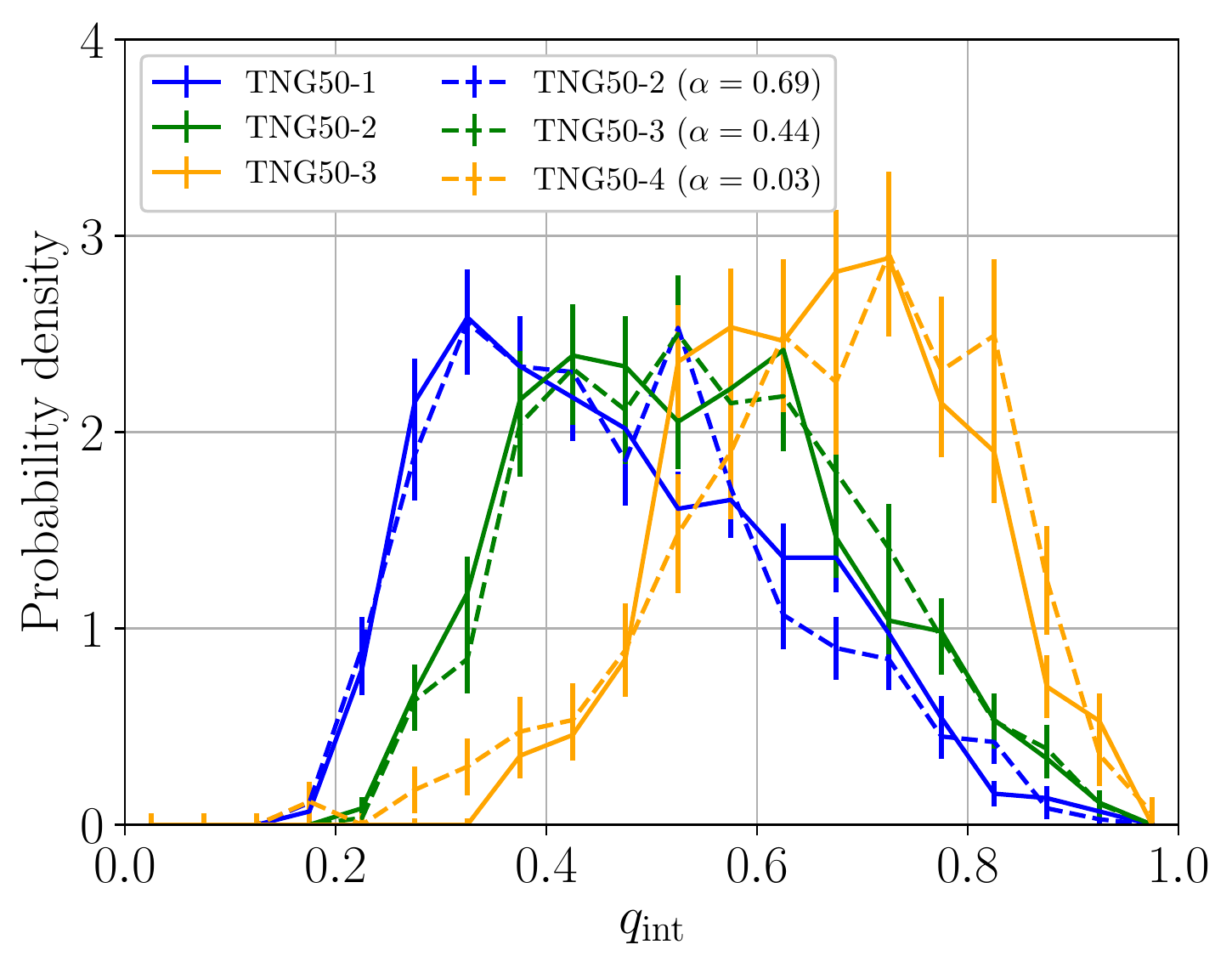}
		\includegraphics[width=8.5cm]{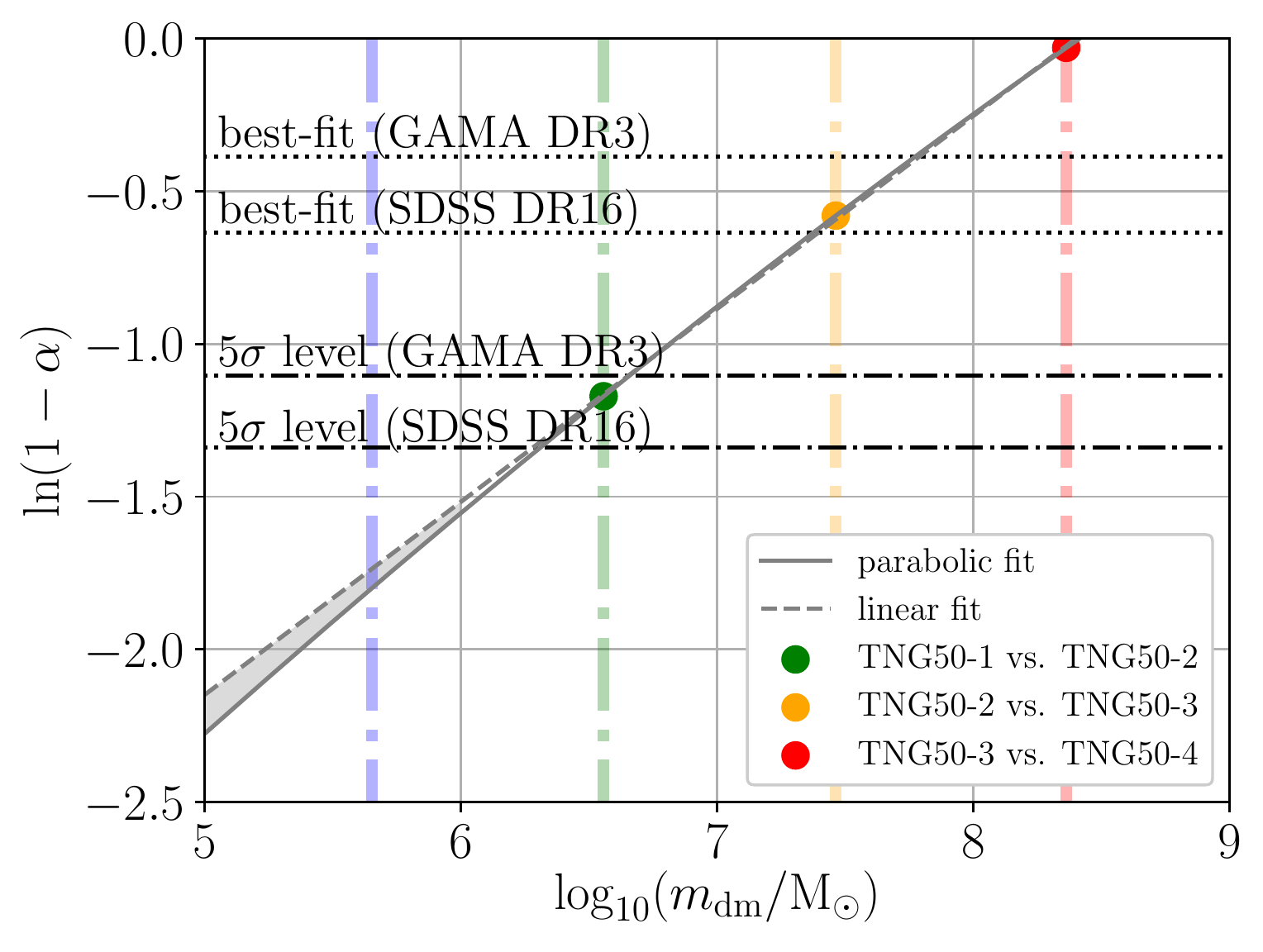}
	\end{center}
	\caption{\emph{Left:} the $q_{\mathrm{int}}$ distribution of subhalos with $10.0 < \log_{10}(M_{*}/M_{\odot}) \leq 11.65$ in each TNG50 simulation (solid lines) is compared with the corrected distribution for subhalos in the next lower-resolution simulation, which we show using a dashed line of the same color. The applied correction factors and resulting levels of tension are $\alpha = 0.69$ ($0.80\sigma$), $\alpha = 0.44$ ($5.9 \times 10^{-3}\sigma$), and $\alpha = 0.03$ ($0.85\sigma$) for subhalos in TNG50-2, TNG50-3, and TNG50-4, respectively. These corrections minimize the tension with the uncorrected $q_{\mathrm{int}}$ distribution of the next higher-resolution simulation. \emph{Right:} checking for convergence of the TNG50 $q_{\mathrm{int}}$ distribution with respect to the dark matter mass resolution $m_{\mathrm{dm}}$ for galaxies with $10.0 < \log_{10}(M_{*}/M_\odot) \leq 11.65$. The above-mentioned $\alpha$ values are plotted in terms of $\ln(1-\alpha)$ against $\log_{10}(m_{\mathrm{dm}}/M_{\odot})$. The vertical lines mark the mass of a dark matter particle in the TNG50 runs as listed in Table~\ref{tab:simulations_parameters}. The dotted-dashed horizontal lines mark $\alpha = 0.668$ and $\alpha = 0.738$, which are required for the $q_{\mathrm{sky}}$ distribution of the TNG50-1 run to match that of GAMA DR3 and SDSS DR16, respectively, at the $5\sigma$ confidence level. The dotted lines represent $\alpha = 0.32$ and $0.47$ because these values would minimize the tension of TNG50-1 with GAMA and SDSS, respectively. The dashed (solid) gray line is a linear (parabolic) fit. The gray shaded region indicates how much the $q_{\mathrm{int}}$ distribution could differ in an even higher-resolution run than TNG50-1. We estimate that using an eight-times lower dark matter particle mass than TNG50-1 (dotted-dashed blue vertical line) is equivalent to scaling galaxies in it by $\alpha$ in the range between $0.834$ (parabolic fit) and $0.824$ (linear fit).}
	\label{fig:TNG_prediction}
\end{figure*}

The tension between the observed and rescaled simulated $q_{\mathrm{sky}}$ distributions is shown in Figure~\ref{fig:TNG50_GAMA_SDSS_alpha} for different resolution realizations of the TNG50 simulation. As expected from our previous analysis, the tension systematically decreases with higher resolution. The tension between TNG50-1 and the GAMA survey (SDSS) reaches the $5\sigma$ confidence level if galaxies in the TNG50-1 run are corrected by $\alpha = 0.668$ ($\alpha = 0.738$). The tension between TNG50-1 and GAMA (SDSS) is minimized for a correction factor of $\alpha = 0.32$ ($\alpha = 0.47$), the tension in this case being only $4.6 \times 10^{-3}\sigma$ ($1.2\times10^{-2}\sigma$). This demonstrates that our simple parametric correction with a single parameter $\alpha$ (Equation~\ref{eq:alpha_prediction}) can make the simulations agree very well with observations.

We can use this procedure to quantify how improving the resolution affects the simulated aspect ratio distribution, which then allows us to quantify if the simulations are numerically converged. To do this, we first find which $\alpha$ value must be applied to a TNG50 run to minimize the tension with the uncorrected $q_{\mathrm{int}}$ distribution of the next higher resolution realization with an $8 \times$ higher mass resolution. We find that the tensions between the $q_{\mathrm{int}}$ distributions of the rescaled TNG50-4 and TNG50-3, rescaled TNG50-3 and TNG50-2, and rescaled TNG50-2 and TNG50-1 runs become minimal for $\alpha$ values of $0.03$ ($0.85\sigma$), $0.44$ ($5.9 \times 10^{-3}\sigma$), and $0.69$ ($0.80\sigma$) applied to the TNG50-4, TNG50-3, and TNG50-2 runs, respectively. The true $q_{\mathrm{int}}$ distribution of each TNG50 run and the so-corrected distribution of the next lower resolution TNG50 run is compared in the left panel of Figure~\ref{fig:TNG_prediction}, demonstrating that our parametric correction to e.g. TNG50-2 galaxy shapes reproduces quite well the actual $q_{\mathrm{int}}$ distribution of TNG50-1. The required $\alpha$ value increases with higher resolution, but we cannot confirm that the TNG50 simulations have numerically converged, as that would be achieved if $\alpha = 1$.

In a second step, we extrapolate the $\alpha$ values by plotting $\ln \left( 1 - \alpha \right)$ against the logarithmic dark matter particle mass of the corresponding TNG50 run (right panel of Figure~\ref{fig:TNG_prediction}). The $\alpha$ value required for galaxies in the TNG50-1 run to mimic the result of an eight-times-lower dark matter particle mass simulation is estimated using a parabolic and a linear fit in the $\log_{10} \left( m_{\mathrm{dm}}/M_{\odot} \right)$ vs. $\ln \left( 1 - \alpha \right)$ diagram of Figure~\ref{fig:TNG_prediction}. This predicts $\alpha = 0.824$ (0.834) for a linear (parabolic) fit. Being more conservative, we apply the linearly extrapolated result that $\alpha = 0.824$ to the TNG50-1 run, which yields an $8.68\sigma$ ($8.71\sigma$) tension with GAMA (SDSS). Thus, an eight-times higher-resolution realization of the TNG50-1 run would almost certainly not reduce the here reported tension below the $5\sigma$ confidence level.

In the third and final step, we estimate the tension with observations by extrapolating the data further to five more refinement levels than in the TNG50-1 run. Therefore, we extract the $\alpha$ values not only for an eight-times-lower $m_{\mathrm{dm}}$ but also for an $8^2\times$, $8^3\times$, $8^4\times$, and $8^5\times$ lower $m_{\mathrm{dm}}$ than in the TNG50-1 run. These so-obtained $\alpha$ values for the five different resolution levels are listed in Table~\ref{tab:results_resolution_extrapolation} for the parabolic and linear extrapolations. The $\alpha$ values are then successively applied to estimate the intrinsic aspect ratio of each subhalo in an $8^5\times$ higher-resolution realization than TNG50-1. In other words, the aspect ratio distribution for an $8^2\times$ higher-resolution run than TNG50-1 is obtained by correcting the aspect ratios of subhalos in the eight-times higher-resolution run by applying Equation~\ref{eq:alpha_prediction} with $\alpha = 0.915$ (parabolic fit) or $\alpha = 0.900$ (linear fit). This procedure is repeated until we reach an $8^5\times$ higher refinement level, which on the last step requires applying $\alpha = 0.991$ (parabolic fit) or $\alpha = 0.982$ (linear fit) to the $q_{\mathrm{int}}$ distribution of the $8^4\times$ higher-resolution realization than TNG50-1. Since $\alpha$ is now close to unity, this almost reaches full convergence in $q_{\mathrm{int}}$. Further improvements to the resolution can be expected to have sub-percent level effects on the intrinsic aspect ratios.

\begin{table*}
  \centering
    \resizebox{\linewidth}{!}{
    \begin{tabular}{lllllll}
	\hline
    & \multicolumn{2}{c}{$\alpha$ (Equation~\ref{eq:alpha_prediction})} & \multicolumn{2}{c}{$\chi^2$ and Tension with GAMA DR3} & \multicolumn{2}{c}{$\chi^2$ and Tension with SDSS DR16} \\
    Resolution & Parabolic Fit & Linear Fit & Parabolic Fit & Linear Fit & Parabolic Fit & Linear Fit \\
    \hline
    TNG50-1 & $-$ & $-$ & $220.69$ ($12.52\sigma$) & $220.69$ ($12.52\sigma$) & $358.36$ ($16.89\sigma$) & $358.36$ ($16.89\sigma$) \\
    $8\times$ & $0.834$ & $0.824$ & $134.17$ ($8.94\sigma$) & $128.96$ ($8.68\sigma$) & $140.34$ ($9.23\sigma$) & $129.60$ ($8.71\sigma$) \\
    $8^2\times$ & $0.915$ & $0.900$ & $99.29$ ($7.12\sigma$) & $90.08$ ($6.59\sigma$) & $77.38$ ($5.72\sigma$) & $63.95$ ($4.78\sigma$) \\
    $8^3\times$ & $0.958$ & $0.944$ & $84.94$ ($6.18\sigma$) & $71.83$ ($5.34\sigma$) & $57.06$ ($4.26\sigma$) & $41.14$ ($2.91\sigma$) \\
    $8^4\times$ & $0.980$ & $0.968$ & $78.27$ ($5.77\sigma$) & $63.12$ ($4.72\sigma$) & $48.57$ ($3.57\sigma$) & $32.31$ ($2.05\sigma$) \\
    $8^5\times$ & $0.991$ & $0.982$ & $75.37$ ($5.58\sigma$) & $58.36$ ($4.36\sigma$) & $45.12$ ($3.27\sigma$) & $28.06$ ($ 1.61\sigma$) \\ \hline 
	\end{tabular}}
	\caption{Statistical comparison of the observed sky-projected aspect ratio distributions from GAMA DR3 (fourth and fifth columns) and SDSS DR16 (sixth and seventh columns) with the results of five more refinement levels (first column) than in the TNG50-1 run. The second and third columns list the $\alpha$ values obtained by extrapolating the TNG50 runs to an $8\times$, $8^2\times$, $8^3\times$, $8^4\times$, and $8^5\times$ lower dark matter particle mass than in TNG50-1 using a parabolic or a linear fit in the $\log_{10} \left( m_{\mathrm{dm}}/M_{\odot} \right)$ vs. $\ln \left( 1 - \alpha \right)$ diagram (right panel of Figure~\ref{fig:TNG_prediction}). These so-obtained $\alpha$ values have been successively applied to the intrinsic aspect ratios of subhalos in TNG50-1 (see the text).}
  \label{tab:results_resolution_extrapolation}
\end{table*}

\begin{figure*}
	\begin{center}
		\includegraphics[width=8.5cm]{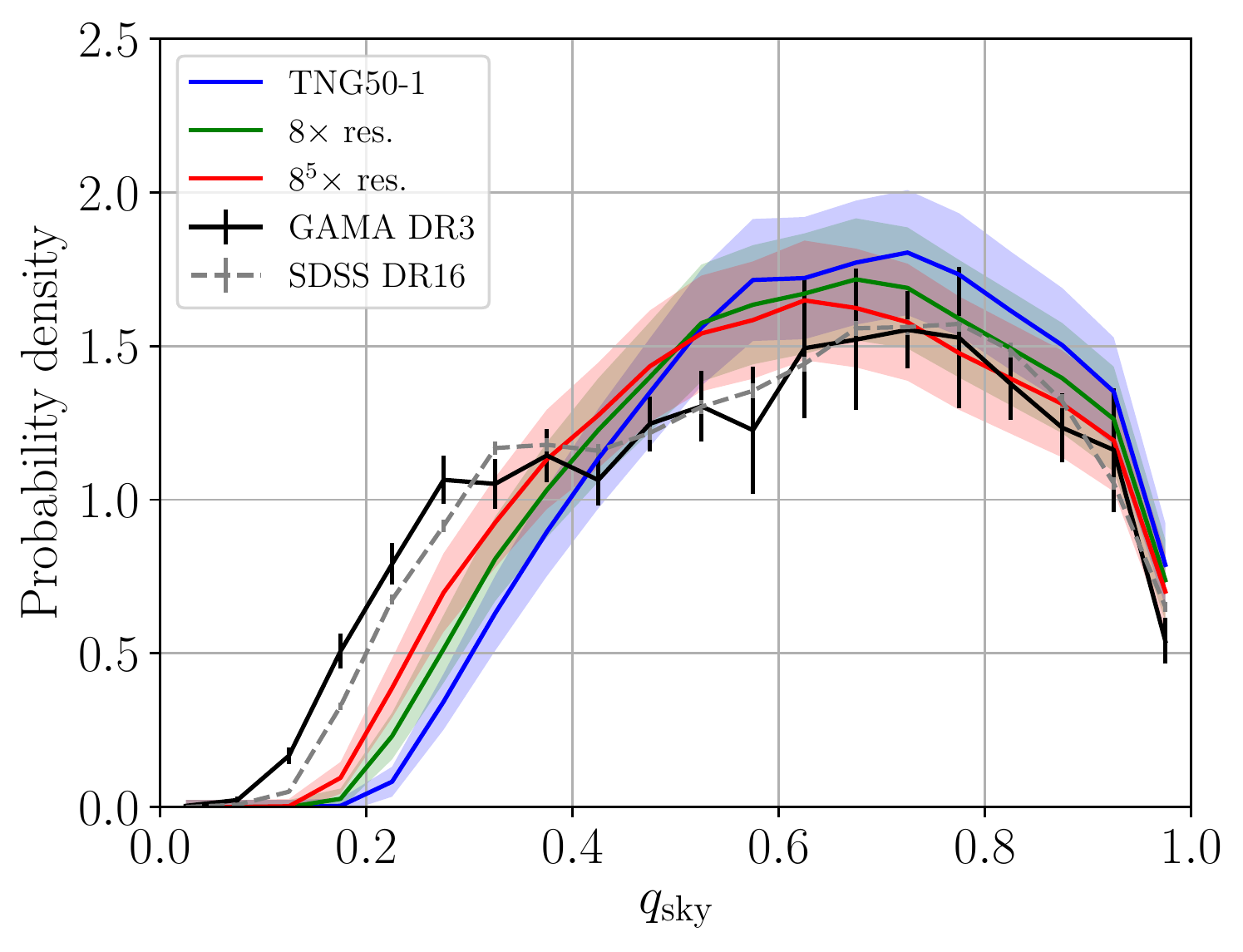}
		\includegraphics[width=8.5cm]{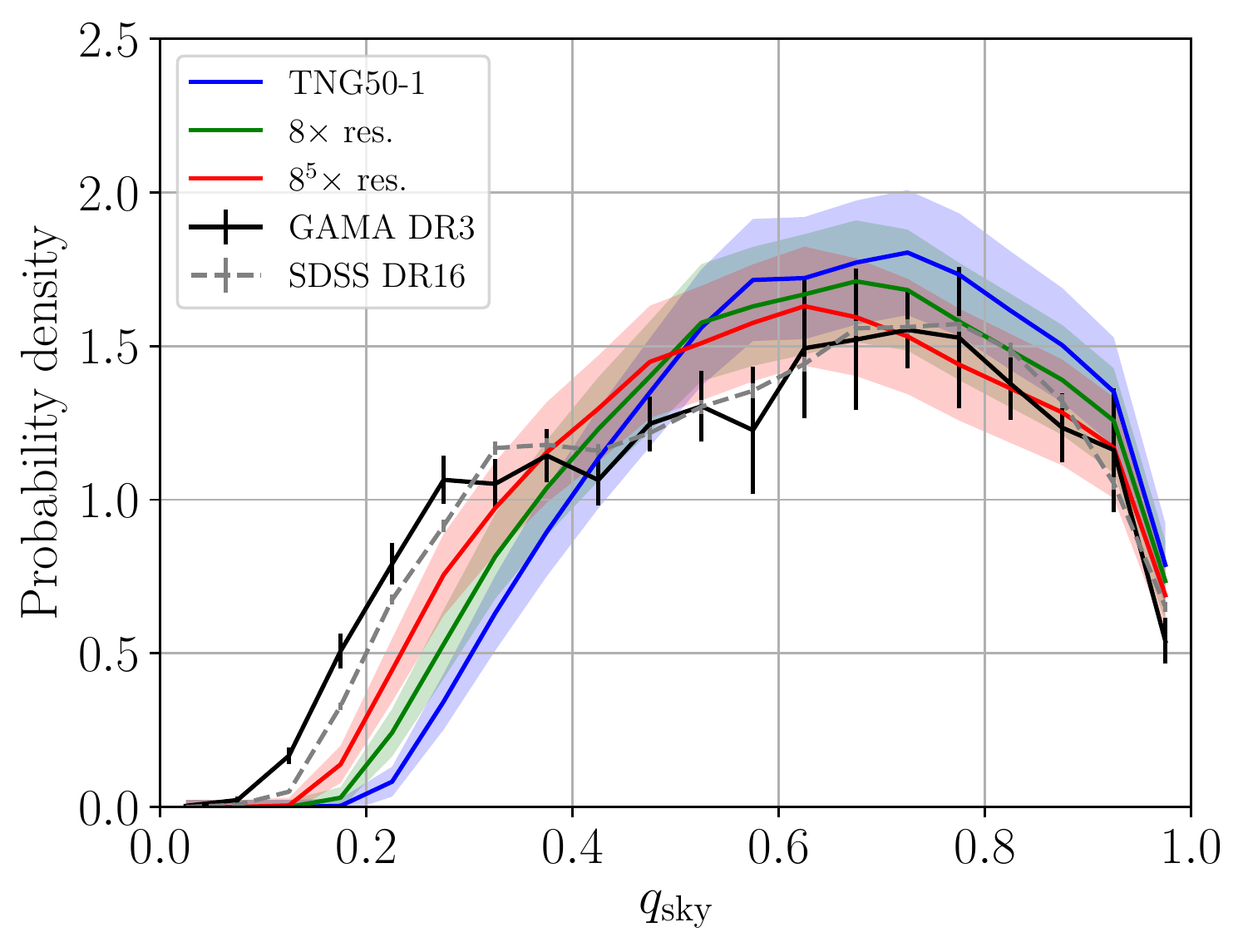}
	\end{center}
	\caption{Similar to Figure~\ref{fig:skyprojected_aspect_ratio}, but for an $8\times$ higher (green line) and $8^5\times$ higher (red line) resolution run than TNG50-1 (blue line). The colored shaded regions highlight the uncertainties given by Equation~\ref{eq:chi_sq_sigma_model}. The simulated $q_{\mathrm{sky}}$ distributions for a higher resolution run than TNG50-1 have been estimated with a parabolic (\emph{left panel}) or linear (\emph{right panel}) extrapolation in the $\log_{10} \left( m_{\mathrm{dm}}/M_{\odot} \right)$ vs. $\ln \left( 1 - \alpha \right)$ diagram as shown in the right panel of Figure~\ref{fig:TNG_prediction} (see the text). The total $\chi^{2}$ (Equation~\ref{eq:chi_sq_model_observations}) between the observed and simulated distribution for different refinement levels is reported in Table~\ref{tab:results_resolution_extrapolation} along with the corresponding level of tension.}
	\label{fig:TNG_extrapolation_qsky}
\end{figure*}

The $q_{\mathrm{sky}}$ distributions for an $8\times$ and $8^5\times$ higher dark matter mass resolution than the TNG50-1 run derived from a parabolic (linear) extrapolation are presented in the left (right) panel of Figure~\ref{fig:TNG_extrapolation_qsky}. Raising the resolution of the dark matter mass in TNG50-1 by a factor of $8^5$ yields an estimated tension of $5.58\sigma$ (parabolic fit) and $4.36\sigma$ (linear fit) with GAMA. In contrast, the discrepancy is significantly alleviated with respect to the older SDSS dataset $-$ we estimate a tension of $3.27\sigma$ with a parabolic fit and only a $1.61\sigma$ tension with a linear fit. The difference between the two observational samples is mainly because the fraction of galaxies with $q_{\mathrm{sky}} \la 0.3$ is higher in the GAMA survey compared to SDSS (Figure \ref{fig:TNG_extrapolation_qsky}). The total $\chi^{2}$ values and corresponding levels of tension with the GAMA and SDSS surveys for different resolution realizations and extrapolation methods are summarized in Table~\ref{tab:results_resolution_extrapolation}.

Since the parabolic fit considers more information than the linear fit and exactly matches the data, we consider it as the nominal case in our main analysis. The linear fit is still shown in order to illustrate the effect of uncertainties in the extrapolation procedure. Therefore, our estimate is that increasing the mass resolution of TNG50-1 by $8^5\times$ still leaves a significant tension of $5.58\sigma$ with GAMA, exceeding the $5\sigma$ plausibility threshold. The SDSS dataset is then in $3.27\sigma$ tension, which still points to an underestimated fraction of very thin galaxies in the $\Lambda$CDM simulations (see Figure~\ref{fig:TNG_extrapolation_qsky}).

Finally, we note that our approach is very conservative with respect to the $\Lambda$CDM framework. First of all, the extrapolation to five more refinement levels than TNG50-1 is based only on the four resolution realizations of the TNG50 simulation. The TNG50-4 simulation has $m_{\mathrm{dm}} = 2.3 \times 10^{8} \, M_{\odot}$, which is a factor of $8^8 = 1.7 \times 10^{7}$ more massive than dark matter particles in an $8^5\times$ higher-resolution realization than the TNG50-1 run. The parabolic and linear fits have been derived from simulations with $4.5 \times 10^5 \, (\text{TNG50-1}) \leq m_{\mathrm{dm}}/M_{\odot} \leq 2.3 \times 10^{8} \, (\text{TNG50-4})$. This introduces additional uncertainties because the shape of the relation between $\log_{10} \left( m_\mathrm{dm}/M_{\odot} \right)$ and $\ln \left( 1 - \alpha \right)$ may deviate significantly from the derived polynomial fits at lower $m_{\mathrm{dm}}$. Secondly, we performed the extrapolation to very high-resolution realizations without assuming that numerical convergence is reached between two different refinement levels. It is not impossible that numerical convergence ($\alpha = 1$) is already attained before reaching the $8^5\times$ higher-resolution level. Consequently, higher-resolution runs of self-consistent $\Lambda$CDM simulations are still required to test when numerical convergence is reached and if the tension with observations is then alleviated.

\subsection{Photometric Bulge/Total Ratios of Illustris-1 Subhalos}
\label{subsec:Photometric bulge/total ratio of subhalos in the Illustris simulation}

In contrast to our findings, \citet{BOTTRELL_2017b} found a significant deficit of bulge-dominated galaxies based on the photometric bulge/total ratios of subhalos in the Illustris-1 simulation compared to SDSS. In particular, \citet{BOTTRELL_2017a} photometrically derived $\left( B/T \right)_{\mathrm{phot}}$ ratios of subhalos with $M_{*}>10^{10}\,M_\odot$ at $z=0$ in the Illustris-1 simulation by performing a bulge-disk decomposition of the surface brightness profile with fixed S\'{e}rsic indices $n_{\mathrm{b}} = 4$ for the bulge and $n_{\mathrm{d}} = 1$ for the disk (see their section~3.2). By applying the same decomposition analysis to observed SDSS galaxies, they found a significant deficit of bulge-dominated galaxies with $M_*/M_\odot = 10^{10} - 10^{11}$ in the Illustris-1 simulation \citep[see figures~4 and 6 in][]{BOTTRELL_2017b}.

\begin{figure}
	\begin{center}
		\includegraphics[width=\linewidth]{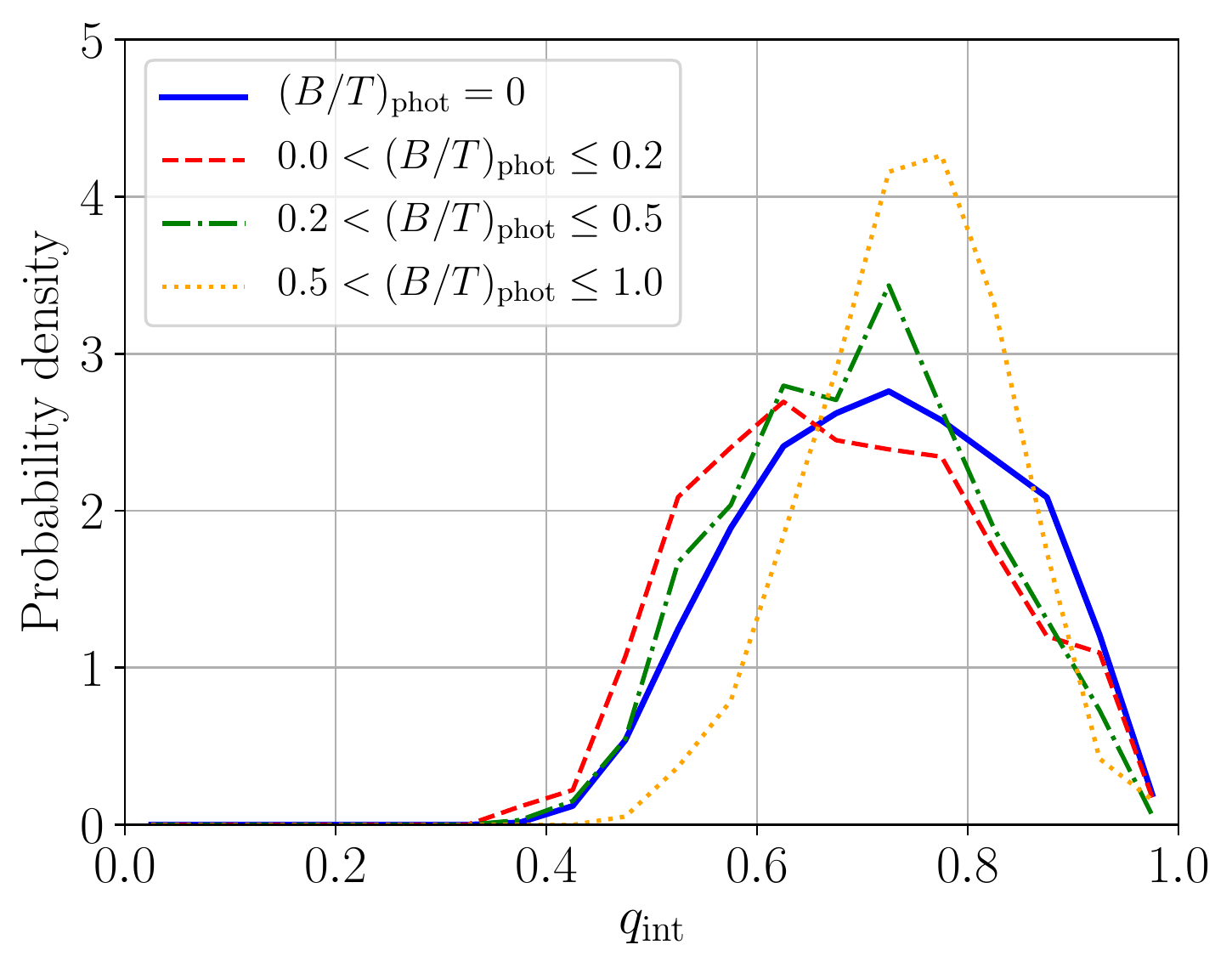}
	\end{center}
	\caption{The $q_{\mathrm{int}}$ distribution of subhalos with $M_{*} > 10^{10}\,M_\odot$ in different photometric $\left( B/T \right)_{\mathrm{phot}}$ bins in the Illustris-1 simulation as derived by \protect\citet{BOTTRELL_2017a} for camera angle $0$. The solid blue, dashed red, dotted-dashed green, and dotted yellow lines refer to galaxy subsamples with $\left( B/T \right)_{\mathrm{phot}} = 0$ (3999 subhalos), $0 < \left( B/T \right)_{\mathrm{phot}} \leq 0.2$ (1715 subhalos), $0.2 < \left( B/T \right)_{\mathrm{phot}} \leq 0.5$ (658 subhalos), and $0.5 < \left( B/T \right)_{\mathrm{phot}} \leq 1.0$ (380 subhalos), respectively.}
	\label{fig:Illustris_BOTTRELL_comparison}
\end{figure}

If the Illustris-1 simulation indeed lacks bulge-dominated galaxies, this deficit should also be evident in the $q_{\mathrm{int}}$ parameter (Section~\ref{subsec:Intrinsic aspect ratio distribution}). To test if $\left(B/T\right)_{\mathrm{phot}}$ reflects the shapes of simulated galaxies as quantified by $q_{\mathrm{int}}$, we show in Figure~\ref{fig:Illustris_BOTTRELL_comparison} its distribution for four simulated galaxy subsamples of \citet{BOTTRELL_2017a} with different photometric morphologies, i.e. $\left( B/T \right)_{\mathrm{phot}}$: 0, (0,0.2], (0.2,0.5], and (0.5,1].\footnote{We use their DISTINCT catalog, which contains the galaxy sample analyzed by \citet{Torrey_2015}.} Throughout this work, we use the $\left( B/T \right)_{\mathrm{phot}}$ ratio derived for the $r$ band from camera angle 0. Remarkably, all four galaxy subsamples have a very similar aspect ratio distribution, with the peak between $q_{\mathrm{int}} = 0.6 - 0.8$. The distributions have very few galaxies with $q_{\mathrm{int}} < 0.5$: the proportions for the above-mentioned $\left( B/T \right)_{\mathrm{phot}}$ bins are only 3.4\%, 7.1\%, 3.6\%, and 0.26\%, respectively, implying that the sample of \citet{BOTTRELL_2017a} is actually bulge-dominated. 

\begin{figure*}
	\begin{center}
		\includegraphics[width=\linewidth]{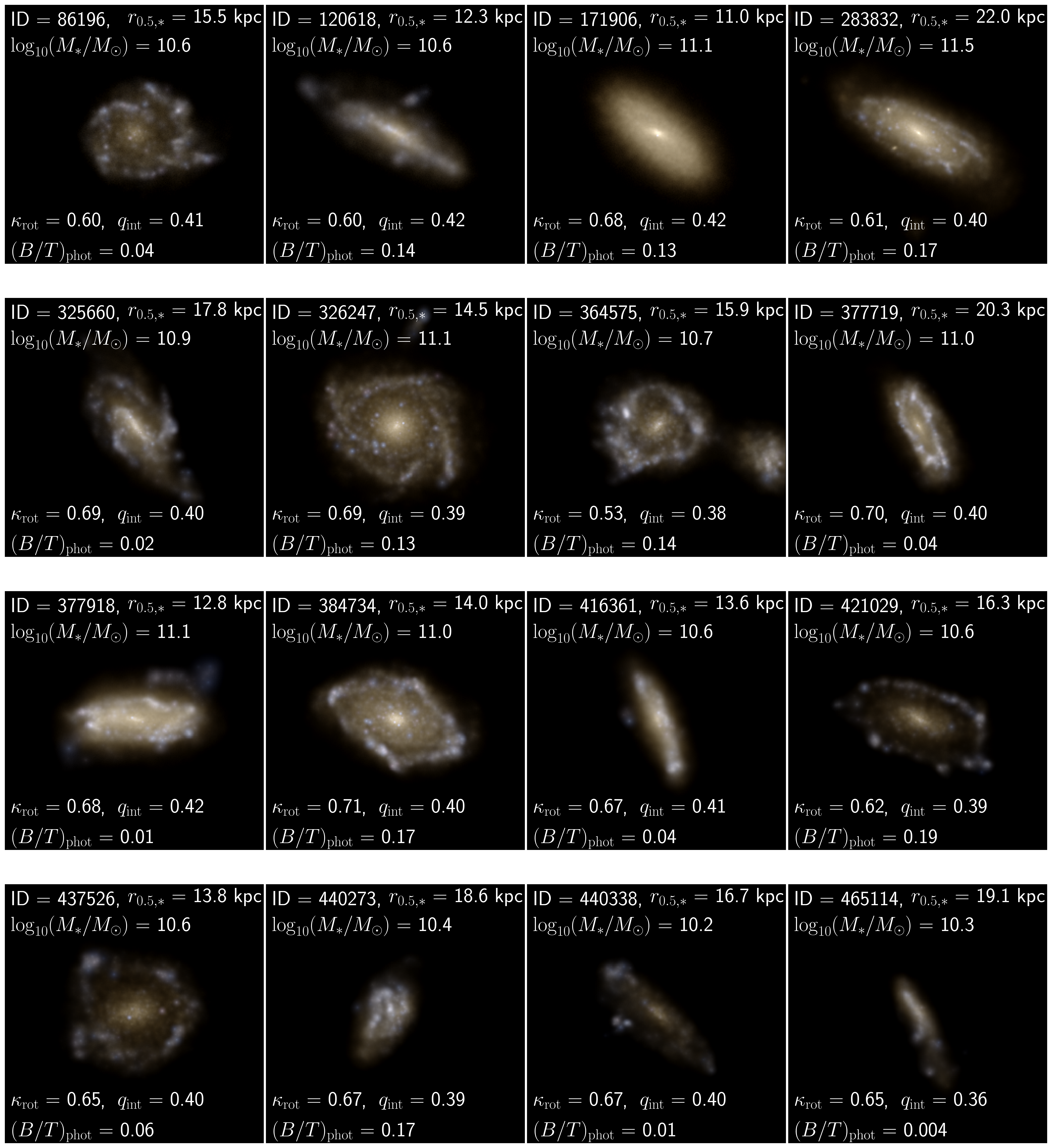}
	\end{center}
	\caption{Images of Illustris-1 galaxies in subhalos with $0 < \left( B/T \right)_{\mathrm{phot}} < 0.2$ from camera angle 0 that have the lowest $q_{\mathrm{int}}$ values in the sample of \citet{BOTTRELL_2017a}. These galaxies are rotation-dominated ($\kappa_{\mathrm{rot}} > 0.5$) and have spiral features, with $q_{\mathrm{int}} \approx 0.4$. The FoF images were downloaded from the Illustris Galaxy Observatory tool: \url{https://www.illustris-project.org/galaxy\_obs}. The field of view is ten stellar half-mass radii of the shown subhalo. These are genuinely disk-dominated galaxies.}
	\label{fig:Illustris_latetypes}
\end{figure*}

In order to understand this mismatch between $\left(B/T\right)_{\mathrm{phot}}$ and $q_{\mathrm{int}}$, we begin by presenting color images\footnote{downloaded [26.09.2020] from the Illustris-1 webpage using the Illustris Galaxy Observatory tool: \url{https://www.illustris-project.org/galaxy\_obs}.} generated from the Friends-of-Friends (FoF) halo finder. Subhalos with low $0 < \left( B/T \right)_{\mathrm{phot}} < 0.2$ and $q_{\mathrm{int}} < 0.4$ indeed have spiral structures and a disk (Figure~\ref{fig:Illustris_latetypes}). The $\kappa_{\mathrm{rot}}$ morphological parameter \citep[equation~2 in][]{RodriguezGomez_2015} indicates that these subhalos are rotation-dominated ($\kappa_{\mathrm{rot}} > 0.5$).

\begin{figure*}
	\begin{center}
		\includegraphics[width=\linewidth]{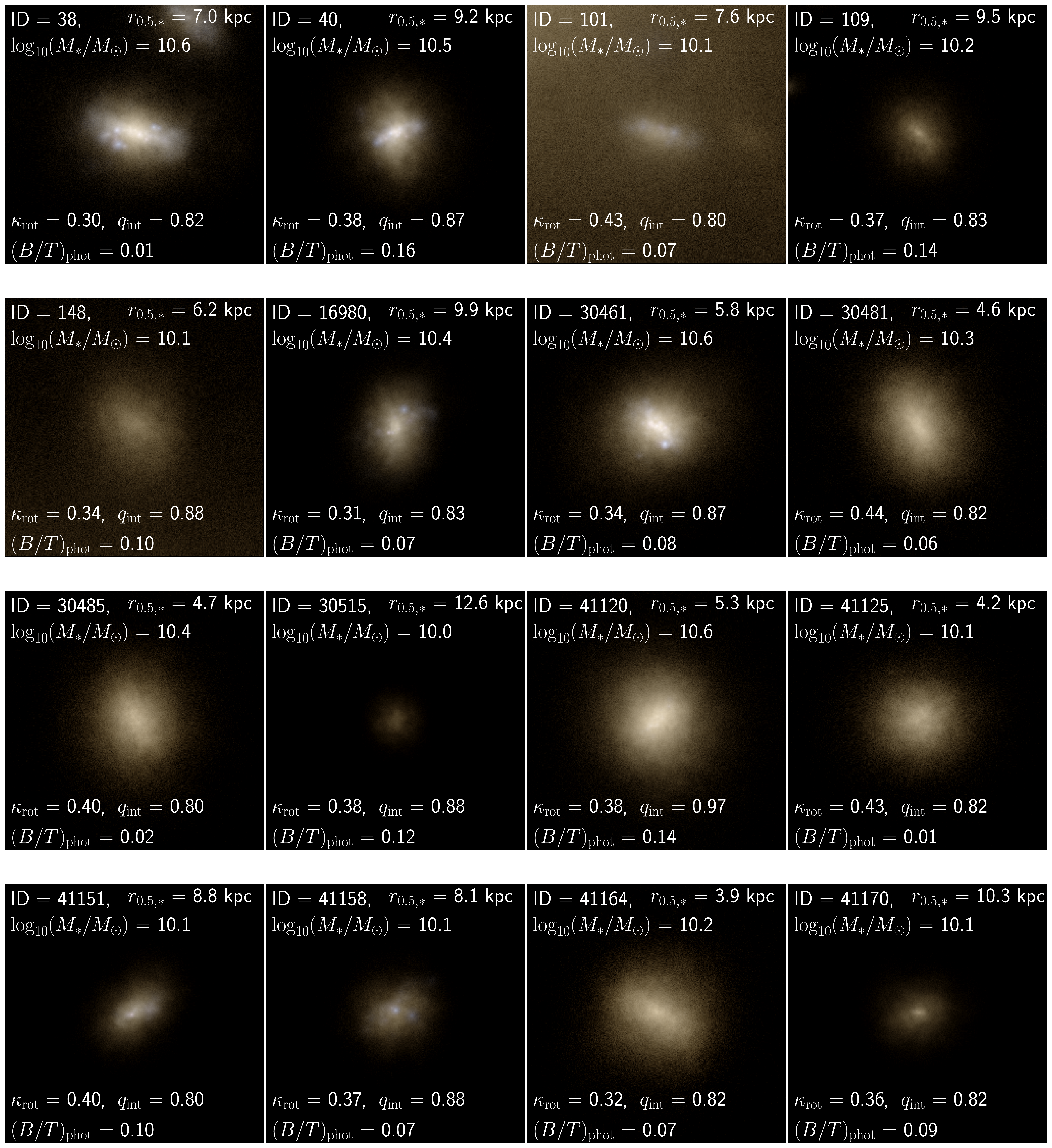}
	\end{center}
    \caption{Similar to Figure~\ref{fig:Illustris_latetypes}, but for galaxies with high $q_{\mathrm{int}} > 0.8$ despite a low $\left( B/T \right)_{\mathrm{phot}} < 0.2$. These are dispersion-dominated ($\kappa_{\mathrm{rot}} < 0.5$) featureless-looking galaxies. They are best understood as early-type galaxies where the photometric classification failed by assigning a low $\left( B/T \right)_{\mathrm{phot}}$.}
	\label{fig:Illustris_earlytypes}
\end{figure*}

The mismatch between the morphological parameters becomes evident in Figure~\ref{fig:Illustris_earlytypes}, which shows the significant fraction of subhalos (13.1\%) with low $0 < \left( B/T \right)_{\mathrm{phot}} < 0.2$ but high $q_{\mathrm{int}} > 0.8$ for their stellar component.\footnote{Subhalos with $\left( B/T \right)_{\mathrm{phot}} = 0$ are probably erroneous \citep{BOTTRELL_2017a} and are excluded from our analysis except in Figure~\ref{fig:Illustris_BOTTRELL_comparison}.} These are featureless-looking dispersion-dominated ($\kappa_{\mathrm{rot}} < 0.5$) galaxies without a pronounced disk or spiral structures. Based on these images and given that the intrinsic aspect ratio is a very robust parameter to quantify the actual shape of a galaxy, we conclude that the 2D parametric surface brightness decomposition applied by \citet{BOTTRELL_2017a} is inadequate if applied to simulated galaxies.

\begin{figure}
	\begin{center}
		\includegraphics[width=8.5cm]{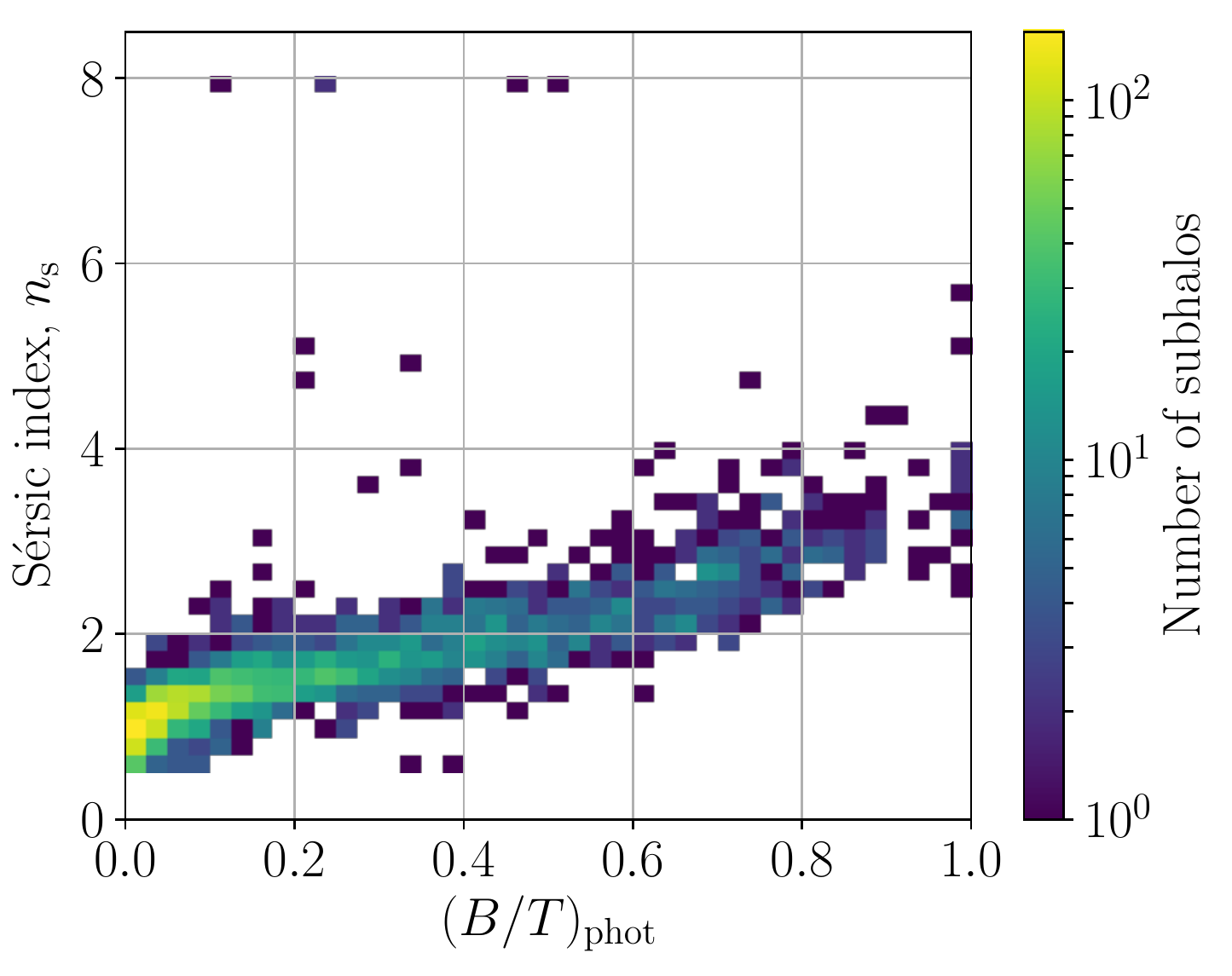}
	\end{center}
	\caption{Correlation between the photometric $\left( B/T \right)_{\mathrm{phot}}$ ratio and the S\'{e}rsic index $n_s$ for the Illustris-1 simulated subhalo sample analyzed by \protect\citet{BOTTRELL_2017a}. The parameters shown here are from catalogs provided in their tables A1 and A2 for camera angle $0$, with different camera orientations leading to the same results (not shown). Notice the tight correlation between $\left( B/T \right)_{\mathrm{phot}}$ and $n_s$ but a lack of correlation between $\left( B/T \right)_{\mathrm{phot}}$ and $q_{\mathrm{int}}$ (see Figure~\ref{fig:Illustris_BOTTRELL_comparison}).}
	\label{fig:BT_SersicIndex_Bottrell}
\end{figure}

In addition to a bulge-disk decomposition, \citet{BOTTRELL_2017a} also applied a pure S\'{e}rsic model to the galaxies in subhalos by varying the S\'{e}rsic index between 0.5 and 8 (see their section~3.2). Figure~\ref{fig:BT_SersicIndex_Bottrell} shows a significant correlation between $\left( B/T \right)_{\mathrm{phot}}$ and the S\'{e}rsic index, highlighting the inevitable confusion between an elliptical galaxy with $n_s \approx 1$ and a thin exponential disk if only considering the surface brightness profile. 
Choosing a different camera angle leads to the same results. In combination with Figure~\ref{fig:Illustris_BOTTRELL_comparison}, this shows that neither $n_{\mathrm{s}}$ nor $\left( B/T \right)_{\mathrm{phot}}$ correlates with $q_{\mathrm{int}}$.

This might be due to the resolution of the Illustris-1 simulation not being sufficient to apply a photometric decomposition. Importantly, we have shown in this contribution that the observed and simulated galaxy morphology distributions differ significantly (Table~\ref{tab:results}). The vast majority of observed galaxies are spirals \citep[e.g.][]{Loveday_1996,DelgadoSerrano_2010}, while the $\Lambda$CDM simulations form a far too large fraction of bulge-dominated galaxies (Figures~\ref{fig:3D_aspect_ratio} and \ref{fig:skyprojected_aspect_ratio}). Thus, assuming that all galaxies with low $\left( B/T \right)_{\mathrm{phot}}$ are intrinsically thin is quite accurate in the real universe where spirals are quite common, but not in a $\Lambda$CDM universe where they are rare (Section~\ref{subsec:Sky-projected aspect ratio distribution and comparison with observations}). In the Illustris-1 simulation, there are so few disk galaxies that galaxies with low $\left( B/T \right)_{\mathrm{phot}}$ are nearly always ellipticals with an exponential-like surface brightness profile (S\'{e}rsic index close to $1$).

These problems with the $\left( B/T \right)_{\mathrm{phot}}$ parameter are avoided by using the sky-projected aspect ratio (Figure~\ref{fig:skyprojected_aspect_ratio}), which is closely linked to the intrinsic aspect ratio and thus yields a more robust measurement of the galaxy morphology. Since $q_{\mathrm{sky}}$ is observable, it allows for a much more direct test of the model, provided the sample size is sufficient to statistically sample over projection effects. This is true in our case because the TNG50-1, GAMA DR3, and SDSS DR16 samples contain 882, 5304, and $232,128$ galaxies, respectively, in the range $10.0 < \log_{10}(M_{*}/M_{\odot}) \leq 11.65$ used for our comparisons.

\subsection{Impact of the Merger History}
\label{subsec:Simulated aspect ratio distribution in dependence of the merger history}

The $\Lambda$CDM theory strictly implies a hierarchical merger-driven build-up of the galaxy population. Galaxy mergers are driven by dynamical friction on the extended dark matter halos of interacting galaxies \citep[e.g.][]{Kroupa_2015}. Mergers grow the bulge component and thicken the stellar disk. By studying galaxy-galaxy interactions in $N$-body/hydrodynamical simulations, \citet{Hwang_2021} showed that the disk angular momentum of the late-type galaxy decreases by about $15\%-20\%$ after a prograde collision. Thus, galaxies with a quiescent merger history are expected to have lower bulge fractions than galaxies that have undergone a major merger \citep[see also, e.g.][]{Bournaud_2005,DOnghia_2006}. Since most observed galaxies are late types \citep[e.g.][]{DelgadoSerrano_2010} and $\approx 50\%$ of these have no classical bulge \citep{Graham_2008, Kormendy_2010}, we might expect that galactic mergers are less frequent in the universe than in $\Lambda$CDM simulations \citep{Disney_2008,Stewart_2008, Fakhouri_2010, Kroupa_2015, Wu_2015}. This may underlie the tension between the observed and simulated sky-projected aspect ratio distributions (Figure~\ref{fig:skyprojected_aspect_ratio}).

We therefore investigate if galaxies with a quiescent merger history in the $\Lambda$CDM framework are typically much thinner. We focus on the TNG50-1 run, as it has the highest resolution (Table~\ref{tab:simulations_parameters}) and yields the lowest tension in the here analyzed simulations (Table~\ref{tab:results}). The merger trees of the Illustris and TNG projects can be downloaded from their webpages\footnote{\url{https://www.illustris-project.org} [21.07.2020] and \url{https://www.tng-project.org} [21.07.2020]}, with a detailed description available in \citet{RodriguezGomez_2015}. We quantify the merger history of a galaxy by considering the total mass ratio $\mu \leq 1$ of the two progenitors and the maximum lookback time $t_{\mathrm{max}}$ during which we consider mergers. If there are multiple mergers in this timeframe, we use the most major merger, i.e. that with the highest $\mu$. This $\mu$ value is used to construct two galaxy samples that differ according to whether the galaxy had at least one major merger with $\mu \geq 1/12$ within a lookback time of $t_{\mathrm{max}} = 12 \, \rm{Gyr}$ ($z = 3.7$). We restrict ourselves to galaxies with $M_{*}>10^{10}\,M_\odot$ and $M_{\mathrm{dm}}/M_{*}>1$ at $z = 0$. The constraint on the $M_{\mathrm{dm}}/M_{*}$ ratio is applied to exclude dark matter-poor galaxies, which cannot be traced back accurately. This excludes $26$ galaxies from the original sample. About $89\%$ of all subhalos at $z = 0$ have undergone at least one major merger, which is broadly consistent with other $\Lambda$CDM simulations \citep{Stewart_2008, Fakhouri_2010}. The remaining $11\%$ of all subhalos had at most only minor merger(s) with $\mu < 1/12$ during the past $12\,\rm{Gyr}$.

\begin{figure*}
    \begin{center}
        \includegraphics[width=\columnwidth]{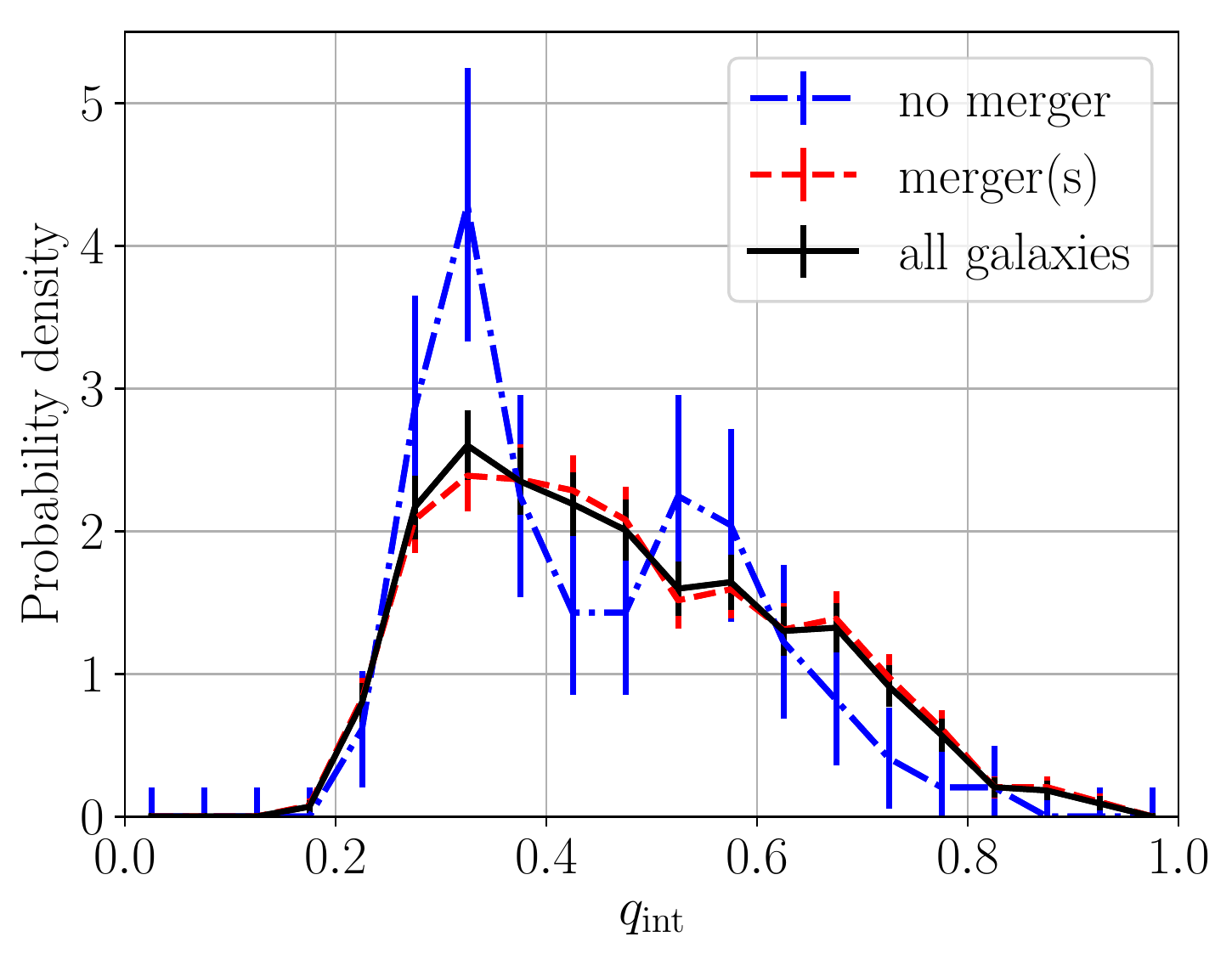}
        \includegraphics[width=\columnwidth]{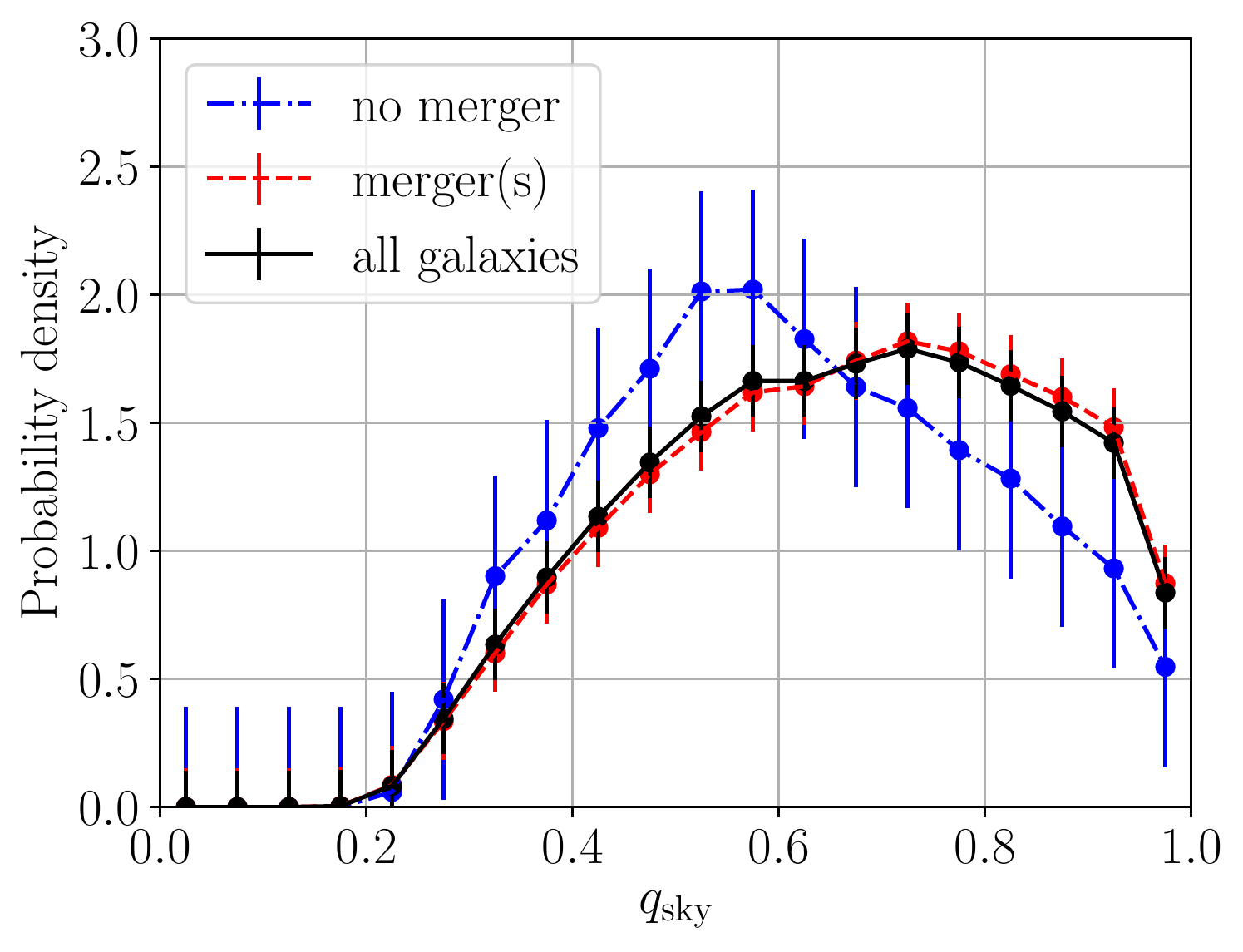}
    \end{center}
    \caption{Intrinsic (\emph{left}) and sky-projected (\emph{right}) aspect ratio distribution of subhalos in the TNG50-1 simulation with $M_{*} > 10^{10}\,M_\odot$ and $M_{\mathrm{dm}}/M_{*}>1$ with an active (dashed red) and quiescent (dotted-dashed blue) merger history, defined according to whether the galaxy had at least one major merger with total mass ratio $\mu \geq 1/12$ within the last 12~Gyr. The intrinsic aspect ratio distribution of the active (quiescent) sample, which contains 779 (98) galaxies, has a median of $q_{\mathrm{int}} = 0.40 \, (0.45)$. Results for all 877 galaxies (regardless of merger history) are shown by the solid black lines. The proportion of galaxies with $q_{\mathrm{int}} < 0.4$ is $39\% \pm 2\%$ and $50\% \pm 7\%$ in the sample with an active and a quiescent merger history, respectively. If instead we require $q_{\mathrm{sky}} < 0.4$, these proportions become $9.5\% \pm 1.1\%$ (active) and $12.5\% \pm 3.7\%$ (quiescent).}
    \label{fig:TNG50_merger_histories_distributions}
\end{figure*}

As expected from zoom-in simulations of galaxies in underdense environments, intrinsically thin galaxies are more frequent in the sample of galaxies with a quiescent merger history compared to that with an active merger history (see the left panel of Figure~\ref{fig:TNG50_merger_histories_distributions}). In particular, $39\% \pm 2\%$ of galaxies with an active merger history have $q_{\mathrm{int}} < 0.4$, but this rises to $50\% \pm 7\%$ for galaxies with a quiescent merger history. The aspect ratio of the thinnest such galaxy ($q_{\mathrm{int}} = 0.22$) is similar to that in the sample with an active merger history ($q_{\mathrm{int}} = 0.19$). Additionally, $9.5\% \pm 1.1\%$ of galaxies with an active merger history have $q_{\mathrm{sky}} < 0.4$, but this rises to $12.5\% \pm 3.7\%$ for the galaxies with a quiescent merger history (right panel of Figure~\ref{fig:TNG50_merger_histories_distributions}). 

Thus, the lack of intrinsically thin galaxies in $\Lambda$CDM could be partly due to major mergers. More likely, it is due to a combination of major and minor mergers, which are unavoidable in $\Lambda$CDM, as galaxies grow their mass through mergers. We emphasize that minor mergers are expected to be much less frequent in an alternative framework where galaxies lack extended dark matter halos, as occurs in MOND \citep{Milgrom_1983}. In addition, secular processes like disk-halo angular momentum exchange could also drive the formation of significant bars and bulges in $\Lambda$CDM \citep{Athanassoula_2002, Sellwood_2019}, but perhaps not in the real universe \citep{Banik_2020_M33, Roshan_2021} where the fraction of bars differs substantially from $\Lambda$CDM expectations \citep{Reddish_2021}. There is also a highly significant discrepancy between the pattern speeds of bars in observations and in $\Lambda$CDM simulations, where the results seem to have converged with respect to the ratio of bar length to corotation radius \citep{Roshan_2021b}. The tension is caused by the fact that bars are expected to be slowed down by dynamical friction with the dark matter halo, but observed bars are fast. This is another indication against dynamical friction from massive dark matter halos.

\subsection{Feedback}
\label{subsec:Feedback}

In the previous section, we discussed that the disagreement between the observed and $\Lambda$CDM simulated galaxy shapes could be partly due to the frequency of mergers being too high in this framework. Another possibility is that the tension is caused by the feedback description used in the simulations. In particular, \citet{Lagos_2018} discussed the link between loss/gain of angular momentum and dry/wet mergers. Using the EAGLE and HYDRANGEA \citep{Bahe_2017, Barnes_2017} hydrodynamical simulations, they showed that dry mergers typically decrease the stellar spin parameter while wet mergers increase it. Therefore, galaxies that further accrete cold gas are able to reform their disks. This process is sensitive to the implemented feedback description.

As shown in Sections~\ref{subsec:Intrinsic aspect ratio distribution} and \ref{subsec:Comparison of different TNG50 and EAGLE resolution realizations}, the TNG50-1 and EAGLE simulations produce very similar aspect ratio distributions despite relying on different sub-grid models. This is a strong indication that the tension is likely not caused by the implemented sub-grid feedback models. Even so, improved models might yet alleviate or resolve the tension. For example, the feedback recipe in both the TNG and EAGLE projects could be too strong to allow the (re)formation of disks. In this case, the shape of the galaxies would be set by the merger rate independently of whether the mergers are dry or wet. However, a weaker feedback description could potentially increase the efficiency of disk formation. We note that strong feedback is required in $\Lambda$CDM simulations for them to explain why the Newtonian dynamical mass of a galaxy or galaxy group often greatly exceeds $6.4\times$ its baryonic mass \citep[e.g.][]{Muller_2022}, even though $\Lambda$CDM needs this to be the cosmic ratio between baryonic and total mass \citep{Planck_2020}. A strong feedback prescription is also needed to solve the missing satellite galaxy problem \citep[e.g.][]{Brooks_2013}.


\subsection{Disk Galaxies in MOND}
\label{subsec:disk galaxies in Milgromian dynamics}

\begin{figure*}
	\begin{center}
		\includegraphics[width=8.5cm]{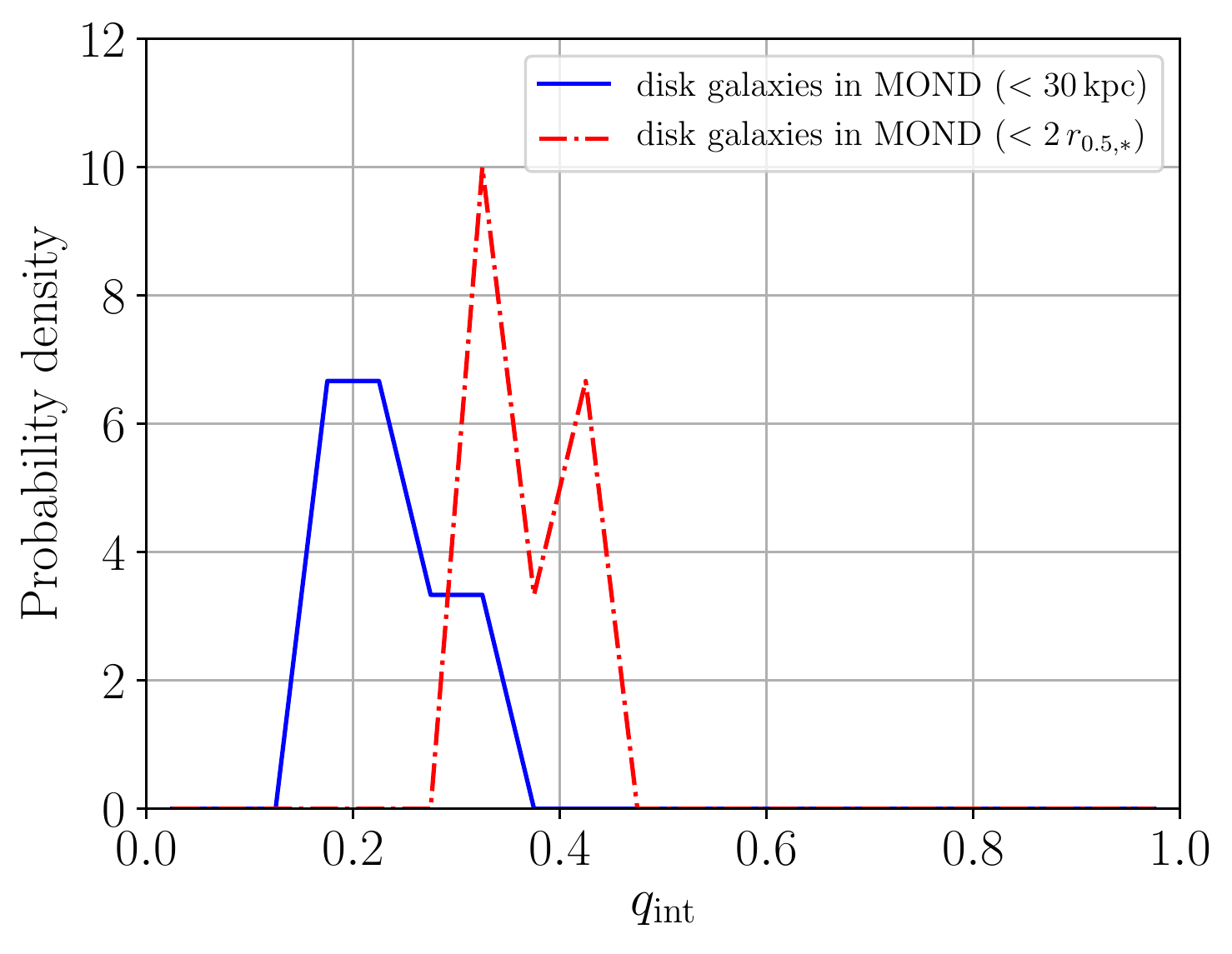}
		\includegraphics[width=8.5cm]{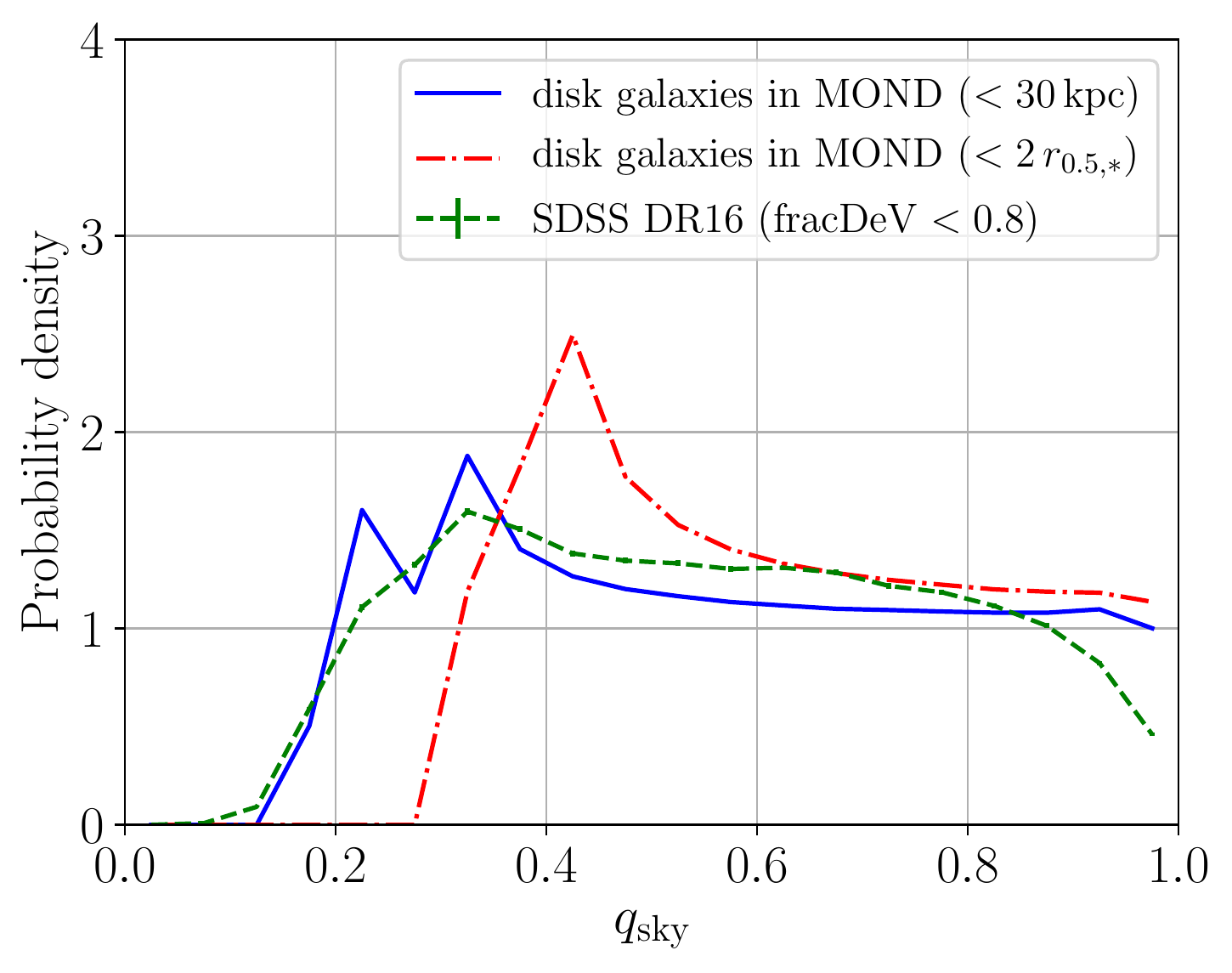}
	\end{center}
	\caption{Distribution of the intrinsic (\emph{left}) and sky-projected (\emph{right}) aspect ratio for six disk galaxies with $10^{10} < M_{*}/M_{\odot} \leq 9.56 \times 10^{10}$ formed in MOND simulations conducted by \protect\citet{Wittenburg_2020}. The solid blue (dotted-dashed red) line refers to the aspect ratio derived from the mass tensor of particles within a sphere of radius 30~kpc ($2 \, r_{0.5,*}$). These MOND results are shown for illustrative purposes only $-$ they are not directly comparable to the $\Lambda$CDM simulations because the Milgromian galaxies are not formed in a self-consistent cosmological simulation. The dashed green line in the right panel shows the $q_{\mathrm{sky}}$ distribution derived from an exponential fit to SDSS galaxies with fracDeV $< 0.8$ and $10^{10} < M_{*}/M_{\odot} \leq 9.56 \times 10^{10}$. We do not apply an $M_{*}$-weighting to the observed sample as we only have a small number of simulated MONDian galaxies.}
	\label{fig:MOND_aspect_ratio_distribution}
\end{figure*}

The difficulties faced by $\Lambda$CDM with regards to the high fraction of thin disk galaxies motivate us to consider MOND, one of the main alternative frameworks that is generally considered to perform better on galaxy scales \citep[for reviews, see, e.g.][]{Famaey_McGaugh_2012, Kroupa_2015, Banik_Zhao_2022}. For illustrative purposes, we show in the left panel of Figure~\ref{fig:MOND_aspect_ratio_distribution} the present-day $q_{\mathrm{int}}$ distribution of six disk galaxies with $ 10^{10} < M_{*}/M_\odot \leq 9.56 \times 10^{10}$ formed in hydrodynamical MOND simulations of collapsing gas clouds conducted by \citet{Wittenburg_2020} using the \textsc{phantom of ramses} code \citep{Lueghausen_2015}, an adaptation of the $N$-body and hydrodynamics solver \textsc{ramses} \citep{Teyssier_2002} for MOND gravity \citep[for a user guide, see][]{Nagesh_2021}. These non-cosmological simulations have a box size of 960~kpc per side, a minimum stellar mass of $M_{*} \approx 3\times10^4\,M_\odot$, and depending on the run a maximum grid cell resolution of 117.19~pc, 234.38~pc, or 468.75~pc. The temperature floor for the gas is set to 10~kK \citep[for further information on the initial conditions and numerical parameters of the individual simulations, see table~1 of][]{Wittenburg_2020}. The Milgromian disk galaxies are systematically thinner than in the here analyzed $\Lambda$CDM simulations, as evidenced by their $q_{\mathrm{int}} = 0.18 - 0.30 \, \left( 0.31 - 0.54 \right)$ for $r < 30$~kpc ($r < 2 \, r_{0.5,*}$), with the global peak at $q_{\mathrm{int}} \approx 0.2 \, \left( 0.3 \right)$. While these isolated MOND results cannot yet be directly compared with self-consistent cosmological $\Lambda$CDM simulations that allow for galaxy interactions and mergers, we note that neglecting mergers may be a good approximation in MOND as mergers are expected to be rare due to the absence of dynamical friction with the dark matter halo \citep*{Renaud_2016}.

Interestingly, the right panel of Figure~\ref{fig:MOND_aspect_ratio_distribution} shows that the sky-projected aspect ratio distribution of these MONDian disk galaxies is very consistent with that of SDSS spiral galaxies (fracDeV $<0.8$; see Section~\ref{subsec:SDSS}). This also demonstrates that the distinction of spiral from elliptical galaxies based on the linear combination of the exponential and de Vaucouleurs models \citep{Abazajian_2004} succeeds in SDSS, in contrast to the bulge-disk decomposition applied to the Illustris-1 simulation (possible reasons were discussed in Section~\ref{subsec:Photometric bulge/total ratio of subhalos in the Illustris simulation}). In the future, whether a self-consistent cosmological MOND simulation would be able to reproduce the observed fraction of spiral and elliptical galaxies needs to be explicitly shown. Such MOND simulations are not available at the moment, but are underway in the Bonn-Prague group based on the promising cosmological MOND framework detailed in \citet{Haslbauer_2020}. Given the problems in reproducing the observed distribution of galactic morphologies with $\Lambda$CDM simulations, it is often argued that these simulations depend sensitively on the implemented feedback model and that the problem is likely to be alleviated once the correct feedback implementation has been found. Hydrodynamical MOND simulations of galaxy formation out of post-Big Bang gas clouds, on the other hand, naturally lead to realistic galaxies similar to the observed ones with the available feedback implementations \citep[][Eappen et al. 2022, submitted]{Wittenburg_2020}.

\section{Conclusions}
\label{sec:Conclusion}

In this contribution, we considered the distribution of galaxy morphologies in state-of-the-art cosmological $\Lambda$CDM simulations. The present-day sky-projected aspect ratio distribution of galaxies in the TNG50-1 (EAGLE50) simulation disagrees with the GAMA survey and SDSS at $\geq 12.52\sigma$ ($\geq 14.82\sigma$) confidence (Section~\ref{subsec:Sky-projected aspect ratio distribution and comparison with observations}). The lowest tension is obtained when comparing GAMA with the TNG50-1 simulation run, which has the highest resolution of the TNG project. The main reason for this mismatch is that the $\Lambda$CDM simulations significantly underproduce galaxies with $q_{\mathrm{sky}} < 0.4$ (Table~\ref{tab:results_fraction}), making it difficult for the latest $\Lambda$CDM simulations to form thin disk galaxies like the Milky Way \citep[with a ratio of scale height to scale length of $h/l \approx 0.07 - 0.21$;][]{BlandHawthorn_2016} or M31 \citep[with a sky-projected aspect ratio of $0.27\pm0.03$;][]{Courteau_2011}. The intrinsic aspect ratio distribution of the highest-resolution models conflicts with the best-observed local galaxy sample at $5.42\sigma$ significance if we use the recently released TNG50-1 run (Section~\ref{subsec:Intrinsic aspect ratio distribution}), confirming this discrepancy independently of the GAMA and SDSS datasets. The advantage of the LV sample is the higher spatial resolution of the galaxy images compared to SAMI, GAMA, and SDSS.

Our results agree with other recent studies \citep[e.g.][]{Lagos_2018, vandenSande_2019, Peebles_2020}. We therefore disagree that the angular momentum problem has been resolved in the Illustris-1 simulation as concluded by \citet{Vogelsberger_2014}. The loss of angular momentum remains a significant problem in the latest cosmological $\Lambda$CDM simulations, as quantified in the present work.

The aspect ratio distribution has numerically converged in the EAGLE simulations \citep[Section~\ref{subsec:Comparison of different TNG50 and EAGLE resolution realizations}; see also][]{Lagos_2018}. However, convergence cannot be confirmed for the TNG50 simulation, where the tension reported here could be related to numerical heating of the stellar particles by the coarse-grained implementation of dark matter halos \citep{Ludlow_2021}. Therefore, we apply a parametric correction to the $q_{\mathrm{int}}$ of each simulated galaxy in TNG50 (Equation~\ref{eq:alpha_prediction}). Depending on the extrapolation, we estimate that galaxies in an eight-times-higher particle resolution realization compared to the TNG50-1 run would be thinned by a factor in the range $0.824 \leq \alpha \leq 0.834$, where $\alpha = 1.0$ would imply numerical convergence in $q_{\mathrm{int}}$. Applying the lower limit and being therewith more conservative with respect to the $\Lambda$CDM framework, the here reported tension would decrease to the $8.68\sigma$ ($8.71\sigma$) confidence level for GAMA DR3 (SDSS DR16). Extrapolating the TNG50 results to $8^5\times$ better dark matter mass resolution than TNG50-1, the tension with GAMA DR3 (SDSS DR16) becomes $5.58\sigma$ ($3.27\sigma$) for the parabolic extrapolation, which better fits the available data (the tension is slightly lower with a linear extrapolation; see Table \ref{tab:results_resolution_extrapolation}). However, such an extrapolation to much higher resolution than the TNG50 runs introduces additional uncertainties, so it is not yet clear if an arbitrarily higher-resolution realization than TNG50-1 can indeed resolve the here reported tension.

Thus, additional self-consistent cosmological $\Lambda$CDM simulations are useful to test if higher-resolution realizations and further improved sub-grid models for the interstellar medium can resolve the tension \citep[e.g.][]{Trayford_2017, Lagos_2018, vandenSande_2019}. This long-standing problem is likely not caused by limitations of the sub-grid model because the latest EAGLE and TNG simulations that are based on very different computational algorithms and baryonic feedback coding lead to galaxy populations whose sky-projected aspect ratio distributions agree with each other (Table~\ref{tab:results_convergence}). Moreover, isolated CDM simulations with a similar temperature floor of about 10~kK in the gas are able to produce thin disk galaxies \citep{Sellwood_2019} $-$ as indeed are EAGLE and TNG50. 

An uncertainty that has not been elaborated on in our work is how observed aspect ratios could be affected by the difference in stellar ages between the typically older bulges in spiral galaxies and their typically younger disks, which thus end up with a lower mass-to-light ratio. This could cause a difference between the mass-weighted aspect ratios obtained from simulations and the luminosity-weighted aspect ratios obtained from observations. However, the high fraction of bulgeless disk galaxies locally \citep{Kormendy_2010} suggests that this issue is not by itself sufficient to reconcile $\Lambda$CDM with the observed galaxy population. Observational systematics can be expected to differ between, e.g., GAMA and the high-resolution LV observations (Section~\ref{subsubsec:The Catalog of Neighboring Galaxies}), but even here there is still a significant $5.42\sigma$ tension with TNG50.

It appears to be impossible to form as many thin disk galaxies as observed. Consequently, the angular momentum problem persists in the hierarchical cosmological $\Lambda$CDM framework and is unlikely to be solved by improving the resolution. This conclusion can also be reached independently through observed galaxy bars being fast with no sign of slowdown through Chandrasekhar dynamical friction on the hypothetical dark matter halo \citep{Roshan_2021b}. $\Lambda$CDM also faces many problems on other scales \citep{Kroupa_2012, Kroupa_2015, Pawlowski_2021, Banik_Zhao_2022}. Almost 50 yr after dark matter halos were first postulated to surround galaxies \citep{Ostriker_1973}, numerical implementations of this model still cannot explain the observed fraction of early- and late-type galaxies, so the observed galaxy population continues to pose a severe challenge to this framework.

If better resolved and/or improved sub-grid models of the interstellar medium in self-consistent $\Lambda$CDM cosmological simulations cannot resolve this tension, the loss of angular momentum would question the hierarchical merger-driven build-up of galaxies in this paradigm. We showed that intrinsically thin galaxies are more frequent in a $\Lambda$CDM galaxy sample selected to have a very quiescent merger history compared to one with an active merger history. However, a sample with a quiescent merger history cannot resolve the tension. 

This leaves open the possibility that the tension we reported here is due to minor mergers and/or disk-halo angular momentum exchange \citep[see also][]{Sellwood_2019, Banik_2020_M33, Roshan_2021, Roshan_2021b}. If so, the angular momentum problem might be alleviated in a cosmological MOND framework \citep{Milgrom_1983} due to the absence of dynamical friction on extended dark matter halos reducing the merger rate \citep{Kroupa_2015, Renaud_2016}. In MOND, cluster-scale and early-universe observables can be explained within the neutrino hot dark matter ($\nu$HDM) model \citep{Angus_2009, Haslbauer_2020, Asencio_2021}. In particular, we emphasize that the statistically significant Hubble tension does not appear in this model \citep{Haslbauer_2020} but does exist in the standard $\Lambda$CDM model, independently showing $\Lambda$CDM to be invalid \citep{Riess_2021}. Thus, the Hubble tension is naturally solved by using the standard MOND cosmological model without tuning any theoretical parameter. The sky-projected aspect ratio distribution of disk galaxies formed in hydrodynamical MOND simulations of collapsing post-Big Bang gas clouds \citep{Wittenburg_2020} is very consistent with observed SDSS spiral galaxies (Section~\ref{subsec:disk galaxies in Milgromian dynamics}). Self-consistent cosmological MOND simulations underway in Bonn and in Prague will allow us to determine the entire galactic morphological distribution for comparison with observations, enabling the same tests as documented here for $\Lambda$CDM.

\begin{acknowledgments}

I.B. is supported by Science and Technology Facilities Council grant ST/V000861/1. He acknowledges support from an Alexander von Humboldt Foundation postdoctoral research fellowship (2018-2020) and the University of Bonn ``Pathways to Research'' program. The authors are grateful to Veselina Kalinova for valuable comments, Karl Menten for his support, Sylvia Pl\"{o}ckinger for assistance with the EAGLE database, Sree Oh for explanations on the Data Central platform (\url{https://datacentral.org.au/}), and Edward Taylor for clarifications on the stellar masses of GAMA DR3 galaxies. The authors are very grateful to the anonymous referee for helpful suggestions that significantly improved this publication. They also thank the GAMA and SDSS teams for providing their data and useful related discussions.

GAMA is a joint European-Australasian project based around a spectroscopic campaign using the Anglo-Australian Telescope. The GAMA input catalog is based on data taken from the Sloan Digital Sky Survey and the UKIRT Infrared Deep Sky Survey. Complementary imaging of the GAMA regions is being obtained by a number of independent survey programs including GALEX MIS, VST KiDS, VISTA VIKING, WISE, Herschel-ATLAS, GMRT, and ASKAP, providing UV to radio coverage. GAMA is funded by the STFC (UK), the ARC (Australia), the AAO, and the participating institutions. The GAMA website is \url{http://www.gama-survey.org/}.

The SAMI Galaxy Survey is based on observations made at the Anglo-Australian Telescope. The Sydney-AAO Multi-object Integral field spectrograph (SAMI) was developed jointly by the University of Sydney and the Australian Astronomical Observatory. The SAMI input catalog is based on data taken from the Sloan Digital Sky Survey, the GAMA Survey, and the VST ATLAS Survey. The SAMI Galaxy Survey is funded by the Australian Research Council Centre of Excellence for All-sky Astrophysics (CAASTRO), through project No. CE110001020, and other participating institutions. The SAMI Galaxy Survey website is \url{http://sami-survey.org/}.

Funding for the SDSS and SDSS-II has been provided by the Alfred P. Sloan Foundation, the Participating Institutions, the National Science Foundation, the U.S. Department of Energy, the National Aeronautics and Space Administration, the Japanese Monbukagakusho, the Max Planck Society, and the Higher Education Funding Council for England. The SDSS is managed by the Astrophysical Research Consortium for the Participating Institutions. The Participating Institutions are the American Museum of Natural History, Astrophysical Institute Potsdam, University of Basel, University of Cambridge, Case Western Reserve University, University of Chicago, Drexel University, Fermilab, the Institute for Advanced Study, the Japan Participation Group, Johns Hopkins University, the Joint Institute for Nuclear Astrophysics, the Kavli Institute for Particle Astrophysics and Cosmology, the Korean Scientist Group, the Chinese Academy of Sciences (LAMOST), Los Alamos National Laboratory, the Max-Planck-Institute for Astronomy (MPIA), the Max-Planck-Institute for Astrophysics (MPA), New Mexico State University, Ohio State University, University of Pittsburgh, University of Portsmouth, Princeton University, the United States Naval Observatory, and the University of Washington. The SDSS website is \url{http://www.sdss.org/}.

\end{acknowledgments}




\begin{appendix}

\section{Statistical significance of extreme events}
\label{subsec:Statistical significance of extreme events} 

If the $\chi^2$ value is particularly large, the $P$-value is too low for a finite element computer to handle. We therefore make an analytic approximation for the integrals of the $\chi^2$ and Gaussian distributions. In both cases, as the integrand declines very rapidly, we locally approximate it as declining exponentially. This allows us to approximate the integral out to infinity. Using this approach, we need to solve for $x$ based on the known value of $\chi^2$ using
\begin{eqnarray}
    \sqrt{\frac{2}{\mathrm{\pi}}} \frac{\exp \left( -x^2 \right)}{x} ~=~ \frac{\left( \chi^2 \right)^{\frac{n}{2} - 1} \exp \left( -\frac{\chi^2}{2} \right)}{2^\frac{n}{2} \left( \frac{n}{2} - 1 \right)! \left( \frac{1}{2} - \frac{\frac{n}{2} - 1}{\chi^2} \right)} \, .
    \label{P_chi_approximation}
\end{eqnarray}
Factorials of non-integer numbers are defined using the $\Gamma$ function.

To minimize numerical errors, we set up Equation~\ref{P_chi_approximation} as an equality between the logarithms of both sides. We then solve for $x$ using the Newton-Raphson algorithm. The statistical significance of the result is approximately $x$ standard deviations, with the approximation becoming very accurate for $x > 7$. This is fortunate as lower values of $x$ allow Equation~\ref{P_chi} to be solved directly without the approximation of Equation~\ref{P_chi_approximation}. We checked that both methods give similar results for $x = 5 - 7$, but numerical difficulties mean that Equation~\ref{P_chi_approximation} is necessary for higher~$x$.

\end{appendix}

\newpage

\bibliographystyle{aasjournal}
\bibliography{Galaxy_Morphology_bbl}

\begin{thebibliography}{124}
\expandafter\ifx\csname natexlab\endcsname\relax\def\natexlab#1{#1}\fi

\bibitem[{{Abazajian} {et~al.}(2004){Abazajian}, {Adelman-McCarthy},
  {Ag{\"u}eros}, {Allam}, {Anderson}, {Anderson}, {Annis}, {Bahcall}, {Baldry},
  {Bastian}, {Berlind}, {Bernardi}, {Blanton}, {Bochanski}, {Boroski},
  {Briggs}, {Brinkmann}, {Brunner}, {Budav{\'a}ri}, {Carey}, {Carliles},
  {Castander}, {Connolly}, {Csabai}, {Doi}, {Dong}, {Eisenstein}, {Evans},
  {Fan}, {Finkbeiner}, {Friedman}, {Frieman}, {Fukugita}, {Gal}, {Gillespie},
  {Glazebrook}, {Gray}, {Grebel}, {Gunn}, {Gurbani}, {Hall}, {Hamabe},
  {Harris}, {Harris}, {Harvanek}, {Heckman}, {Hendry}, {Hennessy}, {Hindsley},
  {Hogan}, {Hogg}, {Holmgren}, {Ichikawa}, {Ichikawa}, {Ivezi{\'c}}, {Jester},
  {Johnston}, {Jorgensen}, {Kent}, {Kleinman}, {Knapp}, {Kniazev}, {Kron},
  {Krzesinski}, {Kunszt}, {Kuropatkin}, {Lamb}, {Lampeitl}, {Lee}, {Leger},
  {Li}, {Lin}, {Loh}, {Long}, {Loveday}, {Lupton}, {Malik}, {Margon},
  {Matsubara}, {McGehee}, {McKay}, {Meiksin}, {Munn}, {Nakajima}, {Nash},
  {Neilsen}, {Newberg}, {Newman}, {Nichol}, {Nicinski}, {Nieto-Santisteban},
  {Nitta}, {Okamura}, {O'Mullane}, {Ostriker}, {Owen}, {Padmanabhan},
  {Peoples}, {Pier}, {Pope}, {Quinn}, {Richards}, {Richmond}, {Rix}, {Rockosi},
  {Schlegel}, {Schneider}, {Scranton}, {Sekiguchi}, {Seljak}, {Sergey},
  {Sesar}, {Sheldon}, {Shimasaku}, {Siegmund}, {Silvestri}, {Smith},
  {Smol{\v{c}}i{\'c}}, {Snedden}, {Stebbins}, {Stoughton}, {Strauss},
  {SubbaRao}, {Szalay}, {Szapudi}, {Szkody}, {Szokoly}, {Tegmark}, {Teodoro},
  {Thakar}, {Tremonti}, {Tucker}, {Uomoto}, {Vanden Berk}, {Vandenberg},
  {Vogeley}, {Voges}, {Vogt}, {Walkowicz}, {Wang}, {Weinberg}, {West}, {White},
  {Wilhite}, {Xu}, {Yanny}, {Yasuda}, {Yip}, {Yocum}, {York}, {Zehavi},
  {Zibetti}, \& {Zucker}}]{Abazajian_2004}
{Abazajian}, K., {Adelman-McCarthy}, J.~K., {Ag{\"u}eros}, M.~A., {et~al.}
  2004, \href{http://dx.doi.org/10.1086/421365}{\color{magenta}\aj},
  \href{https://ui.adsabs.harvard.edu/abs/2004AJ....128..502A}{128, 502}

\bibitem[{{Adelman-McCarthy} {et~al.}(2008){Adelman-McCarthy}, {Ag{\"u}eros},
  {Allam}, {Allende Prieto}, {Anderson}, {Anderson}, {Annis}, {Bahcall},
  {Bailer-Jones}, {Baldry}, {Barentine}, {Bassett}, {Becker}, {Beers}, {Bell},
  {Berlind}, {Bernardi}, {Blanton}, {Bochanski}, {Boroski}, {Brinchmann},
  {Brinkmann}, {Brunner}, {Budav{\'a}ri}, {Carliles}, {Carr}, {Castander},
  {Cinabro}, {Cool}, {Covey}, {Csabai}, {Cunha}, {Davenport}, {Dilday}, {Doi},
  {Eisenstein}, {Evans}, {Fan}, {Finkbeiner}, {Friedman}, {Frieman},
  {Fukugita}, {G{\"a}nsicke}, {Gates}, {Gillespie}, {Glazebrook}, {Gray},
  {Grebel}, {Gunn}, {Gurbani}, {Hall}, {Harding}, {Harvanek}, {Hawley},
  {Hayes}, {Heckman}, {Hendry}, {Hindsley}, {Hirata}, {Hogan}, {Hogg}, {Hyde},
  {Ichikawa}, {Ivezi{\'c}}, {Jester}, {Johnson}, {Jorgensen}, {Juri{\'c}},
  {Kent}, {Kessler}, {Kleinman}, {Knapp}, {Kron}, {Krzesinski}, {Kuropatkin},
  {Lamb}, {Lampeitl}, {Lebedeva}, {Lee}, {French Leger}, {L{\'e}pine}, {Lima},
  {Lin}, {Long}, {Loomis}, {Loveday}, {Lupton}, {Malanushenko}, {Malanushenko},
  {Mandelbaum}, {Margon}, {Marriner}, {Mart{\'\i}nez-Delgado}, {Matsubara},
  {McGehee}, {McKay}, {Meiksin}, {Morrison}, {Munn}, {Nakajima}, {Neilsen},
  {Newberg}, {Nichol}, {Nicinski}, {Nieto-Santisteban}, {Nitta}, {Okamura},
  {Owen}, {Oyaizu}, {Padmanabhan}, {Pan}, {Park}, {Peoples}, {Pier}, {Pope},
  {Purger}, {Raddick}, {Re Fiorentin}, {Richards}, {Richmond}, {Riess}, {Rix},
  {Rockosi}, {Sako}, {Schlegel}, {Schneider}, {Schreiber}, {Schwope}, {Seljak},
  {Sesar}, {Sheldon}, {Shimasaku}, {Sivarani}, {Allyn Smith}, {Snedden},
  {Steinmetz}, {Strauss}, {SubbaRao}, {Suto}, {Szalay}, {Szapudi}, {Szkody},
  {Tegmark}, {Thakar}, {Tremonti}, {Tucker}, {Uomoto}, {Vanden Berk},
  {Vandenberg}, {Vidrih}, {Vogeley}, {Voges}, {Vogt}, {Wadadekar}, {Weinberg},
  {West}, {White}, {Wilhite}, {Yanny}, {Yocum}, {York}, {Zehavi}, \&
  {Zucker}}]{AdelmanMcCarthy_2008}
{Adelman-McCarthy}, J.~K., {Ag{\"u}eros}, M.~A., {Allam}, S.~S., {et~al.} 2008,
  \href{http://dx.doi.org/10.1086/524984}{\color{magenta}\apjs},
  \href{https://ui.adsabs.harvard.edu/abs/2008ApJS..175..297A}{175, 297}

\bibitem[{{Ahumada} {et~al.}(2020){Ahumada}, {Prieto}, {Almeida}, {Anders},
  {Anderson}, {Andrews}, {Anguiano}, {Arcodia}, {Armengaud}, {Aubert}, \&
  et~al.}]{Ahumada_2020}
{Ahumada}, R., {Prieto}, C.~A., {Almeida}, A., {et~al.} 2020,
  \href{http://dx.doi.org/10.3847/1538-4365/ab929e}{\color{magenta}\apjs},
  \href{https://ui.adsabs.harvard.edu/abs/2020ApJS..249....3A}{249, 3}

\bibitem[{{Angus}(2009)}]{Angus_2009}
{Angus}, G.~W. 2009,
  \href{http://dx.doi.org/10.1111/j.1365-2966.2008.14341.x}{\color{magenta}\mnras},
  \href{https://ui.adsabs.harvard.edu/abs/2009MNRAS.394..527A}{394, 527}

\bibitem[{{Asencio} {et~al.}(2021){Asencio}, {Banik}, \&
  {Kroupa}}]{Asencio_2021}
{Asencio}, E., {Banik}, I., \& {Kroupa}, P. 2021,
  \href{http://dx.doi.org/10.1093/mnras/staa3441}{\color{magenta}\mnras},
  \href{https://ui.adsabs.harvard.edu/abs/2021MNRAS.500.5249A}{500, 5249}

\bibitem[{{Athanassoula}(2002)}]{Athanassoula_2002}
{Athanassoula}, E. 2002,
  \href{http://dx.doi.org/10.1086/340784}{\color{magenta}\apjl},
  \href{https://ui.adsabs.harvard.edu/abs/2002ApJ...569L..83A}{569, L83}

\bibitem[{{Bah{\'e}} {et~al.}(2017){Bah{\'e}}, {Barnes}, {Dalla Vecchia},
  {Kay}, {White}, {McCarthy}, {Schaye}, {Bower}, {Crain}, {Theuns}, {Jenkins},
  {McGee}, {Schaller}, {Thomas}, \& {Trayford}}]{Bahe_2017}
{Bah{\'e}}, Y.~M., {Barnes}, D.~J., {Dalla Vecchia}, C., {et~al.} 2017,
  \href{http://dx.doi.org/10.1093/mnras/stx1403}{\color{magenta}\mnras},
  \href{https://ui.adsabs.harvard.edu/abs/2017MNRAS.470.4186B}{470, 4186}

\bibitem[{{Baldry} {et~al.}(2018){Baldry}, {Liske}, {Brown}, {Robotham},
  {Driver}, {Dunne}, {Alpaslan}, {Brough}, {Cluver}, {Eardley}, {Farrow},
  {Heymans}, {Hildebrandt}, {Hopkins}, {Kelvin}, {Loveday}, {Moffett},
  {Norberg}, {Owers}, {Taylor}, {Wright}, {Bamford}, {Bland-Hawthorn},
  {Bourne}, {Bremer}, {Colless}, {Conselice}, {Croom}, {Davies}, {Foster},
  {Grootes}, {Holwerda}, {Jones}, {Kafle}, {Kuijken}, {Lara-Lopez},
  {L{\'o}pez-S{\'a}nchez}, {Meyer}, {Phillipps}, {Sutherland}, {van Kampen}, \&
  {Wilkins}}]{Baldry_2018}
{Baldry}, I.~K., {Liske}, J., {Brown}, M.~J.~I., {et~al.} 2018,
  \href{http://dx.doi.org/10.1093/mnras/stx3042}{\color{magenta}\mnras},
  \href{https://ui.adsabs.harvard.edu/abs/2018MNRAS.474.3875B}{474, 3875}

\bibitem[{{Banik} {et~al.}(2020){Banik}, {Thies}, {Candlish}, {Famaey},
  {Ibata}, \& {Kroupa}}]{Banik_2020_M33}
{Banik}, I., {Thies}, I., {Candlish}, G., {et~al.} 2020,
  \href{http://dx.doi.org/10.3847/1538-4357/abc623}{\color{magenta}ApJ},
  \href{https://ui.adsabs.harvard.edu/abs/2020ApJ...905..135B}{905, 135}

\bibitem[{{Banik} \& {Zhao}(2018)}]{Banik_2018_anisotropy}
{Banik}, I. \& {Zhao}, H. 2018,
  \href{http://dx.doi.org/10.1093/mnras/stx2596}{\color{magenta}\mnras},
  \href{https://ui.adsabs.harvard.edu/abs/2018MNRAS.473.4033B}{473, 4033}

\bibitem[{{Banik} \& {Zhao}(2022)}]{Banik_Zhao_2022}
{Banik}, I. \& {Zhao}, H. 2022, ArXiv e-prints, Arxiv
  \href{https://ui.adsabs.harvard.edu/abs/2021arXiv211006936B}{[\eprint[arXiv]{2110.06936}]}

\bibitem[{{Barnes} {et~al.}(2017){Barnes}, {Kay}, {Bah{\'e}}, {Dalla Vecchia},
  {McCarthy}, {Schaye}, {Bower}, {Jenkins}, {Thomas}, {Schaller}, {Crain},
  {Theuns}, \& {White}}]{Barnes_2017}
{Barnes}, D.~J., {Kay}, S.~T., {Bah{\'e}}, Y.~M., {et~al.} 2017,
  \href{http://dx.doi.org/10.1093/mnras/stx1647}{\color{magenta}\mnras},
  \href{https://ui.adsabs.harvard.edu/abs/2017MNRAS.471.1088B}{471, 1088}

\bibitem[{{Binney} \& {Tremaine}(2008)}]{Binney_2008}
{Binney}, J. \& {Tremaine}, S. 2008, {Galactic Dynamics: Second Edition}
  (Princeton University Press)

\bibitem[{{Bland-Hawthorn} \& {Gerhard}(2016)}]{BlandHawthorn_2016}
{Bland-Hawthorn}, J. \& {Gerhard}, O. 2016,
  \href{http://dx.doi.org/10.1146/annurev-astro-081915-023441}{\color{magenta}\araa},
  \href{https://ui.adsabs.harvard.edu/abs/2016ARA\&A..54..529B}{54, 529}

\bibitem[{{Bottrell} {et~al.}(2017{\natexlab{a}}){Bottrell}, {Torrey},
  {Simard}, \& {Ellison}}]{BOTTRELL_2017a}
{Bottrell}, C., {Torrey}, P., {Simard}, L., \& {Ellison}, S.~L.
  2017{\natexlab{a}},
  \href{http://dx.doi.org/10.1093/mnras/stx017}{\color{magenta}\mnras},
  \href{https://ui.adsabs.harvard.edu/abs/2017MNRAS.467.1033B}{467, 1033}

\bibitem[{{Bottrell} {et~al.}(2017{\natexlab{b}}){Bottrell}, {Torrey},
  {Simard}, \& {Ellison}}]{BOTTRELL_2017b}
{Bottrell}, C., {Torrey}, P., {Simard}, L., \& {Ellison}, S.~L.
  2017{\natexlab{b}},
  \href{http://dx.doi.org/10.1093/mnras/stx276}{\color{magenta}\mnras},
  \href{https://ui.adsabs.harvard.edu/abs/2017MNRAS.467.2879B}{467, 2879}

\bibitem[{{Bournaud} {et~al.}(2005){Bournaud}, {Jog}, \&
  {Combes}}]{Bournaud_2005}
{Bournaud}, F., {Jog}, C.~J., \& {Combes}, F. 2005,
  \href{http://dx.doi.org/10.1051/0004-6361:20042036}{\color{magenta}\aap},
  \href{https://ui.adsabs.harvard.edu/abs/2005A\&A...437...69B}{437, 69}

\bibitem[{{Brooks} {et~al.}(2013){Brooks}, {Kuhlen}, {Zolotov}, \&
  {Hooper}}]{Brooks_2013}
{Brooks}, A.~M., {Kuhlen}, M., {Zolotov}, A., \& {Hooper}, D. 2013,
  \href{http://dx.doi.org/10.1088/0004-637X/765/1/22}{\color{magenta}\apj},
  \href{https://ui.adsabs.harvard.edu/abs/2013ApJ...765...22B}{765, 22}

\bibitem[{{Bryant} {et~al.}(2015){Bryant}, {Owers}, {Robotham}, {Croom},
  {Driver}, {Drinkwater}, {Lorente}, {Cortese}, {Scott}, {Colless}, {Schaefer},
  {Taylor}, {Konstantopoulos}, {Allen}, {Baldry}, {Barnes}, {Bauer},
  {Bland-Hawthorn}, {Bloom}, {Brooks}, {Brough}, {Cecil}, {Couch}, {Croton},
  {Davies}, {Ellis}, {Fogarty}, {Foster}, {Glazebrook}, {Goodwin}, {Green},
  {Gunawardhana}, {Hampton}, {Ho}, {Hopkins}, {Kewley}, {Lawrence},
  {Leon-Saval}, {Leslie}, {McElroy}, {Lewis}, {Liske}, {L{\'o}pez-S{\'a}nchez},
  {Mahajan}, {Medling}, {Metcalfe}, {Meyer}, {Mould}, {Obreschkow}, {O'Toole},
  {Pracy}, {Richards}, {Shanks}, {Sharp}, {Sweet}, {Thomas}, {Tonini}, \&
  {Walcher}}]{Bryant_2015}
{Bryant}, J.~J., {Owers}, M.~S., {Robotham}, A.~S.~G., {et~al.} 2015,
  \href{http://dx.doi.org/10.1093/mnras/stu2635}{\color{magenta}\mnras},
  \href{https://ui.adsabs.harvard.edu/abs/2015MNRAS.447.2857B}{447, 2857}

\bibitem[{{Chambers} {et~al.}(2016){Chambers}, {Magnier}, {Metcalfe},
  {Flewelling}, {Huber}, {Waters}, {Denneau}, {Draper}, {Farrow}, {Finkbeiner},
  {Holmberg}, {Koppenhoefer}, {Price}, {Rest}, {Saglia}, {Schlafly}, {Smartt},
  {Sweeney}, {Wainscoat}, {Burgett}, {Chastel}, {Grav}, {Heasley}, {Hodapp},
  {Jedicke}, {Kaiser}, {Kudritzki}, {Luppino}, {Lupton}, {Monet}, {Morgan},
  {Onaka}, {Shiao}, {Stubbs}, {Tonry}, {White}, {Ba{\~n}ados}, {Bell},
  {Bender}, {Bernard}, {Boegner}, {Boffi}, {Botticella}, {Calamida},
  {Casertano}, {Chen}, {Chen}, {Cole}, {Deacon}, {Frenk}, {Fitzsimmons},
  {Gezari}, {Gibbs}, {Goessl}, {Goggia}, {Gourgue}, {Goldman}, {Grant},
  {Grebel}, {Hambly}, {Hasinger}, {Heavens}, {Heckman}, {Henderson}, {Henning},
  {Holman}, {Hopp}, {Ip}, {Isani}, {Jackson}, {Keyes}, {Koekemoer}, {Kotak},
  {Le}, {Liska}, {Long}, {Lucey}, {Liu}, {Martin}, {Masci}, {McLean}, {Mindel},
  {Misra}, {Morganson}, {Murphy}, {Obaika}, {Narayan}, {Nieto-Santisteban},
  {Norberg}, {Peacock}, {Pier}, {Postman}, {Primak}, {Rae}, {Rai}, {Riess},
  {Riffeser}, {Rix}, {R{\"o}ser}, {Russel}, {Rutz}, {Schilbach}, {Schultz},
  {Scolnic}, {Strolger}, {Szalay}, {Seitz}, {Small}, {Smith}, {Soderblom},
  {Taylor}, {Thomson}, {Taylor}, {Thakar}, {Thiel}, {Thilker}, {Unger},
  {Urata}, {Valenti}, {Wagner}, {Walder}, {Walter}, {Watters}, {Werner},
  {Wood-Vasey}, \& {Wyse}}]{Chambers_2016}
{Chambers}, K.~C., {Magnier}, E.~A., {Metcalfe}, N., {et~al.} 2016, ArXiv
  e-prints, Arxiv
  \href{https://ui.adsabs.harvard.edu/abs/2016arXiv161205560C}{[\eprint[arXiv]{1612.05560}]}

\bibitem[{{Courteau} {et~al.}(2011){Courteau}, {Widrow}, {McDonald},
  {Guhathakurta}, {Gilbert}, {Zhu}, {Beaton}, \& {Majewski}}]{Courteau_2011}
{Courteau}, S., {Widrow}, L.~M., {McDonald}, M., {et~al.} 2011,
  \href{http://dx.doi.org/10.1088/0004-637X/739/1/20}{\color{magenta}\apj},
  \href{https://ui.adsabs.harvard.edu/abs/2011ApJ...739...20C}{739, 20}

\bibitem[{{Crain} {et~al.}(2015){Crain}, {Schaye}, {Bower}, {Furlong},
  {Schaller}, {Theuns}, {Dalla Vecchia}, {Frenk}, {McCarthy}, {Helly},
  {Jenkins}, {Rosas-Guevara}, {White}, \& {Trayford}}]{Crain_2015}
{Crain}, R.~A., {Schaye}, J., {Bower}, R.~G., {et~al.} 2015,
  \href{http://dx.doi.org/10.1093/mnras/stv725}{\color{magenta}\mnras},
  \href{https://ui.adsabs.harvard.edu/abs/2015MNRAS.450.1937C}{450, 1937}

\bibitem[{{Croom} {et~al.}(2012){Croom}, {Lawrence}, {Bland-Hawthorn},
  {Bryant}, {Fogarty}, {Richards}, {Goodwin}, {Farrell}, {Miziarski}, {Heald},
  {Jones}, {Lee}, {Colless}, {Brough}, {Hopkins}, {Bauer}, {Birchall}, {Ellis},
  {Horton}, {Leon-Saval}, {Lewis}, {L{\'o}pez-S{\'a}nchez}, {Min}, {Trinh}, \&
  {Trowland}}]{Croom_2012}
{Croom}, S.~M., {Lawrence}, J.~S., {Bland-Hawthorn}, J., {et~al.} 2012,
  \href{http://dx.doi.org/10.1111/j.1365-2966.2011.20365.x}{\color{magenta}\mnras},
  \href{https://ui.adsabs.harvard.edu/abs/2012MNRAS.421..872C}{421, 872}

\bibitem[{{Croom} {et~al.}(2021){Croom}, {Owers}, {Scott}, {Poetrodjojo},
  {Groves}, {van de Sande}, {Barone}, {Cortese}, {D'Eugenio}, {Bland-Hawthorn},
  {Bryant}, {Oh}, {Brough}, {Agostino}, {Casura}, {Catinella}, {Colless},
  {Cecil}, {Davies}, {Drinkwater}, {Driver}, {Ferreras}, {Foster},
  {Fraser-McKelvie}, {Lawrence}, {Leslie}, {Liske}, {L{\'o}pez-S{\'a}nchez},
  {Lorente}, {McElroy}, {Medling}, {Obreschkow}, {Richards}, {Sharp}, {Sweet},
  {Taranu}, {Taylor}, {Tescari}, {Thomas}, {Tocknell}, \&
  {Vaughan}}]{Croom_2021}
{Croom}, S.~M., {Owers}, M.~S., {Scott}, N., {et~al.} 2021,
  \href{http://dx.doi.org/10.1093/mnras/stab229}{\color{magenta}\mnras},
  \href{https://ui.adsabs.harvard.edu/abs/2021MNRAS.505..991C}{505, 991}

\bibitem[{{de Vaucouleurs}(1948)}]{deVaucouleurs_1948}
{de Vaucouleurs}, G. 1948, Annales d'Astrophysique,
  \href{https://ui.adsabs.harvard.edu/abs/1948AnAp...11..247D}{11, 247}

\bibitem[{{Delgado-Serrano} {et~al.}(2010){Delgado-Serrano}, {Hammer}, {Yang},
  {Puech}, {Flores}, \& {Rodrigues}}]{DelgadoSerrano_2010}
{Delgado-Serrano}, R., {Hammer}, F., {Yang}, Y.~B., {et~al.} 2010,
  \href{http://dx.doi.org/10.1051/0004-6361/200912704}{\color{magenta}\aap},
  \href{https://ui.adsabs.harvard.edu/abs/2010A\&A...509A..78D}{509, A78}

\bibitem[{{Disney} {et~al.}(2008){Disney}, {Romano}, {Garcia-Appadoo}, {West},
  {Dalcanton}, \& {Cortese}}]{Disney_2008}
{Disney}, M.~J., {Romano}, J.~D., {Garcia-Appadoo}, D.~A., {et~al.} 2008,
  \href{http://dx.doi.org/10.1038/nature07366}{\color{magenta}\nat},
  \href{https://ui.adsabs.harvard.edu/abs/2008Natur.455.1082D}{455, 1082}

\bibitem[{{Dolag} {et~al.}(2016){Dolag}, {Komatsu}, \& {Sunyaev}}]{Dolag_2016}
{Dolag}, K., {Komatsu}, E., \& {Sunyaev}, R. 2016,
  \href{http://dx.doi.org/10.1093/mnras/stw2035}{\color{magenta}\mnras},
  \href{https://ui.adsabs.harvard.edu/abs/2016MNRAS.463.1797D}{463, 1797}

\bibitem[{{Dolag} {et~al.}(2017){Dolag}, {Mevius}, \& {Remus}}]{Dolag_2017}
{Dolag}, K., {Mevius}, E., \& {Remus}, R.-S. 2017,
  \href{http://dx.doi.org/10.3390/galaxies5030035}{\color{magenta}Galaxies},
  \href{https://ui.adsabs.harvard.edu/abs/2017Galax...5...35D}{5, 35}

\bibitem[{{D'Onghia} {et~al.}(2006){D'Onghia}, {Burkert}, {Murante}, \&
  {Khochfar}}]{DOnghia_2006}
{D'Onghia}, E., {Burkert}, A., {Murante}, G., \& {Khochfar}, S. 2006,
  \href{http://dx.doi.org/10.1111/j.1365-2966.2006.10996.x}{\color{magenta}\mnras},
  \href{https://ui.adsabs.harvard.edu/abs/2006MNRAS.372.1525D}{372, 1525}

\bibitem[{{Driver} {et~al.}(2011){Driver}, {Hill}, {Kelvin}, {Robotham},
  {Liske}, {Norberg}, {Baldry}, {Bamford}, {Hopkins}, {Loveday}, {Peacock},
  {Andrae}, {Bland-Hawthorn}, {Brough}, {Brown}, {Cameron}, {Ching}, {Colless},
  {Conselice}, {Croom}, {Cross}, {de Propris}, {Dye}, {Drinkwater}, {Ellis},
  {Graham}, {Grootes}, {Gunawardhana}, {Jones}, {van Kampen}, {Maraston},
  {Nichol}, {Parkinson}, {Phillipps}, {Pimbblet}, {Popescu}, {Prescott},
  {Roseboom}, {Sadler}, {Sansom}, {Sharp}, {Smith}, {Taylor}, {Thomas},
  {Tuffs}, {Wijesinghe}, {Dunne}, {Frenk}, {Jarvis}, {Madore}, {Meyer},
  {Seibert}, {Staveley-Smith}, {Sutherland}, \& {Warren}}]{Driver_2011}
{Driver}, S.~P., {Hill}, D.~T., {Kelvin}, L.~S., {et~al.} 2011,
  \href{http://dx.doi.org/10.1111/j.1365-2966.2010.18188.x}{\color{magenta}\mnras},
  \href{https://ui.adsabs.harvard.edu/abs/2011MNRAS.413..971D}{413, 971}

\bibitem[{{Driver} {et~al.}(2009){Driver}, {Norberg}, {Baldry}, {Bamford},
  {Hopkins}, {Liske}, {Loveday}, {Peacock}, {Hill}, {Kelvin}, {Robotham},
  {Cross}, {Parkinson}, {Prescott}, {Conselice}, {Dunne}, {Brough}, {Jones},
  {Sharp}, {van Kampen}, {Oliver}, {Roseboom}, {Bland-Hawthorn}, {Croom},
  {Ellis}, {Cameron}, {Cole}, {Frenk}, {Couch}, {Graham}, {Proctor}, {De
  Propris}, {Doyle}, {Edmondson}, {Nichol}, {Thomas}, {Eales}, {Jarvis},
  {Kuijken}, {Lahav}, {Madore}, {Seibert}, {Meyer}, {Staveley-Smith},
  {Phillipps}, {Popescu}, {Sansom}, {Sutherland}, {Tuffs}, \&
  {Warren}}]{Driver_2009}
{Driver}, S.~P., {Norberg}, P., {Baldry}, I.~K., {et~al.} 2009,
  \href{http://dx.doi.org/10.1111/j.1468-4004.2009.50512.x}{\color{magenta}Astronomy
  and Geophysics},
  \href{https://ui.adsabs.harvard.edu/abs/2009A&G....50e..12D}{50, 5.12}

\bibitem[{{Dubois} {et~al.}(2014){Dubois}, {Pichon}, {Welker}, {Le Borgne},
  {Devriendt}, {Laigle}, {Codis}, {Pogosyan}, {Arnouts}, {Benabed}, {Bertin},
  {Blaizot}, {Bouchet}, {Cardoso}, {Colombi}, {de Lapparent}, {Desjacques},
  {Gavazzi}, {Kassin}, {Kimm}, {McCracken}, {Milliard}, {Peirani}, {Prunet},
  {Rouberol}, {Silk}, {Slyz}, {Sousbie}, {Teyssier}, {Tresse}, {Treyer},
  {Vibert}, \& {Volonteri}}]{Dubois_2014}
{Dubois}, Y., {Pichon}, C., {Welker}, C., {et~al.} 2014,
  \href{http://dx.doi.org/10.1093/mnras/stu1227}{\color{magenta}\mnras},
  \href{https://ui.adsabs.harvard.edu/abs/2014MNRAS.444.1453D}{444, 1453}

\bibitem[{{Efstathiou} {et~al.}(1990){Efstathiou}, {Sutherland}, \&
  {Maddox}}]{Efstathiou_1990}
{Efstathiou}, G., {Sutherland}, W.~J., \& {Maddox}, S.~J. 1990,
  \href{http://dx.doi.org/10.1038/348705a0}{\color{magenta}Nature},
  \href{https://ui.adsabs.harvard.edu/abs/1990Natur.348..705E}{348, 705}

\bibitem[{{Emsellem} {et~al.}(2011){Emsellem}, {Cappellari}, {Krajnovi{\'c}},
  {Alatalo}, {Blitz}, {Bois}, {Bournaud}, {Bureau}, {Davies}, {Davis}, {de
  Zeeuw}, {Khochfar}, {Kuntschner}, {Lablanche}, {McDermid}, {Morganti},
  {Naab}, {Oosterloo}, {Sarzi}, {Scott}, {Serra}, {van de Ven}, {Weijmans}, \&
  {Young}}]{Emsellem_2011}
{Emsellem}, E., {Cappellari}, M., {Krajnovi{\'c}}, D., {et~al.} 2011,
  \href{http://dx.doi.org/10.1111/j.1365-2966.2011.18496.x}{\color{magenta}\mnras},
  \href{https://ui.adsabs.harvard.edu/abs/2011MNRAS.414..888E}{414, 888}

\bibitem[{{Fakhouri} {et~al.}(2010){Fakhouri}, {Ma}, \&
  {Boylan-Kolchin}}]{Fakhouri_2010}
{Fakhouri}, O., {Ma}, C.-P., \& {Boylan-Kolchin}, M. 2010,
  \href{http://dx.doi.org/10.1111/j.1365-2966.2010.16859.x}{\color{magenta}\mnras},
  \href{https://ui.adsabs.harvard.edu/abs/2010MNRAS.406.2267F}{406, 2267}

\bibitem[{{Famaey} \& {McGaugh}(2012)}]{Famaey_McGaugh_2012}
{Famaey}, B. \& {McGaugh}, S.~S. 2012,
  \href{http://dx.doi.org/10.12942/lrr-2012-10}{\color{magenta}Living Reviews
  in Relativity},
  \href{https://ui.adsabs.harvard.edu/abs/2012LRR....15...10F}{15, 10}

\bibitem[{Fletcher \& Powell(1963)}]{Fletcher_1963}
Fletcher, R. \& Powell, M. J.~D. 1963,
  \href{http://dx.doi.org/10.1093/comjnl/6.2.163}{\color{magenta}The Computer
  Journal}, \href{http://dx.doi.org/10.1093/comjnl/6.2.163}{6, 163}

\bibitem[{{Genel} {et~al.}(2015){Genel}, {Fall}, {Hernquist}, {Vogelsberger},
  {Snyder}, {Rodriguez-Gomez}, {Sijacki}, \& {Springel}}]{Genel_2015}
{Genel}, S., {Fall}, S.~M., {Hernquist}, L., {et~al.} 2015,
  \href{http://dx.doi.org/10.1088/2041-8205/804/2/L40}{\color{magenta}\apjl},
  \href{https://ui.adsabs.harvard.edu/abs/2015ApJ...804L..40G}{804, L40}

\bibitem[{{Giavalisco} {et~al.}(2004){Giavalisco}, {Ferguson}, {Koekemoer},
  {Dickinson}, {Alexander}, {Bauer}, {Bergeron}, {Biagetti}, {Brandt},
  {Casertano}, {Cesarsky}, {Chatzichristou}, {Conselice}, {Cristiani}, {Da
  Costa}, {Dahlen}, {de Mello}, {Eisenhardt}, {Erben}, {Fall}, {Fassnacht},
  {Fosbury}, {Fruchter}, {Gardner}, {Grogin}, {Hook}, {Hornschemeier}, {Idzi},
  {Jogee}, {Kretchmer}, {Laidler}, {Lee}, {Livio}, {Lucas}, {Madau},
  {Mobasher}, {Moustakas}, {Nonino}, {Padovani}, {Papovich}, {Park},
  {Ravindranath}, {Renzini}, {Richardson}, {Riess}, {Rosati}, {Schirmer},
  {Schreier}, {Somerville}, {Spinrad}, {Stern}, {Stiavelli}, {Strolger},
  {Urry}, {Vandame}, {Williams}, \& {Wolf}}]{Giavalisco_2004}
{Giavalisco}, M., {Ferguson}, H.~C., {Koekemoer}, A.~M., {et~al.} 2004,
  \href{http://dx.doi.org/10.1086/379232}{\color{magenta}\apjl},
  \href{https://ui.adsabs.harvard.edu/abs/2004ApJ...600L..93G}{600, L93}

\bibitem[{{Graham} \& {Worley}(2008)}]{Graham_2008}
{Graham}, A.~W. \& {Worley}, C.~C. 2008,
  \href{http://dx.doi.org/10.1111/j.1365-2966.2008.13506.x}{\color{magenta}\mnras},
  \href{https://ui.adsabs.harvard.edu/abs/2008MNRAS.388.1708G}{388, 1708}

\bibitem[{{Grazian} {et~al.}(2006){Grazian}, {Fontana}, {de Santis}, {Nonino},
  {Salimbeni}, {Giallongo}, {Cristiani}, {Gallozzi}, \&
  {Vanzella}}]{Grazian_2006}
{Grazian}, A., {Fontana}, A., {de Santis}, C., {et~al.} 2006,
  \href{http://dx.doi.org/10.1051/0004-6361:20053979}{\color{magenta}\aap},
  \href{https://ui.adsabs.harvard.edu/abs/2006A\&A...449..951G}{449, 951}

\bibitem[{{Guedes} {et~al.}(2011){Guedes}, {Callegari}, {Madau}, \&
  {Mayer}}]{Guedes_2011}
{Guedes}, J., {Callegari}, S., {Madau}, P., \& {Mayer}, L. 2011,
  \href{http://dx.doi.org/10.1088/0004-637X/742/2/76}{\color{magenta}\apj},
  \href{https://ui.adsabs.harvard.edu/abs/2011ApJ...742...76G}{742, 76}

\bibitem[{{Haslbauer} {et~al.}(2020){Haslbauer}, {Banik}, \&
  {Kroupa}}]{Haslbauer_2020}
{Haslbauer}, M., {Banik}, I., \& {Kroupa}, P. 2020,
  \href{http://dx.doi.org/10.1093/mnras/staa2348}{\color{magenta}\mnras},
  \href{https://ui.adsabs.harvard.edu/abs/2020MNRAS.499.2845H}{499, 2845}

\bibitem[{{Hinshaw} {et~al.}(2013){Hinshaw}, {Larson}, {Komatsu}, {Spergel},
  {Bennett}, {Dunkley}, {Nolta}, {Halpern}, {Hill}, {Odegard}, {Page}, {Smith},
  {Weiland}, {Gold}, {Jarosik}, {Kogut}, {Limon}, {Meyer}, {Tucker}, {Wollack},
  \& {Wright}}]{Hinshaw_2013_Illustris}
{Hinshaw}, G., {Larson}, D., {Komatsu}, E., {et~al.} 2013,
  \href{http://dx.doi.org/10.1088/0067-0049/208/2/19}{\color{magenta}\apjs},
  \href{https://ui.adsabs.harvard.edu/abs/2013ApJS..208...19H}{208, 19}

\bibitem[{{Hirschmann} {et~al.}(2014){Hirschmann}, {Dolag}, {Saro}, {Bachmann},
  {Borgani}, \& {Burkert}}]{Hirschmann_2014}
{Hirschmann}, M., {Dolag}, K., {Saro}, A., {et~al.} 2014,
  \href{http://dx.doi.org/10.1093/mnras/stu1023}{\color{magenta}\mnras},
  \href{https://ui.adsabs.harvard.edu/abs/2014MNRAS.442.2304H}{442, 2304}

\bibitem[{{Hoffmann} {et~al.}(2020){Hoffmann}, {Laigle}, {Chisari}, {Tallada},
  {Dubois}, \& {Devriendt}}]{Hoffmann_2020}
{Hoffmann}, K., {Laigle}, C., {Chisari}, N.~E., {et~al.} 2020, ArXiv e-prints,
  Arxiv
  \href{https://ui.adsabs.harvard.edu/abs/2020arXiv201013845H}{[\eprint[arXiv]{2010.13845}]}

\bibitem[{{Hwang} {et~al.}(2021){Hwang}, {Park}, {Nam}, \&
  {Chung}}]{Hwang_2021}
{Hwang}, J.-S., {Park}, C., {Nam}, S.-h., \& {Chung}, H. 2021,
  \href{http://dx.doi.org/10.5303/JKAS.2021.54.2.71}{\color{magenta}JKAS},
  \href{https://ui.adsabs.harvard.edu/abs/2021JKAS...54...71H}{54, 71}

\bibitem[{{Karachentsev} {et~al.}(2018){Karachentsev}, {Kaisina}, \&
  {Makarov}}]{Karachentsev_2018}
{Karachentsev}, I.~D., {Kaisina}, E.~I., \& {Makarov}, D.~I. 2018,
  \href{http://dx.doi.org/10.1093/mnras/sty1774}{\color{magenta}\mnras},
  \href{https://ui.adsabs.harvard.edu/abs/2018MNRAS.479.4136K}{479, 4136}

\bibitem[{{Karachentsev} {et~al.}(2004){Karachentsev}, {Karachentseva},
  {Huchtmeier}, \& {Makarov}}]{Karachentsev_2004}
{Karachentsev}, I.~D., {Karachentseva}, V.~E., {Huchtmeier}, W.~K., \&
  {Makarov}, D.~I. 2004,
  \href{http://dx.doi.org/10.1086/382905}{\color{magenta}\aj},
  \href{https://ui.adsabs.harvard.edu/abs/2004AJ....127.2031K}{127, 2031}

\bibitem[{{Karachentsev} {et~al.}(2013){Karachentsev}, {Makarov}, \&
  {Kaisina}}]{Karachentsev_2013}
{Karachentsev}, I.~D., {Makarov}, D.~I., \& {Kaisina}, E.~I. 2013,
  \href{http://dx.doi.org/10.1088/0004-6256/145/4/101}{\color{magenta}\aj},
  \href{https://ui.adsabs.harvard.edu/abs/2013AJ....145..101K}{145, 101}

\bibitem[{{Katz} \& {Gunn}(1991)}]{Katz_1991}
{Katz}, N. \& {Gunn}, J.~E. 1991,
  \href{http://dx.doi.org/10.1086/170367}{\color{magenta}\apj},
  \href{https://ui.adsabs.harvard.edu/abs/1991ApJ...377..365K}{377, 365}

\bibitem[{{Kauffmann} {et~al.}(2003){Kauffmann}, {Heckman}, {White}, {Charlot},
  {Tremonti}, {Brinchmann}, {Bruzual}, {Peng}, {Seibert}, {Bernardi},
  {Blanton}, {Brinkmann}, {Castander}, {Cs{\'a}bai}, {Fukugita}, {Ivezic},
  {Munn}, {Nichol}, {Padmanabhan}, {Thakar}, {Weinberg}, \&
  {York}}]{Kauffmann_2003}
{Kauffmann}, G., {Heckman}, T.~M., {White}, S. D.~M., {et~al.} 2003,
  \href{http://dx.doi.org/10.1046/j.1365-8711.2003.06291.x}{\color{magenta}\mnras},
  \href{https://ui.adsabs.harvard.edu/abs/2003MNRAS.341...33K}{341, 33}

\bibitem[{{Kautsch} {et~al.}(2006){Kautsch}, {Grebel}, {Barazza}, \&
  {Gallagher}}]{Kautsch_2006}
{Kautsch}, S.~J., {Grebel}, E.~K., {Barazza}, F.~D., \& {Gallagher}, J.~S., I.
  2006,
  \href{http://dx.doi.org/10.1051/0004-6361:20053981}{\color{magenta}A\&A},
  \href{https://ui.adsabs.harvard.edu/abs/2006A\&A...445..765K}{445, 765}

\bibitem[{{Kelvin} {et~al.}(2012){Kelvin}, {Driver}, {Robotham}, {Hill},
  {Alpaslan}, {Baldry}, {Bamford}, {Bland-Hawthorn}, {Brough}, {Graham},
  {H{\"a}ussler}, {Hopkins}, {Liske}, {Loveday}, {Norberg}, {Phillipps},
  {Popescu}, {Prescott}, {Taylor}, \& {Tuffs}}]{Kelvin_2012}
{Kelvin}, L.~S., {Driver}, S.~P., {Robotham}, A. S.~G., {et~al.} 2012,
  \href{http://dx.doi.org/10.1111/j.1365-2966.2012.20355.x}{\color{magenta}\mnras},
  \href{https://ui.adsabs.harvard.edu/abs/2012MNRAS.421.1007K}{421, 1007}

\bibitem[{{Kelvin} {et~al.}(2014){Kelvin}, {Driver}, {Robotham}, {Taylor},
  {Graham}, {Alpaslan}, {Baldry}, {Bamford}, {Bauer}, {Bland-Hawthorn},
  {Brown}, {Colless}, {Conselice}, {Holwerda}, {Hopkins}, {Lara-L{\'o}pez},
  {Liske}, {L{\'o}pez-S{\'a}nchez}, {Loveday}, {Norberg}, {Phillipps},
  {Popescu}, {Prescott}, {Sansom}, \& {Tuffs}}]{Kelvin_2014b}
{Kelvin}, L.~S., {Driver}, S.~P., {Robotham}, A. S.~G., {et~al.} 2014,
  \href{http://dx.doi.org/10.1093/mnras/stu1507}{\color{magenta}\mnras},
  \href{https://ui.adsabs.harvard.edu/abs/2014MNRAS.444.1647K}{444, 1647}

\bibitem[{{Kormendy} {et~al.}(2010){Kormendy}, {Drory}, {Bender}, \&
  {Cornell}}]{Kormendy_2010}
{Kormendy}, J., {Drory}, N., {Bender}, R., \& {Cornell}, M.~E. 2010,
  \href{http://dx.doi.org/10.1088/0004-637X/723/1/54}{\color{magenta}\apj},
  \href{https://ui.adsabs.harvard.edu/abs/2010ApJ...723...54K}{723, 54}

\bibitem[{{Kroupa}(2012)}]{Kroupa_2012}
{Kroupa}, P. 2012,
  \href{http://dx.doi.org/10.1071/AS12005}{\color{magenta}\pasa},
  \href{https://ui.adsabs.harvard.edu/abs/2012PASA...29..395K}{29, 395}

\bibitem[{{Kroupa}(2015)}]{Kroupa_2015}
{Kroupa}, P. 2015,
  \href{http://dx.doi.org/10.1139/cjp-2014-0179}{\color{magenta}Canadian
  Journal of Physics},
  \href{https://ui.adsabs.harvard.edu/abs/2015CaJPh..93..169K}{93, 169}

\bibitem[{{Kroupa} {et~al.}(2020){Kroupa}, {Subr}, {Jerabkova}, \&
  {Wang}}]{Kroupa_2020}
{Kroupa}, P., {Subr}, L., {Jerabkova}, T., \& {Wang}, L. 2020,
  \href{http://dx.doi.org/10.1093/mnras/staa2276}{\color{magenta}\mnras},
  \href{https://ui.adsabs.harvard.edu/abs/2020MNRAS.498.5652K}{498, 5652}

\bibitem[{{Lagos} {et~al.}(2018){Lagos}, {Schaye}, {Bah{\'e}}, {Van de Sande},
  {Kay}, {Barnes}, {Davis}, \& {Dalla Vecchia}}]{Lagos_2018}
{Lagos}, C. d.~P., {Schaye}, J., {Bah{\'e}}, Y., {et~al.} 2018,
  \href{http://dx.doi.org/10.1093/mnras/sty489}{\color{magenta}\mnras},
  \href{https://ui.adsabs.harvard.edu/abs/2018MNRAS.476.4327L}{476, 4327}

\bibitem[{{Lange} {et~al.}(2016){Lange}, {Moffett}, {Driver}, {Robotham},
  {Lagos}, {Kelvin}, {Conselice}, {Margalef-Bentabol}, {Alpaslan}, {Baldry},
  {Bland-Hawthorn}, {Bremer}, {Brough}, {Cluver}, {Colless}, {Davies},
  {H{\"a}u{\ss}ler}, {Holwerda}, {Hopkins}, {Kafle}, {Kennedy}, {Liske},
  {Phillipps}, {Popescu}, {Taylor}, {Tuffs}, {van Kampen}, \&
  {Wright}}]{Lange_2016}
{Lange}, R., {Moffett}, A.~J., {Driver}, S.~P., {et~al.} 2016,
  \href{http://dx.doi.org/10.1093/mnras/stw1495}{\color{magenta}\mnras},
  \href{https://ui.adsabs.harvard.edu/abs/2016MNRAS.462.1470L}{462, 1470}

\bibitem[{{Loveday}(1996)}]{Loveday_1996}
{Loveday}, J. 1996,
  \href{http://dx.doi.org/10.1093/mnras/278.4.1025}{\color{magenta}\mnras},
  \href{https://ui.adsabs.harvard.edu/abs/1996MNRAS.278.1025L}{278, 1025}

\bibitem[{{Ludlow} {et~al.}(2021){Ludlow}, {Fall}, {Schaye}, \&
  {Obreschkow}}]{Ludlow_2021}
{Ludlow}, A.~D., {Fall}, S.~M., {Schaye}, J., \& {Obreschkow}, D. 2021,
  \href{http://dx.doi.org/10.1093/mnras/stab2770}{\color{magenta}MNRAS},
  \href{https://ui.adsabs.harvard.edu/abs/2021MNRAS.508.5114L}{508, 5114}

\bibitem[{{L{\"u}ghausen} {et~al.}(2015){L{\"u}ghausen}, {Famaey}, \&
  {Kroupa}}]{Lueghausen_2015}
{L{\"u}ghausen}, F., {Famaey}, B., \& {Kroupa}, P. 2015,
  \href{http://dx.doi.org/10.1139/cjp-2014-0168}{\color{magenta}Canadian
  Journal of Physics},
  \href{https://ui.adsabs.harvard.edu/abs/2015CaJPh..93..232L}{93, 232}

\bibitem[{{McAlpine} {et~al.}(2016){McAlpine}, {Helly}, {Schaller}, {Trayford},
  {Qu}, {Furlong}, {Bower}, {Crain}, {Schaye}, {Theuns}, {Dalla Vecchia},
  {Frenk}, {McCarthy}, {Jenkins}, {Rosas-Guevara}, {White}, {Baes}, {Camps}, \&
  {Lemson}}]{McAlpine_2016}
{McAlpine}, S., {Helly}, J.~C., {Schaller}, M., {et~al.} 2016,
  \href{http://dx.doi.org/10.1016/j.ascom.2016.02.004}{\color{magenta}Astronomy
  and Computing},
  \href{https://ui.adsabs.harvard.edu/abs/2016A\&C....15...72M}{15, 72}

\bibitem[{{McGaugh} \& {Schombert}(2014)}]{McGaugh_2014}
{McGaugh}, S.~S. \& {Schombert}, J.~M. 2014,
  \href{http://dx.doi.org/10.1088/0004-6256/148/5/77}{\color{magenta}\aj},
  \href{https://ui.adsabs.harvard.edu/abs/2014AJ....148...77M}{148, 77}

\bibitem[{{Milgrom}(1983)}]{Milgrom_1983}
{Milgrom}, M. 1983,
  \href{http://dx.doi.org/10.1086/161130}{\color{magenta}\apj},
  \href{https://ui.adsabs.harvard.edu/abs/1983ApJ...270..365M}{270, 365}

\bibitem[{{Mosenkov} {et~al.}(2010){Mosenkov}, {Sotnikova}, \&
  {Reshetnikov}}]{Mosenkov_2010}
{Mosenkov}, A.~V., {Sotnikova}, N.~Y., \& {Reshetnikov}, V.~P. 2010,
  \href{http://dx.doi.org/10.1111/j.1365-2966.2009.15671.x}{\color{magenta}\mnras},
  \href{https://ui.adsabs.harvard.edu/abs/2010MNRAS.401..559M}{401, 559}

\bibitem[{{M{\"u}ller} {et~al.}(2022){M{\"u}ller}, {Lelli}, {Famaey},
  {Pawlowski}, {Fahrion}, {Rejkuba}, {Hilker}, \& {Jerjen}}]{Muller_2022}
{M{\"u}ller}, O., {Lelli}, F., {Famaey}, B., {et~al.} 2022,
  \href{https://ui.adsabs.harvard.edu/abs/2021arXiv211110306M}{A\&A, in press}

\bibitem[{{Nagesh} {et~al.}(2021){Nagesh}, {Banik}, {Thies}, {Kroupa},
  {Famaey}, {Wittenburg}, {Parziale}, \& {Haslbauer}}]{Nagesh_2021}
{Nagesh}, S.~T., {Banik}, I., {Thies}, I., {et~al.} 2021,
  \href{http://dx.doi.org/10.1139/cjp-2020-0624}{\color{magenta}Canadian
  Journal of Physics},
  \href{https://ui.adsabs.harvard.edu/abs/2021CaJPh..99..607N}{99, 607}

\bibitem[{{Navarro} \& {Benz}(1991)}]{Navarro_1991}
{Navarro}, J.~F. \& {Benz}, W. 1991,
  \href{http://dx.doi.org/10.1086/170590}{\color{magenta}\apj},
  \href{https://ui.adsabs.harvard.edu/abs/1991ApJ...380..320N}{380, 320}

\bibitem[{{Navarro} \& {Steinmetz}(2000)}]{Navarro_2000}
{Navarro}, J.~F. \& {Steinmetz}, M. 2000,
  \href{http://dx.doi.org/10.1086/309175}{\color{magenta}\apj},
  \href{https://ui.adsabs.harvard.edu/abs/2000ApJ...538..477N}{538, 477}

\bibitem[{{Navarro} \& {White}(1994)}]{Navarro_1994}
{Navarro}, J.~F. \& {White}, S. D.~M. 1994,
  \href{http://dx.doi.org/10.1093/mnras/267.2.401}{\color{magenta}\mnras},
  \href{https://ui.adsabs.harvard.edu/abs/1994MNRAS.267..401N}{267, 401}

\bibitem[{{Nelson} {et~al.}(2015){Nelson}, {Pillepich}, {Genel},
  {Vogelsberger}, {Springel}, {Torrey}, {Rodriguez-Gomez}, {Sijacki}, {Snyder},
  {Griffen}, {Marinacci}, {Blecha}, {Sales}, {Xu}, \&
  {Hernquist}}]{Nelson_2015}
{Nelson}, D., {Pillepich}, A., {Genel}, S., {et~al.} 2015,
  \href{http://dx.doi.org/10.1016/j.ascom.2015.09.003}{\color{magenta}Astronomy
  and Computing},
  \href{https://ui.adsabs.harvard.edu/abs/2015A\&C....13...12N}{13, 12}

\bibitem[{{Nelson} {et~al.}(2018){Nelson}, {Pillepich}, {Springel},
  {Weinberger}, {Hernquist}, {Pakmor}, {Genel}, {Torrey}, {Vogelsberger},
  {Kauffmann}, {Marinacci}, \& {Naiman}}]{Nelson_2018}
{Nelson}, D., {Pillepich}, A., {Springel}, V., {et~al.} 2018,
  \href{http://dx.doi.org/10.1093/mnras/stx3040}{\color{magenta}\mnras},
  \href{https://ui.adsabs.harvard.edu/abs/2018MNRAS.475..624N}{475, 624}

\bibitem[{{Nelson} {et~al.}(2019){Nelson}, {Springel}, {Pillepich},
  {Rodriguez-Gomez}, {Torrey}, {Genel}, {Vogelsberger}, {Pakmor}, {Marinacci},
  {Weinberger}, {Kelley}, {Lovell}, {Diemer}, \& {Hernquist}}]{Nelson_2019}
{Nelson}, D., {Springel}, V., {Pillepich}, A., {et~al.} 2019,
  \href{http://dx.doi.org/10.1186/s40668-019-0028-x}{\color{magenta}Computational
  Astrophysics and Cosmology},
  \href{https://ui.adsabs.harvard.edu/abs/2019ComAC...6....2N}{6, 2}

\bibitem[{{Oh} {et~al.}(2020){Oh}, {Colless}, {Barsanti}, {Casura}, {Cortese},
  {van de Sande}, {Owers}, {Scott}, {D'Eugenio}, {Bland-Hawthorn}, {Brough},
  {Bryant}, {Croom}, {Foster}, {Groves}, {Lawrence}, {Richards}, \&
  {Sweet}}]{Oh_2020}
{Oh}, S., {Colless}, M., {Barsanti}, S., {et~al.} 2020,
  \href{http://dx.doi.org/10.1093/mnras/staa1330}{\color{magenta}\mnras},
  \href{https://ui.adsabs.harvard.edu/abs/2020MNRAS.495.4638O}{495, 4638}

\bibitem[{{Ostriker} \& {Peebles}(1973)}]{Ostriker_1973}
{Ostriker}, J.~P. \& {Peebles}, P.~J.~E. 1973,
  \href{http://dx.doi.org/10.1086/152513}{\color{magenta}ApJ},
  \href{https://ui.adsabs.harvard.edu/abs/1973ApJ...186..467O}{186, 467}

\bibitem[{{Ostriker} \& {Steinhardt}(1995)}]{Ostriker_Steinhardt_1995}
{Ostriker}, J.~P. \& {Steinhardt}, P.~J. 1995,
  \href{http://dx.doi.org/10.1038/377600a0}{\color{magenta}Nature},
  \href{http://adsabs.harvard.edu/abs/1995Natur.377..600O}{377, 600}

\bibitem[{{Owers} {et~al.}(2017){Owers}, {Allen}, {Baldry}, {Bryant}, {Cecil},
  {Cortese}, {Croom}, {Driver}, {Fogarty}, {Green}, {Helmich}, {de Jong},
  {Kuijken}, {Mahajan}, {McFarland}, {Pracy}, {Robotham}, {Sikkema}, {Sweet},
  {Taylor}, {Verdoes Kleijn}, {Bauer}, {Bland-Hawthorn}, {Brough}, {Colless},
  {Couch}, {Davies}, {Drinkwater}, {Goodwin}, {Hopkins}, {Konstantopoulos},
  {Foster}, {Lawrence}, {Lorente}, {Medling}, {Metcalfe}, {Richards}, {van de
  Sande}, {Scott}, {Shanks}, {Sharp}, {Thomas}, \& {Tonini}}]{Owers_2017}
{Owers}, M.~S., {Allen}, J.~T., {Baldry}, I., {et~al.} 2017,
  \href{http://dx.doi.org/10.1093/mnras/stx562}{\color{magenta}\mnras},
  \href{https://ui.adsabs.harvard.edu/abs/2017MNRAS.468.1824O}{468, 1824}

\bibitem[{{Owers} {et~al.}(2019){Owers}, {Hudson}, {Oman}, {Bland-Hawthorn},
  {Brough}, {Bryant}, {Cortese}, {Couch}, {Croom}, {van de Sande}, {Federrath},
  {Groves}, {Hopkins}, {Lawrence}, {Lorente}, {McDermid}, {Medling},
  {Richards}, {Scott}, {Taranu}, {Welker}, \& {Yi}}]{Owers_2019}
{Owers}, M.~S., {Hudson}, M.~J., {Oman}, K.~A., {et~al.} 2019,
  \href{http://dx.doi.org/10.3847/1538-4357/ab0201}{\color{magenta}\apj},
  \href{https://ui.adsabs.harvard.edu/abs/2019ApJ...873...52O}{873, 52}

\bibitem[{{Padilla} \& {Strauss}(2008)}]{Padilla_2008}
{Padilla}, N.~D. \& {Strauss}, M.~A. 2008,
  \href{http://dx.doi.org/10.1111/j.1365-2966.2008.13480.x}{\color{magenta}\mnras},
  \href{https://ui.adsabs.harvard.edu/abs/2008MNRAS.388.1321P}{388, 1321}

\bibitem[{{Paturel} {et~al.}(1997){Paturel}, {Andernach}, {Bottinelli}, {di
  Nella}, {Durand}, {Garnier}, {Gouguenheim}, {Lanoix}, {Marthinet}, {Petit},
  {Rousseau}, {Theureau}, \& {Vauglin}}]{Paturel_1997}
{Paturel}, G., {Andernach}, H., {Bottinelli}, L., {et~al.} 1997,
  \href{http://dx.doi.org/10.1051/aas:1997354}{\color{magenta}\aaps},
  \href{https://ui.adsabs.harvard.edu/abs/1997A\&AS..124..109P}{124, 109}

\bibitem[{{Pawlowski}(2021)}]{Pawlowski_2021}
{Pawlowski}, M.~S. 2021,
  \href{http://dx.doi.org/10.1038/s41550-021-01452-7}{\color{magenta}Nature
  Astronomy}, \href{https://ui.adsabs.harvard.edu/abs/2021NatAs...5.1185P}{5,
  1185}

\bibitem[{{Peebles}(2020)}]{Peebles_2020}
{Peebles}, P.~J.~E. 2020,
  \href{http://dx.doi.org/10.1093/mnras/staa2649}{\color{magenta}\mnras},
  \href{https://ui.adsabs.harvard.edu/abs/2020MNRAS.498.4386P}{498, 4386}

\bibitem[{{Peng} {et~al.}(2002){Peng}, {Ho}, {Impey}, \& {Rix}}]{Peng_2002}
{Peng}, C.~Y., {Ho}, L.~C., {Impey}, C.~D., \& {Rix}, H.-W. 2002,
  \href{http://dx.doi.org/10.1086/340952}{\color{magenta}\aj},
  \href{https://ui.adsabs.harvard.edu/abs/2002AJ....124..266P}{124, 266}

\bibitem[{{Peng} {et~al.}(2010){Peng}, {Ho}, {Impey}, \& {Rix}}]{Peng_2010}
{Peng}, C.~Y., {Ho}, L.~C., {Impey}, C.~D., \& {Rix}, H.-W. 2010,
  \href{http://dx.doi.org/10.1088/0004-6256/139/6/2097}{\color{magenta}\aj},
  \href{https://ui.adsabs.harvard.edu/abs/2010AJ....139.2097P}{139, 2097}

\bibitem[{{Pillepich} {et~al.}(2019){Pillepich}, {Nelson}, {Springel},
  {Pakmor}, {Torrey}, {Weinberger}, {Vogelsberger}, {Marinacci}, {Genel}, {van
  der Wel}, \& {Hernquist}}]{Pillepich_2019}
{Pillepich}, A., {Nelson}, D., {Springel}, V., {et~al.} 2019,
  \href{http://dx.doi.org/10.1093/mnras/stz2338}{\color{magenta}\mnras},
  \href{https://ui.adsabs.harvard.edu/abs/2019MNRAS.490.3196P}{490, 3196}

\bibitem[{{Pillepich} {et~al.}(2018){Pillepich}, {Springel}, {Nelson}, {Genel},
  {Naiman}, {Pakmor}, {Hernquist}, {Torrey}, {Vogelsberger}, {Weinberger}, \&
  {Marinacci}}]{Pillepich_2018}
{Pillepich}, A., {Springel}, V., {Nelson}, D., {et~al.} 2018,
  \href{http://dx.doi.org/10.1093/mnras/stx2656}{\color{magenta}\mnras},
  \href{https://ui.adsabs.harvard.edu/abs/2018MNRAS.473.4077P}{473, 4077}

\bibitem[{{Piontek} \& {Steinmetz}(2011)}]{Piontek_2011}
{Piontek}, F. \& {Steinmetz}, M. 2011,
  \href{http://dx.doi.org/10.1111/j.1365-2966.2010.17637.x}{\color{magenta}\mnras},
  \href{https://ui.adsabs.harvard.edu/abs/2011MNRAS.410.2625P}{410, 2625}

\bibitem[{{Planck Collaboration I}(2014)}]{Planck_2014_EAGLE}
{Planck Collaboration I}. 2014,
  \href{http://dx.doi.org/10.1051/0004-6361/201321529}{\color{magenta}\aap},
  \href{https://ui.adsabs.harvard.edu/abs/2014A\&A...571A...1P}{571, A1}

\bibitem[{{Planck Collaboration VI}(2020)}]{Planck_2020}
{Planck Collaboration VI}. 2020,
  \href{http://dx.doi.org/10.1051/0004-6361/201833910}{\color{magenta}A\&A},
  \href{https://ui.adsabs.harvard.edu/abs/2020A\&A...641A...6P}{641, A6}

\bibitem[{{Planck Collaboration XIII}(2016)}]{Planck_2016_IllustrisTNG}
{Planck Collaboration XIII}. 2016,
  \href{http://dx.doi.org/10.1051/0004-6361/201525830}{\color{magenta}\aap},
  \href{https://ui.adsabs.harvard.edu/abs/2016A\&A...594A..13P}{594, A13}

\bibitem[{{Reddish} {et~al.}(2021){Reddish}, {Kraljic}, {Petersen}, {Tep},
  {Dubois}, {Pichon}, {Peirani}, {Bournaud}, {Choi}, {Devriendt}, {Jackson},
  {Martin}, {Park}, {Volonteri}, \& {Yi}}]{Reddish_2021}
{Reddish}, J., {Kraljic}, K., {Petersen}, M.~S., {et~al.} 2021, ArXiv e-prints,
  Arxiv
  \href{https://ui.adsabs.harvard.edu/abs/2021arXiv210602622R}{[\eprint[arXiv]{2106.02622}]}

\bibitem[{{Renaud} {et~al.}(2016){Renaud}, {Famaey}, \& {Kroupa}}]{Renaud_2016}
{Renaud}, F., {Famaey}, B., \& {Kroupa}, P. 2016,
  \href{http://dx.doi.org/10.1093/mnras/stw2331}{\color{magenta}\mnras},
  \href{https://ui.adsabs.harvard.edu/abs/2016MNRAS.463.3637R}{463, 3637}

\bibitem[{{Riess} {et~al.}(2021){Riess}, {Yuan}, {Macri}, {Scolnic}, {Brout},
  {Casertano}, {Jones}, {Murakami}, {Breuval}, {Brink}, {Filippenko},
  {Hoffmann}, {Jha}, {Kenworthy}, {Mackenty}, {Stahl}, \& {Zheng}}]{Riess_2021}
{Riess}, A.~G., {Yuan}, W., {Macri}, L.~M., {et~al.} 2021, ArXiv e-prints,
  Arxiv
  \href{https://ui.adsabs.harvard.edu/abs/2021arXiv211204510R}{[\eprint[arXiv]{2112.04510}]}

\bibitem[{{Rodriguez-Gomez} {et~al.}(2015){Rodriguez-Gomez}, {Genel},
  {Vogelsberger}, {Sijacki}, {Pillepich}, {Sales}, {Torrey}, {Snyder},
  {Nelson}, {Springel}, {Ma}, \& {Hernquist}}]{RodriguezGomez_2015}
{Rodriguez-Gomez}, V., {Genel}, S., {Vogelsberger}, M., {et~al.} 2015,
  \href{http://dx.doi.org/10.1093/mnras/stv264}{\color{magenta}\mnras},
  \href{https://ui.adsabs.harvard.edu/abs/2015MNRAS.449...49R}{449, 49}

\bibitem[{{Rodriguez-Gomez} {et~al.}(2019){Rodriguez-Gomez}, {Snyder}, {Lotz},
  {Nelson}, {Pillepich}, {Springel}, {Genel}, {Weinberger}, {Tacchella},
  {Pakmor}, {Torrey}, {Marinacci}, {Vogelsberger}, {Hernquist}, \&
  {Thilker}}]{RodriguezGomez_2019}
{Rodriguez-Gomez}, V., {Snyder}, G.~F., {Lotz}, J.~M., {et~al.} 2019,
  \href{http://dx.doi.org/10.1093/mnras/sty3345}{\color{magenta}\mnras},
  \href{https://ui.adsabs.harvard.edu/abs/2019MNRAS.483.4140R}{483, 4140}

\bibitem[{{Roshan} {et~al.}(2021{\natexlab{a}}){Roshan}, {Banik}, {Ghafourian},
  {Thies}, {Famaey}, {Asencio}, \& {Kroupa}}]{Roshan_2021}
{Roshan}, M., {Banik}, I., {Ghafourian}, N., {et~al.} 2021{\natexlab{a}},
  \href{http://dx.doi.org/10.1093/mnras/stab651}{\color{magenta}\mnras},
  \href{https://ui.adsabs.harvard.edu/abs/2021MNRAS.tmp..658R}{503, 2833}

\bibitem[{{Roshan} {et~al.}(2021{\natexlab{b}}){Roshan}, {Ghafourian},
  {Kashfi}, {Banik}, {Haslbauer}, {Cuomo}, {Famaey}, \&
  {Kroupa}}]{Roshan_2021b}
{Roshan}, M., {Ghafourian}, N., {Kashfi}, T., {et~al.} 2021{\natexlab{b}},
  \href{http://dx.doi.org/10.1093/mnras/stab2553}{\color{magenta}\mnras},
  \href{https://ui.adsabs.harvard.edu/abs/2021MNRAS.508..926R}{508, 926}

\bibitem[{{Salim} {et~al.}(2007){Salim}, {Rich}, {Charlot}, {Brinchmann},
  {Johnson}, {Schiminovich}, {Seibert}, {Mallery}, {Heckman}, {Forster},
  {Friedman}, {Martin}, {Morrissey}, {Neff}, {Small}, {Wyder}, {Bianchi},
  {Donas}, {Lee}, {Madore}, {Milliard}, {Szalay}, {Welsh}, \&
  {Yi}}]{Salim_2007}
{Salim}, S., {Rich}, R.~M., {Charlot}, S., {et~al.} 2007,
  \href{http://dx.doi.org/10.1086/519218}{\color{magenta}\apjs},
  \href{https://ui.adsabs.harvard.edu/abs/2007ApJS..173..267S}{173, 267}

\bibitem[{{Santini} {et~al.}(2009){Santini}, {Fontana}, {Grazian}, {Salimbeni},
  {Fiore}, {Fontanot}, {Boutsia}, {Castellano}, {Cristiani}, {de Santis},
  {Gallozzi}, {Giallongo}, {Menci}, {Nonino}, {Paris}, {Pentericci}, \&
  {Vanzella}}]{Santini_2009}
{Santini}, P., {Fontana}, A., {Grazian}, A., {et~al.} 2009,
  \href{http://dx.doi.org/10.1051/0004-6361/200811434}{\color{magenta}\aap},
  \href{https://ui.adsabs.harvard.edu/abs/2009A\&A...504..751S}{504, 751}

\bibitem[{{Scannapieco} {et~al.}(2012){Scannapieco}, {Wadepuhl}, {Parry},
  {Navarro}, {Jenkins}, {Springel}, {Teyssier}, {Carlson}, {Couchman}, {Crain},
  {Dalla Vecchia}, {Frenk}, {Kobayashi}, {Monaco}, {Murante}, {Okamoto},
  {Quinn}, {Schaye}, {Stinson}, {Theuns}, {Wadsley}, {White}, \&
  {Woods}}]{Scannapieco_2012}
{Scannapieco}, C., {Wadepuhl}, M., {Parry}, O.~H., {et~al.} 2012,
  \href{http://dx.doi.org/10.1111/j.1365-2966.2012.20993.x}{\color{magenta}\mnras},
  \href{https://ui.adsabs.harvard.edu/abs/2012MNRAS.423.1726S}{423, 1726}

\bibitem[{{Schaye} {et~al.}(2015){Schaye}, {Crain}, {Bower}, {Furlong},
  {Schaller}, {Theuns}, {Dalla Vecchia}, {Frenk}, {McCarthy}, {Helly},
  {Jenkins}, {Rosas-Guevara}, {White}, {Baes}, {Booth}, {Camps}, {Navarro},
  {Qu}, {Rahmati}, {Sawala}, {Thomas}, \& {Trayford}}]{Schaye_2015}
{Schaye}, J., {Crain}, R.~A., {Bower}, R.~G., {et~al.} 2015,
  \href{http://dx.doi.org/10.1093/mnras/stu2058}{\color{magenta}\mnras},
  \href{https://ui.adsabs.harvard.edu/abs/2015MNRAS.446..521S}{446, 521}

\bibitem[{{SDSS Collaboration}(2000)}]{York_2000}
{SDSS Collaboration}. 2000,
  \href{http://dx.doi.org/10.1086/301513}{\color{magenta}\aj},
  \href{https://ui.adsabs.harvard.edu/abs/2000AJ....120.1579Y}{120, 1579}

\bibitem[{{Sellwood} {et~al.}(2019){Sellwood}, {Shen}, \& {Li}}]{Sellwood_2019}
{Sellwood}, J.~A., {Shen}, J., \& {Li}, Z. 2019,
  \href{http://dx.doi.org/10.1093/mnras/stz1145}{\color{magenta}\mnras},
  \href{https://ui.adsabs.harvard.edu/abs/2019MNRAS.486.4710S}{486, 4710}

\bibitem[{{Springel}(2005)}]{Springel_2005}
{Springel}, V. 2005,
  \href{http://dx.doi.org/10.1111/j.1365-2966.2005.09655.x}{\color{magenta}\mnras},
  \href{https://ui.adsabs.harvard.edu/abs/2005MNRAS.364.1105S}{364, 1105}

\bibitem[{{Springel}(2010)}]{Springel_2010}
{Springel}, V. 2010,
  \href{http://dx.doi.org/10.1111/j.1365-2966.2009.15715.x}{\color{magenta}\mnras},
  \href{https://ui.adsabs.harvard.edu/abs/2010MNRAS.401..791S}{401, 791}

\bibitem[{{Springel} {et~al.}(2005){Springel}, {White}, {Jenkins}, {Frenk},
  {Yoshida}, {Gao}, {Navarro}, {Thacker}, {Croton}, {Helly}, {Peacock}, {Cole},
  {Thomas}, {Couchman}, {Evrard}, {Colberg}, \&
  {Pearce}}]{Springel_2005_Millennium}
{Springel}, V., {White}, S.~D.~M., {Jenkins}, A., {et~al.} 2005,
  \href{http://dx.doi.org/10.1038/nature03597}{\color{magenta}\nat},
  \href{http://adsabs.harvard.edu/abs/2005Natur.435..629S}{435, 629}

\bibitem[{{Stewart} {et~al.}(2008){Stewart}, {Bullock}, {Wechsler}, {Maller},
  \& {Zentner}}]{Stewart_2008}
{Stewart}, K.~R., {Bullock}, J.~S., {Wechsler}, R.~H., {Maller}, A.~H., \&
  {Zentner}, A.~R. 2008,
  \href{http://dx.doi.org/10.1086/588579}{\color{magenta}\apj},
  \href{https://ui.adsabs.harvard.edu/abs/2008ApJ...683..597S}{683, 597}

\bibitem[{{Tamburri} {et~al.}(2014){Tamburri}, {Saracco}, {Longhetti},
  {Gargiulo}, {Lonoce}, \& {Ciocca}}]{Tamburri_2014}
{Tamburri}, S., {Saracco}, P., {Longhetti}, M., {et~al.} 2014,
  \href{http://dx.doi.org/10.1051/0004-6361/201424040}{\color{magenta}\aap},
  \href{https://ui.adsabs.harvard.edu/abs/2014A\&A...570A.102T}{570, A102}

\bibitem[{{Taylor} {et~al.}(2011){Taylor}, {Hopkins}, {Baldry}, {Brown},
  {Driver}, {Kelvin}, {Hill}, {Robotham}, {Bland-Hawthorn}, {Jones}, {Sharp},
  {Thomas}, {Liske}, {Loveday}, {Norberg}, {Peacock}, {Bamford}, {Brough},
  {Colless}, {Cameron}, {Conselice}, {Croom}, {Frenk}, {Gunawardhana},
  {Kuijken}, {Nichol}, {Parkinson}, {Phillipps}, {Pimbblet}, {Popescu},
  {Prescott}, {Sutherland}, {Tuffs}, {van Kampen}, \&
  {Wijesinghe}}]{Taylor_2011}
{Taylor}, E.~N., {Hopkins}, A.~M., {Baldry}, I.~K., {et~al.} 2011,
  \href{http://dx.doi.org/10.1111/j.1365-2966.2011.19536.x}{\color{magenta}\mnras},
  \href{https://ui.adsabs.harvard.edu/abs/2011MNRAS.418.1587T}{418, 1587}

\bibitem[{{Teyssier}(2002)}]{Teyssier_2002}
{Teyssier}, R. 2002,
  \href{http://dx.doi.org/10.1051/0004-6361:20011817}{\color{magenta}\aap},
  \href{https://ui.adsabs.harvard.edu/abs/2002A\&A...385..337T}{385, 337}

\bibitem[{{Thob} {et~al.}(2019){Thob}, {Crain}, {McCarthy}, {Schaller},
  {Lagos}, {Schaye}, {Talens}, {James}, {Theuns}, \& {Bower}}]{Thob_2019}
{Thob}, A. C.~R., {Crain}, R.~A., {McCarthy}, I.~G., {et~al.} 2019,
  \href{http://dx.doi.org/10.1093/mnras/stz448}{\color{magenta}\mnras},
  \href{https://ui.adsabs.harvard.edu/abs/2019MNRAS.485..972T}{485, 972}

\bibitem[{{Torrey} {et~al.}(2015){Torrey}, {Snyder}, {Vogelsberger}, {Hayward},
  {Genel}, {Sijacki}, {Springel}, {Hernquist}, {Nelson}, {Kriek}, {Pillepich},
  {Sales}, \& {McBride}}]{Torrey_2015}
{Torrey}, P., {Snyder}, G.~F., {Vogelsberger}, M., {et~al.} 2015,
  \href{http://dx.doi.org/10.1093/mnras/stu2592}{\color{magenta}\mnras},
  \href{https://ui.adsabs.harvard.edu/abs/2015MNRAS.447.2753T}{447, 2753}

\bibitem[{{Trayford} {et~al.}(2017){Trayford}, {Camps}, {Theuns}, {Baes},
  {Bower}, {Crain}, {Gunawardhana}, {Schaller}, {Schaye}, \&
  {Frenk}}]{Trayford_2017}
{Trayford}, J.~W., {Camps}, P., {Theuns}, T., {et~al.} 2017,
  \href{http://dx.doi.org/10.1093/mnras/stx1051}{\color{magenta}\mnras},
  \href{https://ui.adsabs.harvard.edu/abs/2017MNRAS.470..771T}{470, 771}

\bibitem[{{van de Sande} {et~al.}(2019){van de Sande}, {Lagos}, {Welker},
  {Bland-Hawthorn}, {Schulze}, {Remus}, {Bah{\'e}}, {Brough}, {Bryant},
  {Cortese}, {Croom}, {Devriendt}, {Dubois}, {Goodwin}, {Konstantopoulos},
  {Lawrence}, {Medling}, {Pichon}, {Richards}, {Sanchez}, {Scott}, \&
  {Sweet}}]{vandenSande_2019}
{van de Sande}, J., {Lagos}, C. D.~P., {Welker}, C., {et~al.} 2019,
  \href{http://dx.doi.org/10.1093/mnras/sty3506}{\color{magenta}\mnras},
  \href{https://ui.adsabs.harvard.edu/abs/2019MNRAS.484..869V}{484, 869}

\bibitem[{{van den Bosch}(2001)}]{vandenBosch_2001}
{van den Bosch}, F.~C. 2001,
  \href{http://dx.doi.org/10.1046/j.1365-8711.2001.04861.x}{\color{magenta}\mnras},
  \href{https://ui.adsabs.harvard.edu/abs/2001MNRAS.327.1334V}{327, 1334}

\bibitem[{{V{\'a}zquez-Mata} {et~al.}(2020){V{\'a}zquez-Mata}, {Loveday},
  {Riggs}, {Baldry}, {Davies}, {Robotham}, {Holwerda}, {Brown}, {Cluver},
  {Wang}, {Alpaslan}, {Bland-Hawthorn}, {Brough}, {Driver}, {Hopkins},
  {Taylor}, \& {Wright}}]{VazquezMata_2020}
{V{\'a}zquez-Mata}, J.~A., {Loveday}, J., {Riggs}, S.~D., {et~al.} 2020,
  \href{http://dx.doi.org/10.1093/mnras/staa2889}{\color{magenta}\mnras},
  \href{https://ui.adsabs.harvard.edu/abs/2020MNRAS.499..631V}{499, 631}

\bibitem[{{Vogelsberger} {et~al.}(2014){Vogelsberger}, {Genel}, {Springel},
  {Torrey}, {Sijacki}, {Xu}, {Snyder}, {Bird}, {Nelson}, \&
  {Hernquist}}]{Vogelsberger_2014}
{Vogelsberger}, M., {Genel}, S., {Springel}, V., {et~al.} 2014,
  \href{http://dx.doi.org/10.1038/nature13316}{\color{magenta}\nat},
  \href{http://adsabs.harvard.edu/abs/2014Natur.509..177V}{509, 177}

\bibitem[{{Wetzel} {et~al.}(2016){Wetzel}, {Hopkins}, {Kim},
  {Faucher-Gigu{\`e}re}, {Kere{\v{s}}}, \& {Quataert}}]{Wetzel_2016}
{Wetzel}, A.~R., {Hopkins}, P.~F., {Kim}, J.-h., {et~al.} 2016,
  \href{http://dx.doi.org/10.3847/2041-8205/827/2/L23}{\color{magenta}\apjl},
  \href{https://ui.adsabs.harvard.edu/abs/2016ApJ...827L..23W}{827, L23}

\bibitem[{{Wittenburg} {et~al.}(2020){Wittenburg}, {Kroupa}, \&
  {Famaey}}]{Wittenburg_2020}
{Wittenburg}, N., {Kroupa}, P., \& {Famaey}, B. 2020,
  \href{http://dx.doi.org/10.3847/1538-4357/ab6d73}{\color{magenta}\apj},
  \href{https://ui.adsabs.harvard.edu/abs/2020ApJ...890..173W}{890, 173}

\bibitem[{{Wu} \& {Kroupa}(2015)}]{Wu_2015}
{Wu}, X. \& {Kroupa}, P. 2015,
  \href{http://dx.doi.org/10.1093/mnras/stu2099}{\color{magenta}\mnras},
  \href{https://ui.adsabs.harvard.edu/abs/2015MNRAS.446..330W}{446, 330}

\end{thebibliography}

\end{document}